\documentclass[letterpaper,11pt]{article}
\pdfoutput=1
\usepackage{jheppub}
\usepackage{multirow}
\usepackage{shuffle}

\usepackage{breqn}

\usepackage{subfig}

\usepackage{xspace}
\usepackage[countmax]{subfloat}
\usepackage{enumitem}
\usepackage{amssymb}
\usepackage{amsmath}
\usepackage{cancel}
\usepackage{color}
\usepackage{braket}
\usepackage{graphicx}
\usepackage{multirow}
\usepackage{verbatim}
\usepackage{amsthm}
\usepackage{slashed}
\usepackage{wasysym}
\usepackage{simplewick}
\usepackage{mathtools}
\usepackage{soul}
\usepackage{xspace}

\usepackage{mathrsfs}

\def\beq{\begin{equation}}
\def\eeq{\end{equation}}
\def\bea{\begin{eqnarray}}
\def\eea{\end{eqnarray}}

\def\cC{\mathcal{C}} 

\def\cE{\mathcal{E}}

\def\cG{\mathcal{G}}

\def\cL{\mathcal{L}}

\def\cN{\mathcal{N}}
\def\cO{\mathcal{O}}

\def\nn{{\nonumber}}

\newcommand{\Eq}[1]{Equation~\eqref{#1}}

\DeclareRobustCommand{\Sec}[1]{Sec.~\ref{#1}}

\DeclareRobustCommand{\Fig}[1]{Fig.~\ref{#1}}

\DeclareRobustCommand{\Eq}[1]{Eq.~(\ref{#1})}

\def\be{\begin{equation}}
\def\ee{\end{equation}}

\newcommand{\vev}[1]{\langle #1 \rangle}

\newcommand{\mae}[3]{\langle#1\rvert#2\rvert#3\rangle}

\newcommand{\eps}{\epsilon}

  \newcount\hour \newcount\minute
  \hour=\time \divide \hour by 60 \minute=\time
  \newcommand{\todaytime}{\today \ -- \number\hour :\ifnum \minute<10 0\fi\number\minute}
\def\eps{\epsilon}

\def\spa#1.#2{\left\langle#1\,#2\right\rangle}
\def\spb#1.#2{\left[#1\,#2\right]}

\def\feynsl#1{
  \setbox0=\hbox{/} \setbox1=\hbox{$#1$}
  \dimen0=\wd0 \advance\dimen0 by -\wd1 \divide\dimen0 by 2
  \ifdim\wd0>\wd1 \raise.15ex\copy0\kern-\wd0\kern\dimen0\copy1\kern\dimen0
  \else \kern-\dimen0\raise.15ex\copy0\kern-\dimen0\kern-\wd1\copy1\fi}

\newskip\humongous \humongous=0pt plus 1000pt minus 100pt

\newif\ifdtup

\def\beq{\begin{equation}}
\def\eeq{\end{equation}}
\def\beeq{\begin{eqnarray}}
\def\eeeq{\end{eqnarray}}

\def\Je2e{J_{E^2 E}}

\newcommand{\zb}{\bar{z}}

\newcommand{\xb}{\bar{x}}

\newcommand{\rb}{\rangle}
\newcommand{\lb}{\langle}

\newcommand{\dDisc}{\text{dDisc}}
\newcommand{\inv}{\text{INV}}

\usepackage{xcolor}

\DeclareMathOperator*{\sumint}{%
\mathchoice%
  {\ooalign{$\displaystyle\sum$\cr\hidewidth$\displaystyle\int$\hidewidth\cr}}
  {\ooalign{\raisebox{.14\height}{\scalebox{.7}{$\textstyle\sum$}}\cr\hidewidth$\textstyle\int$\hidewidth\cr}}
  {\ooalign{\raisebox{.2\height}{\scalebox{.6}{$\scriptstyle\sum$}}\cr$\scriptstyle\int$\cr}}
  {\ooalign{\raisebox{.2\height}{\scalebox{.6}{$\scriptstyle\sum$}}\cr$\scriptstyle\int$\cr}}
}
\title{Celestial Blocks and Transverse Spin in the\\ Three-Point Energy Correlator}

\author[1]{Hao Chen,}
\author[2]{Ian Moult,}
\author[3]{Joshua Sandor,}
\author[1]{and Hua Xing Zhu}

\affiliation[1]{Zhejiang Institute of Modern Physics, Department of Physics, Zhejiang University, Hangzhou, Zhejiang 310027, China}
\affiliation[2]{Department of Physics, Yale University, New Haven, CT 06511, USA\vspace{0.5ex}}
\affiliation[3]{Stanford Institute for Theoretical Physics, Stanford University, Stanford, CA 94305, USA\vspace{0.5ex}}
\emailAdd{chenhao201224@zju.edu.cn}
\emailAdd{ian.moult@yale.edu}
\emailAdd{jsandor@stanford.edu}
\emailAdd{zhuhx@zju.edu.cn}

\abstract{Quantitative theoretical techniques for understanding the substructure of jets at the LHC enable new insights into the dynamics of QCD, and novel approaches to search for new physics.
Recently, there has been a program to reformulate jet substructure in terms of correlation functions, $\langle \cE(\vec n_1) \cE(\vec n_2) \cdots \cE(\vec n_k) \rangle$, of light-ray operators, $\cE(\vec n)$, allowing the application of techniques developed in the study of Conformal Field Theories (CFTs).
In this paper we further develop these techniques in the particular context of the three-point correlator $\langle \cE(\vec n_1) \cE(\vec n_2) \cE(\vec n_3) \rangle$, using recently computed perturbative data in both QCD and $\cN=4$ sYM.
We derive the celestial blocks appearing in the light-ray operator product expansion (OPE) of the three-point correlator,  and use the Lorentzian inversion formula to extract the spectrum of light-ray operators appearing in the expansion, showing, in particular, that the OPE data is analytic in transverse spin.  
Throughout our presentation, we highlight the relation between the OPE approach, and more standard splitting function based approaches of perturbative QCD, emphasizing the utility of the OPE approach for incorporating symmetries in jet substructure calculations.
We hope that our presentation introduces a number of new techniques to the jet substructure community, and also illustrates the phenomenological relevance of the study of light-ray operators in the OPE limit to the CFT community.
 }
\begin{document} 

\maketitle

\newpage

%%%%%%%%%%%%%%%%%%%%%%%%%%%%%%%%%%%%%%%%%%%%%%%%%%%%%%%%%%%%%%%%%%%%%%%%%%%%%%%%
\section{Introduction}\label{sec:intro}
%%%%%%%%%%%%%%%%%%%%%%%%%%%%%%%%%%%%%%%%%%%%%%%%%%%%%%%%%%%%%%%%%%%%%%%%%%%%%%%%

At collider experiments such as the LHC, the microscopic physics of the underlying collision is encoded in the pattern of radiation in the detectors at infinity. In the last decade there has been tremendous progress in our ability to exploit this information, driven by the introduction of robust jet algorithms \cite{Cacciari:2005hq,Cacciari:2008gp,Cacciari:2011ma}, which has given rise to an emerging field referred to as jet substructure.\footnote{For early applications see \cite{Butterworth:2008iy,Kaplan:2008ie,Krohn:2009th}, for reviews see \cite{Larkoski:2017jix,Marzani:2019hun,Cunqueiro:2021wls}} Jet substructure has provided many new ways to search for potential new physics, as well as to study the dynamics of QCD at high energies. 

From a field theoretic perspective, jet substructure is the study of correlation functions, $\langle \cE(\vec n_1) \cE(\vec n_2) \cdots \cE(\vec n_k) \rangle$, of a specific class of light-ray operators, known as energy flow operators\footnote{More generally, charges other than energy flow can be studied (see e.g. \cite{Hofman:2008ar,Belitsky:2013xxa,Li:2021zcf,Chicherin:2020azt}). However, particularly for applications to QCD, one typically restricts to energy flow since it can be computed in perturbation theory.}
\cite{Sveshnikov:1995vi,Tkachov:1995kk,Korchemsky:1999kt,Bauer:2008dt,Hofman:2008ar,Belitsky:2013xxa,Belitsky:2013bja,Kravchuk:2018htv}
\begin{align}\label{eq:ANEC_op}
\mathcal{E}(\vec n) = \lim_{r\to \infty}  \int\limits_0^\infty dt~ r^2 n^i T_{0i}(t,r \vec n)\,,
\end{align}
in the limit that all the light-ray operators are nearly collinear, namely $\vec{n}_i \cdot \vec{n}_j  \to  1$. Physically, this corresponds to the limit where all the correlators lie inside a single high energy jet, as shown schematically in \Fig{fig:intro_a}. Throughout the text we will interchangeably refer to this limit as the collinear limit, the operator product expansion (OPE) limit or the jet substructure limit. Recently there has been a program \cite{Chen:2020vvp} to reformulate questions of physical interest in jet substructure purely in terms of these correlation functions, as opposed to more traditional jet shapes.\footnote{Note that jet shapes (even those with similar names to the energy correlators \cite{Larkoski:2013eya,Larkoski:2014gra,Moult:2016cvt}) cannot be expressed as a finite sum of $n$-point correlators, but are instead infinite sums over all $n$-point correlators \cite{Chen:2020vvp}. This is easily understood, since they are not local on the celestial sphere. This makes jet shapes significantly more complex than their correlator counterparts. In particular, the techniques of this paper utilize the specific symmetry properties of correlators for a fixed value of $n$, which are mixed together for the case of jet shapes.} This has lead to many exciting developments, including factorization theorems describing the behavior in the collinear limits \cite{Dixon:2019uzg,Chen:2020vvp}, the calculation of multi-point correlators \cite{Chen:2019bpb}, calculations incorporating tracking information \cite{Chen:2020vvp,Li:2021zcf,Jaarsma:2022kdd}, the study of spin interference effects \cite{Chen:2020adz,Chen:2021gdk}, applications to top quark mass measurements \cite{Holguin:2022epo}, and the first analysis of these correlators on real LHC data \cite{Komiske:2022enw}!\footnote{For applications of the energy correlators to the Sudakov limit, see \cite{Moult:2018jzp,Gao:2019ojf,Moult:2019vou,Ebert:2020sfi,Li:2021txc,Li:2020bub}.} This wealth of phenomenological applications motivates further developing the theoretical understanding of these correlation functions.

Apart from the phenomenological interest for understanding the dynamics of jets at the LHC, the study of light-ray operators in the OPE limit is also of theoretical interest, since one generically expects universal behavior as operators are brought together, as described by an OPE \cite{Wilson:1969zs,Kadanoff:1969zz}. However, unlike for the well studied case of the OPE local operators, the case of interest relevant for understanding jet substructure is the OPE of non-local light-ray operators. See \Fig{fig:intro_b} for a Penrose diagram of the spacetime structure of the measurement of three light-ray operators. In \cite{Hofman:2008ar}, it was argued that an OPE for light-ray operators should exist, and most excitingly, that one could understand the dynamics of jets purely from  the structure of this OPE.\footnote{For early applications of light-ray operators in QCD, see \cite{Balitsky:1987bk,Balitsky:1988fi,Balitsky:1990ck}.} This OPE was systematized into a rigorous expansion in \cite{Kologlu:2019mfz,1822249}. In particular, \cite{1822249} emphasized the important role of transverse spin in the OPE, which will be central to this paper. From a phenomenological perspective, the light-ray OPE is particularly interesting in that it opens the door to the use of many techniques used to study correlation functions (conformal blocks, Casimir differential equations, Lorentzian inversion, ...) for answering questions of phenomenological importance in jet substructure. 

Unlike the Euclidean OPE of local operators, the light-ray OPE is much less well understood, and also much less well tested. This is in part due to a lack of concrete perturbative data for higher point correlation functions of light-ray operators. While the two-point correlator is known analytically to three-loop order in $\cN=4$ super Yang-Mills~\cite{Belitsky:2013ofa,Henn:2019gkr} and two-loop order in QCD~\cite{Dixon:2018qgp,Luo:2019nig,Gao:2020vyx}, and has provided a useful testing ground, higher point correlators are particularly interesting since they probe a wider range of light-ray operators, in particular those with transverse spin. 

Recently, the perturbative three-point correlator of energy flow operators, was computed analytically in both QCD and $\cN=4$ sYM in  \cite{Chen:2019bpb} at leading order in the collinear limit, and was shown to take a simple form, expressed in terms of two conformal cross ratios $G(z, \bar z)$. The leading twist analysis of the OPE including logarithmic resummation for this correlator was performed in \cite{Chen:2021gdk}, and has also been used to study transverse spin effects in \cite{Chen:2020adz}. This has enabled, for example, tests of transverse spin in parton showers \cite{Karlberg:2021kwr}.

Motivated by the desire to better understand the structure of the light-ray OPE, and ultimately to allow it to be used as a practical tool for calculations in QCD, in this paper we analyze the perturbative data for the three-point correlator from \cite{Chen:2019bpb} from the perspective of the light-ray OPE.  The goal of this paper is two-fold. First, for an audience that is familiar with the perturbative calculation of jet substructure observables, we wish to show that formulating questions in terms of energy correlators opens the door to a number of sophisticated techniques developed in the context of conformal field theories (CFTs), and we wish to introduce these techniques in a concrete perturbative setting, and show how they interplay with standard perturbative calculations. This includes in particular the use of celestial blocks for incorporating symmetries \cite{Kologlu:2019mfz}, Casimir equations \cite{Dolan:2003hv}, and the Lorentzian inversion formula \cite{Caron-Huot:2017vep,Simmons-Duffin:2017nub}. We believe that these techniques can have a much broader impact in perturbative QCD calculations, once questions of phenomenological interest are rephrased in a language in which they can be applied.\footnote{We note that of course the use of conformal symmetry has a long history in QCD (see e.g. \cite{Braun:2003rp} for a review, and \cite{Braun:2020zjm} recent applications), with very earlier applications to ``jets" \cite{Polyakov:1970lyy,Polyakov:1971gx}, but unfortunately has been less applied in jet physics/jet substruture in modern times.} For the CFT audience that is already familiar with these techniques, we want to highlight why the ``jet substructure" limit is of particular phenomenological interest, how the light-ray OPE relates to the perturbative language of splitting functions, and how perturbative calculations can be used to investigate the structure of the light-ray OPE. 

Although techniques for perturbative calculations in QCD are extremely advanced, one aspect of these calculations that has received less attention, particularly in jet substructure, is the use of symmetries. This is where we believe the techniques developed in the context of CFT, where exploiting symmetries to maximal impact has been the central focus since very early on \cite{Ferrara:1973yt,Polyakov:1974gs}, can have the most impact.\footnote{Due to the rapid recent progress in the conformal bootstrap, there exist many excellent reviews on the topic. See e.g. \cite{Rychkov:2016iqz,Poland:2016chs,Simmons-Duffin:2016gjk,Poland:2018epd}} In particular, in analogy with multi-point correlation functions of local operators, multi-point correlation functions of light-ray operators can be expanded in conformal blocks on the celestial sphere, which are referred to as celestial blocks \cite{Kologlu:2019mfz}.  In this paper we derive the relevant celestial blocks for the three-point function, and show that in the collinear limit, these are simply conformal blocks of two-dimensional Euclidean CFT.  We derive this result from two-distinct perspectives to clarify the relation between the standard perturbative QCD language, and the language more typically used in CFTs. We attempt to provide a pedagogical introduction to the techniques used to derive these celestial blocks, with the hope that they can be more widely used in perturbative QCD. We then show in simple cases where the intermediate states in the OPE are perturbative quark or gluon states, how the OPE approach relates to a more standard splitting function approach, and how the celestial blocks incorporate higher power corrections from expanding perturbative propagators.

Due to the simple structure of the three-point energy correlator in the collinear limit, in particular the fact that it behaves mathematically as a four point function of local operators, we are able to directly use the Lorentzian inversion formula \cite{Caron-Huot:2017vep} to extract the spectrum of higher twist light-ray operators, providing interesting insight into the higher twist structure of the collinear limit. Furthermore, quite remarkably, this shows that the OPE data for the three-point correlator is analytic in transverse spin, much in analogy with analyticity in spin found in studying correlation functions of local operators. This further highlights the crucial role that  tranvserse spin plays in the consistency of the light-ray OPE, as was emphasized in \cite{1822249} in the two-point correlator case, and also motivates further studies of the high transverse-spin limit, generalizing the by now familiar studies of the high-spin limit.

Finally, we briefly discuss in the simplified setting of $\phi^4$ theory some interesting features of the light-ray OPE in perturbation theory that first appear in the three-point correlator. In particular, we carefully track the origin of logarithms and derivatives of celestial blocks appearing in the twist expansion, and show that they are related to infrared behavior (zero-modes) in perturbation theory. This is in contrast to the appearance of logarithms and derivatives of conformal blocks in the perturbative expansion of correlation functions of local operators. This provides some further intuition for the behavior of the light-ray OPE in perturbation theory. We illustrate this using a particularly simple example of a number correlator, which we believe will be useful for future concrete studies of subleading twist light-ray operators.

An outline of this paper is as follows. In \Sec{sec:pQCD} we provide a review of energy correlator observables intended for those with a background in perturbative QCD. In particular, we show how questions in jet substructure can be rephrased in the language of correlation functions, how these can be computed perturbatively from splitting functions, and we highlight the interesting symmetry structure of the three-point correlator that will be explained in the rest of the paper. For those with a CFT background, we hope that this section can provide some motivation as to why techniques for studying the OPE limit of light-ray operators have more general phenomenological applicability, including in the complicated LHC environment.
 In \Sec{sec:jss_limit} we provide a review of light-ray operators and the light-ray OPE from a perturbative perspective, highlighting the symmetry properties of the light-ray operators, as well as the connection between calculations using the light-ray OPE, and the perturbative calculations of \Sec{sec:pQCD}.
 In \Sec{sec:blocks} we then derive the relevant celestial blocks describing the OPE of the three-point energy correlator. This is done pedagogically from two different perspectives to highlight the relation between the language used in the perturbative QCD literature and that used in the CFT literature.
 In \Sec{sec:spin2} we illustrate these celestial blocks in a particularly simple example where the OPE limit isolates a physical perturbative state, namely a gluon, to make clear the relation to the factorization onto collinear splitting functions, as well as the relationship between conformal blocks and kinematic power corrections. 
In \Sec{sec:analyticity} we introduce the technique of the Lorentzian inversion formula to QCD audiences, and show how it enables us to derive the structure of higher twist contributions in the collinear limit. For the CFT audience, the main result of this section is that we show that the OPE data for correlation functions of light-ray operators (at least at this order in perturbation theory) is analytic in transverse spin.
In \Sec{sec:QCD_application}  we numerically study the convergence of the celestial block expansion throughout the phase space of the leading order~(LO) three-point correlator in QCD, and find that a few orders in the twist expansion provides a good approximation throughout most of the phase space.
In \Sec{sec:zero} we briefly discuss some issues associated with zero-modes, and their appearance in the OPE limit. 
We conclude in \Sec{sec:conc}.

\vspace{0.5cm}

{\bf{Note Added:}} This paper will appear simultaneously with a paper by Cyuan-Han Chang and David Simmons-Duffin, that also studies the structure of the three-point correlator of light-ray operators from the perspective of the light-ray OPE. We thank these authors for coordination of the submission.

%%%%%%%%%%%%%%%%%%%%%%%%%%%%%%%%%%%%%%%%%%%%%%%%%%%%%%%%%%%%%%%%%%%%%%%%%%%%%%%%
\section{Energy Correlators from a Perturbative QCD Perspective}\label{sec:pQCD}
%%%%%%%%%%%%%%%%%%%%%%%%%%%%%%%%%%%%%%%%%%%%%%%%%%%%%%%%%%%%%%%%%%%%%%%%%%%%%%%%

We begin by reviewing energy correlators from the perspective of perturbative QCD calculations. As mentioned above, the goal of this section is two-fold. First, for those familiar with standard jet substructure calculations in perturbative QCD, we wish to highlight the simplicity of the energy correlator observables, and their relation to integrals of the splitting functions which can be efficiently computed using modern integration techniques.   We will then show that their expansion in squeezed limits have interesting symmetries that are hidden from the splitting function perspective, motivating studying the symmetry properties of these correlators at the operator level.
Second, for those familiar with the use of symmetry based techniques for studying correlation functions, we wish to emphasize why these techniques can be applied in the case of jet substructure at the LHC, which is naively complicated by the use of jet algorithms and a complicated initial state.

The energy correlators are ensemble averaged observables, $\langle \Psi | \cE(\vec n_1) \cE(\vec n_2) \cdots \cE(\vec n_k)| \Psi \rangle$, which are functions of the angles between the energy flow operators (equivalently distances on the celestial sphere)
\begin{align}
\zeta_{ij}=\frac{1-\cos \theta_{ij}}{2}\,.
\end{align} 
They were first introduced to study jets in $e^+e^-$ colliders \cite{Basham:1979gh,Basham:1978zq,Basham:1978bw,Basham:1977iq}, with a particular focus on the one- and two-point correlators.  The case of $e^+e^-$ colliders is the simplest theoretically, since the state $\Psi$ is produced by a local operator. After a gap of 30 years, interest in the energy correlators was rejuvenated by the work of Hofman and Maldacena \cite{Hofman:2008ar}, which highlighted the interesting properties of these observables in CFTs, and their relation to (the light transform) of correlation functions of local operators.\footnote{This paper also highlighted the interesting role of Lorentzian observables in constraining the space CFTs, known as the ``conformal collider bounds" \cite{Hofman:2008ar,Hofman:2016awc,Cordova:2017zej}. } This led to significant progress in the understanding of these observables in CFTs, primarily focusing on the case where they are measured on states produced by local operators \cite{Belitsky:2014zha,Korchemsky:2015ssa,Belitsky:2013xxa,Belitsky:2013bja,Belitsky:2013ofa,Chicherin:2020azt,Henn:2019gkr,Kologlu:2019bco,Kologlu:2019mfz,1822249,Korchemsky:2019nzm,Korchemsky:2021okt,Korchemsky:2021htm,Chicherin:2020azt}. For a recent study of an energy flow operator in a bi-local state, see \cite{Poland:2021xjs}. 

\begin{figure}
%[htp]
\begin{center}
\subfloat[]{
\includegraphics[scale=0.26]{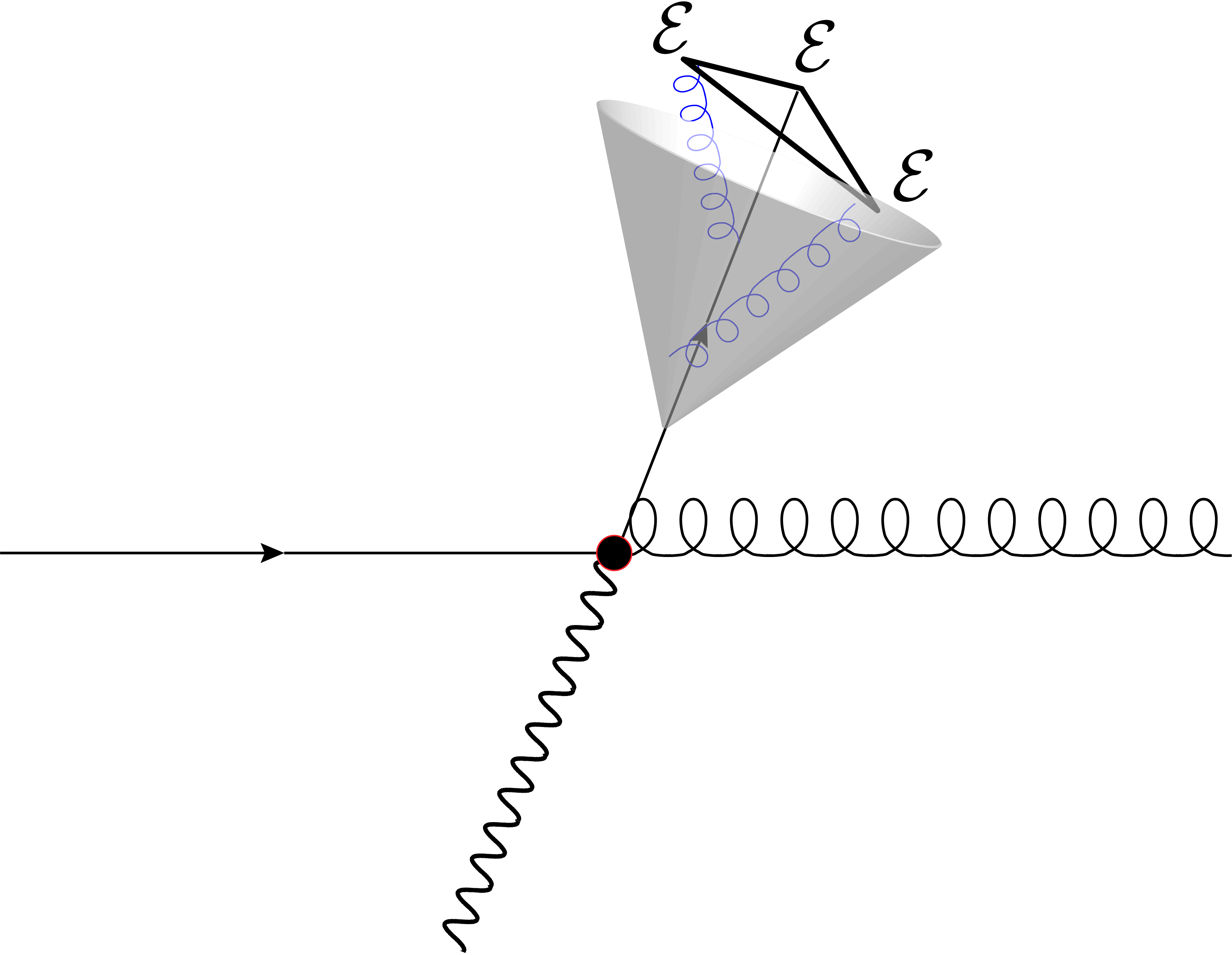}\label{fig:intro_a}
}
\subfloat[]{
\includegraphics[scale=0.45]{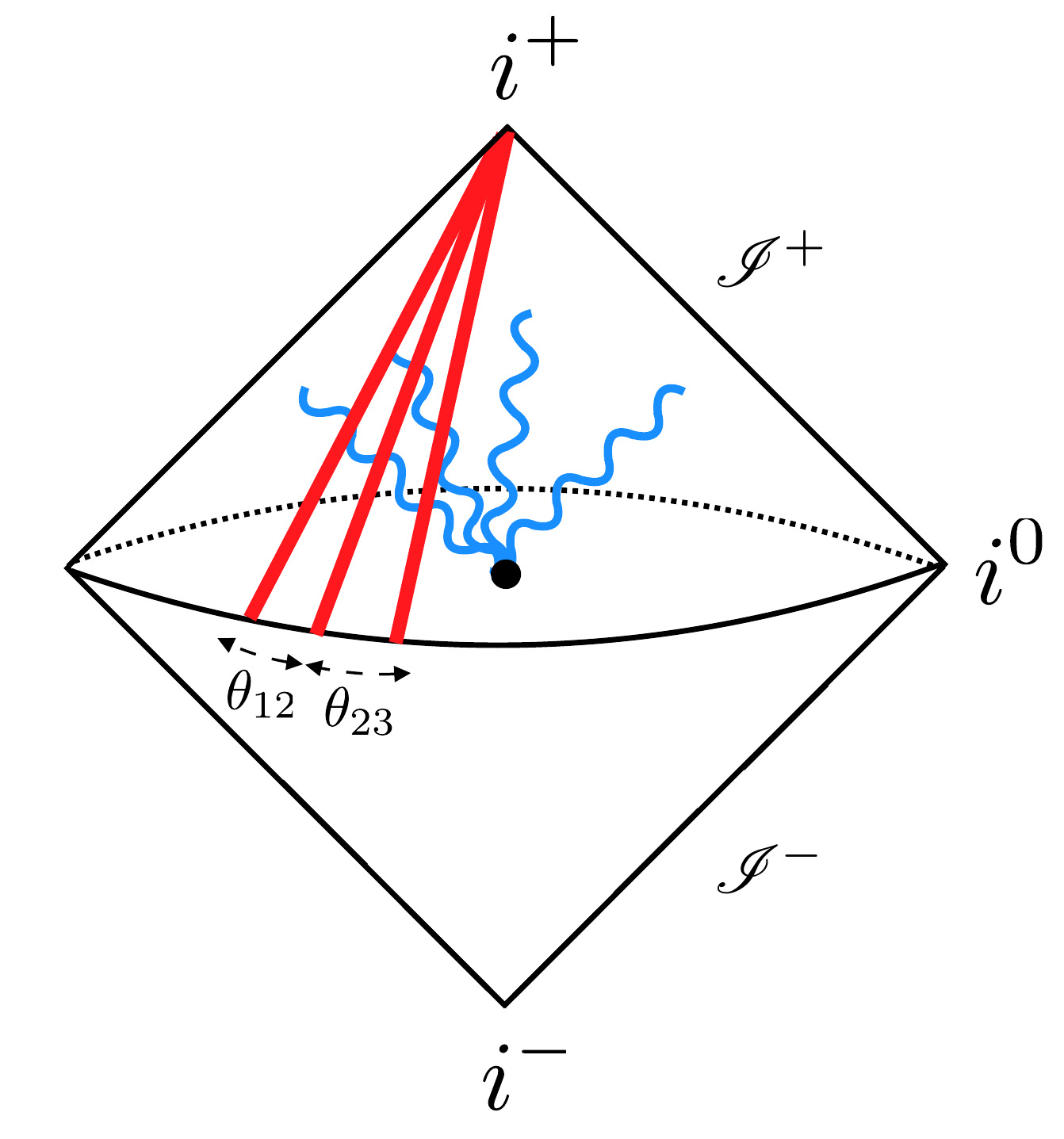}\label{fig:intro_b}
}\qquad
%\captionsetup{font={footnotesize}}
\end{center}
\caption{(a) Energy correlators measured inside an identified jet at the LHC. (b) A Penrose diagram of the spacetime structure of three light-ray operators measured on a state produced by a local operator. Due to the universality of the OPE limit, symmetry techniques applicable to (b) can be used to understand (a). }
\label{fig:collision_schematic}
\end{figure}

To emphasize the more general phenomenological applicability of these techniques, beyond the case of states produced by local operators, in this section we will focus on high $p_T$ jet production, which is one of the most frequent hard scattering processes at the Large Hadron Collider~(LHC). It is of particular interest for the study of jet substructure~\cite{Larkoski:2017jix,Marzani:2019hun}, and is used both for beyond the Standard Model searches, as well as to probe QCD dynamics in vacuum and medium. In this case we measure the energy correlators on an ensemble of high-$p_T$ jets, identified with some jet algorithm (most commonly anti-$k_T$ \cite{Cacciari:2008gp}). This situation is shown schematically in \Fig{fig:intro_a}.  Jet algorithms are complicated to implement in perturbative calculations, and will in general destroy any symmetries of the energy correlators. Furthermore, in proton-proton collisions, the initial state cannot be described by a local operator. However, what makes jet substructure of particular theoretical interest, is that it is naturally interested in the limit that $\zeta_{ij}\ll 1$, where all the correlators lie within a single jet. We will refer to this as the jet substructure, or collinear limit. In the collinear limit, we have a factorization theorem\footnote{This factorization theorem is an extension of the factorization theorems for inclusive particle production proved as part of the Collins-Soper-Sterman program \cite{Collins:1981ta,Bodwin:1984hc,Collins:1985ue,Collins:1988ig,Collins:1989gx,Collins:2011zzd,Nayak:2005rt,Mitov:2012gt}. The use of a jet instead of an identified particle does not modify the infrared state. See e.g. \cite{Kang:2016ehg,Kang:2016mcy,Kang:2017frl} for applications in the context of jet substructure, and \cite{Aversa:1988fv,Aversa:1988mm,Aversa:1990uv,Aversa:1989xw,Aversa:1988vb,Czakon:2021ohs} for results for the hard functions.} 
\begin{equation}\label{eq:fact_pp}
\frac{d \sigma}{dp_T dY d \{ \zeta_{ij} \}} = H(p_T, Y, R) \otimes J( \{ \zeta_{ij} \}) +\cO(\zeta_{ij}/R) \,,
\end{equation}
where $H(p_T, Y, R)$ describes the production of the jet state in the hadron collider environment with transverse moment, $p_T$, rapidity $Y$ and with a jet algorithm of radius $R$, and $J( \{ \zeta_{ij} \})$. Importantly, $H(p_T, Y, R)$ carries all the dependence on the jet algorithm and the complicated state, including parton distribution functions. On the other hand, the function $J( \{ \zeta_{ij} \})$ is universal, and is the same function that would appear in the idealized situation of light-ray operators measured on a state produced by a local operator without any jet algorithm, see \Fig{fig:intro_b} (We will give a more precise definition of the jet function below.). This universality of the collinear limit opens the door to the use of elegant symmetry techniques to study real world jet substructure. Most excitingly to us, these techniques apply directly to the physical observable measured by experimentalists, namely the energy pattern on the calorimeter!

In the context of jet substructure at the LHC, the techniques of this paper are most interesting for multi-point correlation functions.  In this case, it is convenient to define $x_L=\text{max}( \zeta_{ij})$ and $u_{ij} = \zeta_{ij}/x_L$. The variable $x_L$ describes the overall size dependence, while $u_{ij}$ are cross ratios that describe the shape dependence (This will be described in detail for the three-point correlator below). The factorization formula in \Eq{eq:fact_pp} shows that the shape dependence at leading order in the $x_L$ expansion is universal, and therefore for applications to the LHC, we will be primarily interested in developing techniques to understand this leading shape dependence. It is for this reason that the first calculation of the three-point correlator \cite{Chen:2019bpb} focused on the collinear limit.
Although for applications to the LHC we can only consider the leading expansion in $x_L$, multi-point correlation functions exhibit squeezed limits $u_{ij} \ll 1$, which are universal to all orders in the twist expansion. Therefore, symmetry based techniques to study the shape dependence of multipoint correlators are of great practical use at the LHC.

We should emphasize that in the case of $e^+e^-$ colliders where no jet algorithms are required, the techniques of this paper can be used to systematically expand in $x_L$, and we will indeed derive the subleading blocks in Sec.~\ref{sec:blocks}. However, the primary focus of this paper, and the perturbative results to which we will compare are all expanded to leading order in $x_L$, and therefore can be applied at the LHC. 

Having set up the general philosophy, and emphasized why the jet substructure limit is particularly interesting for the application of symmetry based techniques due to the universal nature of collinear factorization, we now review the perturbative structure of the two- and three-point correlators in more detail.

We start from the simplest case of two-point energy correlator~(EEC) produced by a local spinless source, $\langle {\cal E}(n_1) {\cal E}(n_2) \rangle$, where $n_i^\mu = (1, \vec{n}_i)$ and $\vec{n}_i$ is a 3D unit vector specifying the direction of the energy detector.  Ignoring the spin dependence of the source, the two-point correlator has no non-trivial cross ratios, and is a function of a single scaling variable $\zeta = (1 - \vec{n}_1 \cdot \vec{n}_2)/2$. We are interested in the limit where $\vec{n}_1 \to \vec{n}_2$. In this limit the leading power factorization theorem for the EEC has the following explicit form~\cite{Dixon:2019uzg}\footnote{For early studies of the collinear limit at leading-logarithm, see \cite{Konishi:1979cb}.}
\begin{equation}
\frac{d \sigma}{d \zeta} = \int dE_J\, E_J^2 H(E_J) J_{\rm EEC}(\zeta, E_J)  + \text{power corrections}\,,
\end{equation}
where $H(E_J)$ can be interpreted as the probability of producing a hard parton with energy $E_J$, which seeds the jet. As emphasized above, it encodes the detailed information of how the jet is produced, but not any details about the jet substructure. Since it is process dependent, we will not discuss it further.  Our primary focus will be on the jet function, $J_{\rm EEC}(\zeta, E_J)$, which  describes how the parent parton evolves into a jet with the prescribed EEC distribution.  It contains only collinear dynamics and is universal. It can be defined in terms of a matrix element of gauge invariant collinear quark or gluon fields in Soft-Collinear Effective Theory~(SCET)~\cite{Bauer:2000ew,Bauer:2000yr,Bauer:2001ct,Bauer:2001yt}. For example, for a quark jet, the EEC jet function is defined as
\begin{equation}
J_{\rm EEC}^q(\zeta, E_J) = \frac{1}{2 E_J} \sumint_{X_c} \, 
\mathrm{Tr}  \langle \Omega  | \chi_n  \frac{\slashed{n}}{2} |  X_c \rangle  \sum_{a, b \in X_c}\frac{E_a E_b}{E_J^2} \delta\left( \zeta - \frac{s_{ab}}{4 E_a E_b} \right) \langle X_c | \bar{\chi}_n | \Omega \rangle  \,,
\label{eq:two_point}
\end{equation}
where we sum and integrate over collinear states $X_c$ with the collinear phase space $\Phi_c$.

At leading order,  the jet function is simply obtained from the time-like splitting kernel and the two-particle collinear phase space in $d = 4 - 2 \epsilon$ dimension,
\begin{equation}
P_{qq}(\xi) = C_F \left(  \frac{1 + \xi^2}{1 - \xi}  - \epsilon ( 1 - \xi) \right) \,,
\qquad 
d\Phi_c^{(2)} = \int ds d\xi \,  \frac{(\xi (1 - \xi) s)^{-\epsilon}}{(4 \pi)^{2 - \epsilon} \Gamma(1 - \epsilon) } \,.
\end{equation}
Eq.~\eqref{eq:two_point} gives the bare one-loop jet function in the $\overline{\text{MS}}$ scheme,
\begin{align}
J_{\rm EEC}^{q, (1)}(\zeta, E_J, \epsilon) =&\  \left(\frac{ \mu^2 e^{\gamma_E} }{4 \pi}\right)^\epsilon \int d\Phi_c^{(2)} 2 \xi (1-\xi) \frac{g^2}{s}  P_{qq}( \xi)  \delta \left(\zeta - \frac{s}{4 E_J^2 \xi ( 1 - \xi) } \right)
\nn
\\
= &\
\frac{g^2}{16 \pi^2} \left(\frac{\mu^2 e^{ \gamma_E}}{(2 E_J)^2}  \right)^{\epsilon} 
 C_F \frac{  (9 -  2 \epsilon) }{\zeta^{1 + \epsilon} } \frac{\Gamma(3 - 2 \epsilon) \Gamma(-2 \epsilon)}{ \Gamma(4 - 4 \epsilon) \Gamma(- \epsilon) }
\nn\\
 = &\ \frac{\alpha_s}{4 \pi} \left[ - \frac{3}{2 \epsilon} \delta(\zeta) + \frac{3}{2 [\zeta]_+}  + \delta(\zeta) \left( \frac{3}{2} \log\frac{(2 E_J)^2}{\mu^2} - \frac{37}{6} \right)  \right] + {\cal O}(\epsilon) \,.
 \label{eq:tree_qjet_res}
\end{align}
Here we will consider only the non-contact term $\xi >0$ and will set the dimensional regulator $\epsilon = 0$. Furthermore, $1/[\zeta]_+$ indicates a standard plus distribution. The two-loop jet functions for quark and gluon  jets have been obtained in \cite{Dixon:2019uzg} using sum rules for the EEC.

We can now extend this discussion to the three-point correlation function in the collinear limit. In this case, we have a factorization formula similar to the case of EEC
\begin{equation}
\frac{d \sigma}{d \zeta_{12} d \zeta_{23} d \zeta_{31} }
= \int dE_J \, E_J^3 H(E_J) J_{\rm EEEC} ( \zeta_{12}, \zeta_{23}, \zeta_{31}, E_J) + \text{power corrections} \,.
\end{equation}
The hard function is identical to the case of the two-point correlator, since it is independent of the observable, while the jet function now has a non-trivial shape dependence, described by $\zeta_{ij} = (1 - \vec{n}_i \cdot \vec{n}_j)/2$. Explicitly, for a quark jet, the jet function is given by
\begin{align}
J_{\rm EEC}^q(\zeta_{12}, \zeta_{23}, \zeta_{31}, E_J) = &\ \frac{1}{2 E_J} \sumint_{X_c} \, 
\mathrm{Tr}  \langle \Omega  | \chi_n  \frac{\slashed{n}}{2} |  X_c \rangle 
 \sum_{a, b, c \in X_c}\frac{E_a E_b E_c}{E_J^3} \delta\left( \zeta_{12} - \frac{s_{ab}}{4 E_a E_b} \right)
\nn
\\
&\  \cdot \delta\left( \zeta_{23} - \frac{s_{bc}}{4 E_b E_c} \right)
  \delta\left( \zeta_{31} - \frac{s_{ca}}{4 E_c E_a} \right)
 \langle X_c | \bar{\chi}_n | \Omega \rangle  \,,
\label{eq:three_point}
\end{align}
with a similar definition for gluon jets. 

At leading order, the EEEC jet function can be calculated using the tree-level $1 \to 3$ splitting function $P_{abc}^{(0)}$, which have been obtained for QCD in \cite{Campbell:1997hg,Catani:1998nv}. We will consider the non-contact terms only, $\zeta_{ij}>0$, for which the LO jet function is $E_J$ independent
\begin{align}
J_{\rm EEEC}^{q, (1)}(\zeta_{12}, \zeta_{23}, \zeta_{31}) = &\ \sum_{a,b,c} \int d\Phi_c^{(3)} \frac{4 g^4}{s^2_{abc}} P_{a b c}^{(0)} \frac{E_a E_b E_c}{(E_J)^3} 
\nn\\
&\ \cdot \delta\left(\zeta_{12}-\frac{s_{a b}}{4 E_a E_b}\right) \delta\left(\zeta_{23}-\frac{s_{b c}}{4 E_b E_c}\right)\delta\left(\zeta_{13}-\frac{s_{a c}}{4 E_a E_c}\right)\,,
\label{eq:def}
\end{align}
where $\Phi_c^{(3)} $ is the three-particle collinear phase space~\cite{Gehrmann-DeRidder:1997fom,Ritzmann:2014mka}, and $s_{abc} = (p_a + p_b + p_c)^2$. The energy weighting factor $E_a E_b E_c$ in \eqref{eq:def} is necessary for ensuring collinear and soft safety of the measurement\footnote{At tree-level the observable is still finite without the energy weighting. However, starting from one-loop order, the energy weighting is necessary to obtain a finite result.}. Because of the energy weighting factor, energy correlators should be regarded as weighted cross sections, rather than observables in the usual sense~\cite{Chen:2020vvp}.   
From the perspective of modern integration techniques, these observables are particularly convenient, since one integrates over energy fractions, leaving angles fixed. 
This makes the integrals closely related to Feynman parameter integrals \cite{Chen:2019bpb}. Indeed, it was shown in  \cite{Chen:2019bpb} that one can obtain analytic results for these observables using known one-loop Feynman integrals. These analytic results provide the perturbative data for testing the light-ray OPE.

\begin{figure}
%[htp]
\includegraphics[]{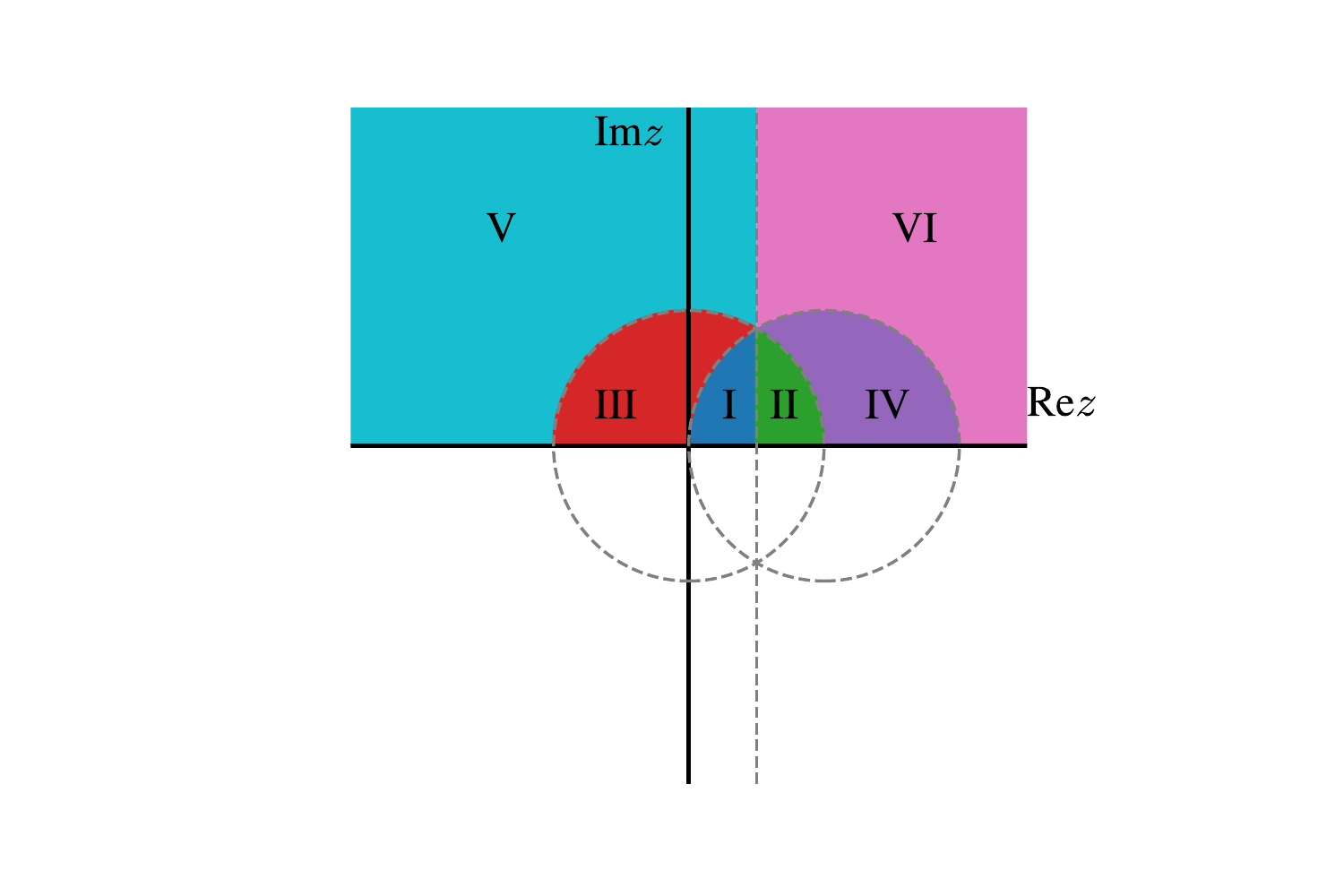}
\caption{The complex $z$ plane divides into $12$ region under the symmetry $S_3 \times Z_2$ of the three-point energy correlator. In this paper we will focus on region I, where the squeezed limit corresponds to $z \to 0$.}
\label{fig:region}
\end{figure}

For example, for a quark or gluon jet, the EEEC in the collinear limit can be written at leading order as
\beq
J_{\rm EEEC}^{q, (1)}(\zeta_{12}, \zeta_{23}, \zeta_{31}) = \frac{g^4}{32 \pi^5 \zeta_{12} \zeta_{23} \zeta_{31}} \frac{|z|^2 |1-z|^2}{|z - \bar z|} G_{q,g}(z) \,,
\eeq
where we introduced the cross ratios $u = \zeta_{12}/\zeta_{31} = |z|^2$ and $v = \zeta_{23}/\zeta_{31} = |1-z|^2$ and parameterized the shape by a single complex variable $z$. The functions $G_{q,g}(z)$ are single-valued functions defined on the complex plane, whose explicit expression can be found in \cite{Chen:2019bpb}. The EEEC defined in \eqref{eq:def} has an $S_3$ symmetry under the permutation of $\vec{n}_i$, since the energy-flow operator is bosonic. There is an additional $Z_2$ symmetry in QCD due to parity invariance. The $S_3 \times Z_2$ symmetry imposes constraints on the function $G_{q,g}(z)$. The full complex plane is divided into $12$ regions related by the $S_3 \times Z_2$ symmetry, as is shown in Fig.~\ref{fig:region}. In this paper, we will focus on region I, where the squeezed limit (OPE limit) occurs as $z \to 0$. This region is characterized by the ordering $\zeta_{12} \leq \zeta_{23} \leq \zeta_{13}$.

%%%%%%%%%%%%%%%%%%%%
%%%%%%%%%%%%%%%%%%%%

To illustrate the explicit structure of a three-point correlator, we consider for simplicity the case of  $\cN=4$ SYM. Writing the correlator as
\beq\label{crosssec}
J_{\rm EEEC}^{{\cal N}=4, (1)} (\zeta_{12}, \zeta_{23}, \zeta_{31}) = \frac{g^4}{32 \pi^5 \zeta_{12} \zeta_{23} \zeta_{31}} \frac{|z|^2 |1-z|^2}{|z - \bar z|} G_{{\cal N}=4}(z) \,,
\eeq
the function $G_{\cN=4}(z)$ takes the simple form \cite{Chen:2019bpb}
\begin{align}\label{eq:three_point_N4}
&G_{\cN=4}(z)=\frac{1+u+v}{2uv}(1+\zeta_2)-\frac{1+v}{2uv}\log(u)-\frac{1+u}{2uv}\log(v)\nn \\
&-(1+u+v)(\partial_u +\partial_v)\Phi(z)+\frac{(1+u^2+v^2)}{2uv}\Phi(z) +\frac{(z-\bar z)^2(u+v+u^2+v^2+u^2v+uv^2)}{4u^2 v^2}\Phi(z)\nn \\
&+\frac{(u-1)(u+1)}{2u v^2}D_2^+(z) +\frac{(v-1)(v+1)}{2u^2 v}D_2^+(1-z)+\frac{(u-v)(u+v)}{2uv}D_2^+\left( \frac{z}{z-1} \right)\,,
\end{align}
where we have used the well-known one-loop box function
\begin{align}
  \label{eq:1}
  \Phi(z) =\frac{2}{z-\bar z} \left(  {\rm Li}_2(z) - {\rm Li}_2(\bar z) + \frac{1}{2} \left(\log(1-z) - \log(1 - \bar z) \right) \log (z \bar z) \right)\,,
\end{align}
and a transcendental weight-2 function
\begin{align}
  \label{eq:2}
  D_2^+(z) = {\rm Li}_2(1 - |z|^2) + \frac{1}{2} \log(|1-z|^2) \log(|z|^2)\,,
\end{align}
which is even under $z \leftrightarrow \bar z$. Remarkably, this correlator is of the same level of complexity as a four-point correlator of local operators. This is in contrast to the usual multi-jet event shape variables which often exhibit elliptic function even at tree level~\cite{Ellis:1980wv}.  Furthermore, it has been shown that these correlators are of genuine phenomenological interest, and can be directly measured at the LHC, where they exhibit many useful features~\cite{Komiske:2022enw}.

In addition to their simple functional structure arising from their integral definitions in terms of splitting functions, we want to show in this paper how these correlators in fact have a much more rigid structure enforced by the light-ray OPE. As a simple example to motivate the reader that something interesting underlies the energy correlators, we consider the limit as we squeeze two detectors together, $|z| \to 0$. Expanding the gluon-jet result in this limit, and keeping only the $C_A^2$ and $C_A n_f T_F$ color structures, we have~\cite{Chen:2021gdk} 
\beq\label{eq:gluon_result_expand}
{\footnotesize
\begin{split}
&\widetilde{G}_{g}(w)=C_A n_f T_F\left\{
\frac{1}{\left| w\right| ^2}\left[\frac{7}{800}{\color{blue}-\frac{1}{720} \cos (2 \phi )} \right]
+\left[{\color{blue}-\frac{\cos (4 \phi )}{1680} }+\left(\frac{5}{6} \pi ^2-\frac{207251}{25200}\right) \cos (2 \phi )+\frac{\pi^2}{4}-\frac{120899}{50400}\right]\right.\\
& +\left| w\right| ^2 \left[{\color{magenta}\left(-\frac{8}{315}\cos (2 \phi )-\frac{22}{63}\right) \log (2 \left| w\right| )}
{\color{blue}-\frac{\cos (6 \phi )}{3024}}\right. \\
&\qquad\quad \left.+\left(\frac{53}{3} \pi ^2 -\frac{13182121}{75600}\right) \cos (4 \phi)
+\left(\frac{43}{3} \pi ^2 -\frac{224661347}{1587600}\right) \cos(2 \phi )
+\frac{25 \pi^2}{6}-\frac{5317297}{127008}\right]\\
&+\left| w\right| ^4
   \left[{\color{magenta}\left(-\frac{17588 \cos (2 \phi )}{3465}-\frac{122 \cos (4 \phi)}{3465}-\frac{964}{105}\right) \log (2 \left| w\right| )}
   {\color{blue}-\frac{\cos (8 \phi )}{4752}}
   +\left(\frac{799}{6} \pi ^2 -\frac{1092978311}{831600} \right)\cos (6 \phi)\right. \\
&\left.\left.   +\left(\frac{961}{6} \pi ^2-\frac{303716037469}{192099600} \right)\cos (4 \phi)
   +\left(\frac{511}{6} \pi ^2-\frac{162328436041}{192099600}\right) \cos (2\phi )
   +\frac{123 \pi^2}{4}-\frac{714502853}{2328480}\right]
\right\}\\
&+C_A^2\left\{
\frac{1}{\left| w\right| ^2}\left[{\color{blue}\frac{\cos (2 \phi )}{1440}}+\frac{49}{800}\right]
+\left[{\color{blue}\frac{\cos (4 \phi )}{3360}} -\left(\frac{4}{3} \pi ^2-\frac{66881}{5040}\right) \cos (2 \phi )-\frac{5 \pi ^2}{8}+\frac{318193}{50400}\right]\right.\\
&+\left| w\right| ^2
   \left[{\color{magenta}\left(-\frac{73}{630} \cos (2 \phi )-\frac{62}{315}\right) \log (2 \left| w\right| )}
   +{\color{blue}\frac{\cos (6 \phi )}{6048}}\right.\\
&\qquad\quad \left.   -\left(\frac{44}{3} \pi ^2 - \frac{10951547}{75600}\right)\cos (4 \phi )
   -\left(\frac{191}{12} \pi ^2-\frac{125023781}{793800}\right) \cos (2 \phi )
   -6 \pi^2+\frac{7550981}{127008}\right]\\
& +\left| w\right| ^4
   \left[{\color{magenta}\left(\frac{964}{495} \cos (2 \phi )-\frac{599 \cos (4 \phi)}{3465}+\frac{557}{210}\right) \log (2 \left| w\right| )}
   +{\color{blue}\frac{\cos (8 \phi )}{9504}}
   -\left(\frac{256}{3} \pi ^2 -\frac{28018675 }{33264}\right) \cos (6 \phi)\right. \\
&\quad\left.\left. -\left(\frac{1375}{12} \pi ^2-\frac{217341879809 }{192099600} \right)\cos (4 \phi)   
   -\left(\frac{208}{3} \pi ^2-\frac{18842665223}{27442800}\right) \cos (2\phi )
   -\frac{245 \pi^2}{8}+\frac{3544660961}{11642400}\right]
\right\}\\
&+\cO(|w|^6)\,,
\end{split}
}
\eeq
where we have introduced $ z = 2 w /(1 + w)$ and $\arg w = \phi$. The $w$ coordinate is convenient for exhibiting transverse spin effects, which correspond to rotating a pair of energy detector around the jet axis. We have highlighted the terms of highest transverse spin in blue in \eqref{eq:gluon_result_expand}.\footnote{The terms in pink are highlighted since they involve logarithms. Their origin will be discussed in more detail in \Sec{sec:zero}.} The leading terms in the power expansion of $|w|$ have been studied in detail in \cite{Chen:2020adz}. They are interesting as they are the manifestation of gluon spin in jet substructure~\cite{Chen:2020adz}, and also allow one to test parton shower implementations of spin correlations~\cite{Collins:1987cp,Knowles:1988hu,Knowles:1987cu,Knowles:1988vs,Karlberg:2021kwr,Hamilton:2021dyz}. There are also higher harmonic terms in \eqref{eq:gluon_result_expand}, which can not come from gluon spin alone, but also from orbital angular momentum.
Looking closely at the highlighted terms in blue, we note that for both color structures they resum into a simple expression in terms of hypergeometric functions
\begin{align}
\label{eq:res_power}
&\frac{1}{\bar{w}^2}\, _2F_1\left[1,  \frac{3}{2},\frac{7}{2},w^2 \right]+\frac{1}{w^2} \, _2F_1\left[1,  \frac{3}{2},\frac{7}{2},\bar w^2\right] \\
&\hspace{3cm}=\frac{\cos 2\phi}{|w|^2}+\frac{3}{7} \cos 4\phi  +|w|^2 \frac{5}{21}\cos6\phi +|w|^4 \frac{5}{33}\cos8\phi  +\cdots\,. \nn
\end{align}
Note that from perturbative amplitude language, the squeezed limit corresponds to iterated $1 \to 2$ splittings. The resummation of the power corrections terms in \eqref{eq:res_power} is seemingly magical from the perspective of the amplitude style calculation. However, we will see that it is an immediate consequence of symmetries and the light-ray OPE structure governing the correlators. The light-ray OPE allows multipoint correlators of energy flow operators to be analyzed much like correlation functions of local operators, by performing OPEs onto operators with definite celestial dimensions, $\delta$, and transverse spin, $j$, see \Fig{fig: sequential_OPE}. In the particular case illustrated here, we will see that the leading twist collinear limit is governed by the light-ray operators describing the exchanged gluon, and the hypergeometric functions are simply celestial blocks that are completely fixed by the symmetries of the problem. This OPE language provides a complementary perspective for studying energy correlators that is  hidden from the perspective of perturbative amplitude calculations.

%%%%%%%%%%%%%%%%%%%%%%%%%%%%%%%%%%%%%%%%%%%%%%%%%%%%%%%%%%%%%%%%%%%%%%%%%%%%%%%%
\section{Energy Correlators from a Light-Ray Perspective}\label{sec:jss_limit}
%%%%%%%%%%%%%%%%%%%%%%%%%%%%%%%%%%%%%%%%%%%%%%%%%%%%%%%%%%%%%%%%%%%%%%%%%%%%%%%%

In the previous section we have shown how one can compute energy correlators perturbatively using splitting functions. To be able to exploit the underlying symmetry properties of these observables, we must learn to study them in operator language. As emphasized in the introduction, the energy correlators are expressed as matrix elements $\langle \cE(\vec n_1) \cE(\vec n_2) \cdots \cE(\vec n_k) \rangle$ of light-ray operators 
\begin{align}\label{eq:ANEC_op_2}
\mathcal{E}(\vec n) = \lim_{r\to \infty}  \int\limits_0^\infty dt~ r^2 n^i T_{0i}(t,r \vec n)\,.
\end{align} 
In this section we briefly review the general properties of these light-ray operators, their symmetries, and the light-ray OPE. Since we will work in perturbation theory in this paper (as our ultimate goal is QCD, for which we currently are only able to work perturbatively), our presentation of light-ray operators is in a perturbative context. Earlier excellent discussions of the properties of energy flow operators from a perturbative perspective were given in \cite{Belitsky:2013xxa,Belitsky:2013bja}.  For discussions of light-ray operators in CFTs using techniques that are valid non-perturbatively, see \cite{Kravchuk:2018htv,Kologlu:2019mfz,1822249}.

%%%%%%%%%%%%%%%%%%%%%%%%%%%%%%%%%%%%%%%%%%%%%%%%%%%%%%%%%%%%%%%%%%%%%%%%%%%%%%%%
\subsection{Light-Ray Operators}
%%%%%%%%%%%%%%%%%%%%%%%%%%%%%%%%%%%%%%%%%%%%%%%%%%%%%%%%%%%%%%%%%%%%%%%%%%%%%%%%

For a dimension-$\Delta$, spin-$J$, transverse spin-$j$ local primary operator $\mathcal{O}^{\mu_1\dots\mu_J;\nu_1\dots\nu_j}$, we define a corresponding light-ray operator as \cite{Kravchuk:2018htv}
\beq
\mathbb{O}^{[J]}(n,\varepsilon)\equiv \mathbb{O}^{[J]}(\vec{n},\varepsilon)=\lim_{r\to \infty} r^{\Delta-J}\int_{0}^{\infty}\! dt\; \mathcal{O}^{\mu_1\dots\mu_J;\nu_1\dots\nu_j}(t,r\vec{n})\bar{n}_{\mu_1}\dots\bar{n}_{\mu_J}\varepsilon_{\nu_1}\dots\varepsilon_{\nu_j}\,,
\eeq
where $n^\mu=(1,\vec{n}), \bar{n}^\mu=(1,-\vec{n})$, and $\varepsilon$ is a polarization vector that satisfies $\varepsilon^2=\varepsilon\cdot n = \varepsilon\cdot \bar{n}=0$. We lift the unit vector $\vec{n}$ to the null vector $n$ in the argument of the light-ray operator $\mathbb{O}$ because this definition is Lorentz invariant. Perturbatively such an operator can be viewed as measuring some particle state, whose nature depends on the operator $\mathcal{O}$, along the null direction $n^\mu$. The ANEC operator in \Eq{eq:ANEC_op_2} is a particular example of such a light-ray operator.  In QCD, the twist-2 operators for quarks and gluons (which can be viewed as detecting quark and gluon states) are given by
\bea
\mathcal{O}_q^{[J]}&=&\frac{1}{2^J}\bar{\psi}\gamma^{+}(iD^+)^{J-1}\psi\,,\\
\mathcal{O}_{g}^{[J]}&=&-\frac{1}{2^J} F_{c}^{i +}(iD^+)^{J-2}F_{c}^{i +}, \\
\mathcal{O}_{\tilde{g},\lambda}^{[J]}&=&-\frac{1}{2^J} F_{c}^{i +}(iD^+)^{J-2}F_{c}^{j +} \varepsilon_{\lambda,i} \varepsilon_{\lambda,j} \,,
\eea
 and the corresponding light-ray operators in the free theory are
\bea
\mathbb{O}_q^{[J]}(n)&=&\sum_{s}\int\!\! \frac{E^2 dE}{(2\pi)^3 2E}  E^{J-1} \left(b_{p, s}^{\dagger}b_{p,s}+(-1)^J d_{p,s}^{\dagger} d_{p,s} \right)\,, \label{eq: unpolarized_quark}\\
\mathbb{O}_{g}^{[J]}(n)&=&\sum_{\lambda, c} \int\!\! \frac{E^2 dE}{(2\pi)^3 2E} E^{J-1} a^{\dagger}_{p,\lambda,c} a_{p,\lambda,c}\,, \label{eq: unpolarized_gluon}\\
\mathbb{O}_{\tilde{g},\lambda}^{[J]}(n)&=&-\sum_{\lambda} \int\!\! \frac{E^2 dE}{(2\pi)^3 2E}  E^{J-1} a^{\dagger}_{p,\lambda,c} a_{p,-\lambda,c}\,, \label{eq: polarized_gluon}
\eea
where $p^\mu = E n^\mu$ and $\lambda$ is the helicity of the polarization vector $\varepsilon_\lambda$. Here we use the following conventions for the mode expansion of free massless quark and gluon field
\bea
\psi(x)&=&\sum_s \int\frac{dp^+ d^2p_{\perp}}{(2\pi)^3 2p^+} \left(u_s(p) b_{p,s} e^{-i p\cdot x} + v_s(p) d^{\dagger}_{p,s} e^{i p\cdot x}\right)\,,\\
A^{\mu}_{c}(x)&=&\sum_{\lambda} \int\frac{dp^+ d^2p_{\perp}}{(2\pi)^3 2p^+}\left(\varepsilon^{\mu}_{\lambda}(p) a_{p,\lambda,c} e^{-ip\cdot x} +{\varepsilon^{*}_{\lambda}}^{\mu}(p) a^{\dagger}_{p,\lambda,c} e^{ip\cdot x} \right)\,.
\eea
These twist-two operators describe the leading twist scaling behavior of $n$-point energy correlator observables in QCD \cite{Hofman:2008ar,Chen:2021gdk}.

%%%%%%%%%%%%%%%%%%%%%%%%%%%%%%%%%%%%%%%%%%%%%%%%%%%%%%%%%%%%%%%%%%%%%%%%%%%%%%%%
\subsection{Lorentz Transformations on the Celestial Sphere}
%%%%%%%%%%%%%%%%%%%%%%%%%%%%%%%%%%%%%%%%%%%%%%%%%%%%%%%%%%%%%%%%%%%%%%%%%%%%%%%%

Light-ray operators can be viewed as being labelled by a point on the celestial sphere, and therefore behave in many respects as local operators on the celestial sphere.  Correlation functions of light-ray operators transform non-trivially under the action of the Lorentz group on the celestial sphere. Since this symmetry will play an important role in our analysis, we review it here in some detail.

The celestial sphere is the space of all light-rays passing the origin. Recall that the infinitesimal Lorentz transformation corresponding to $M^{\mu\nu}$ on a coordinate $x^\alpha$ is
\beq
x^\alpha \to \left( \delta^{\alpha}_\beta -i \epsilon \left(M^{\mu\nu}\right)^\alpha_\beta\right) x^\beta = x^\alpha +\epsilon \left(g^{\mu \alpha} x^\nu -g^{\nu \alpha} x^\mu\right),
\eeq
where $\left(M^{\mu\nu}\right)_{\alpha \beta}$ is the vector representation matrix $i\left( \delta^\mu_\alpha \delta^\nu_\beta-\delta^\mu_\beta \delta^\nu_\alpha\right)$. Its action on null vector $n^\mu=(1, n^i)$ is 
\beq
M^{0 a}:\; (1,n^i) \to (1+\epsilon n^a, n^i+\epsilon \delta^{i a}), \quad\quad M^{ab}:\; (1,n^i) \to (1, n^i+\epsilon (\delta^{i b} n^a -\delta^{i a} n^b)),
\eeq
where $i,a,b$ are spatial indices. Under the equivalence relation $n^\mu \sim \lambda n^\mu$, we can obtain the Lorentz transformation on the celestial sphere:
\beq \label{eq: inf_transf_unit_vec}
M^{0 a}:\; n^i \to n^i +\epsilon(\delta^{i a}-n^i n^a),\quad \quad M^{ab}:\; n^i \to n^i+\epsilon (\delta^{i b} n^a -\delta^{i a} n^b).
\eeq
The Killing vectors of each generator in terms of spherical coordinates $\vec{n}=(\sin\theta \cos \phi, \sin\theta \sin\phi,\cos \theta)$, are given by
\bea
M^{03}=-\sin\theta \partial_\theta ,\;
M^{01} =\cos\theta \cos\phi \partial_\theta-\frac{\sin\phi}{\sin\theta}\partial_\phi ,\; 
M^{02}=\cos\theta \sin\phi \partial_\theta+ \frac{\cos\phi}{\sin\theta} \partial_\phi, \\
M^{12}=\partial_\phi,\;
M^{31}=\cos\phi \partial_\theta -\frac{\cos\theta}{\sin\theta}\sin\phi \partial_\phi,\;
M^{32}=\sin\phi \partial_\theta +\frac{\cos\theta}{\sin\theta} \cos\phi \partial_\phi.
\eea

\begin{figure}
%[htp]
\begin{center}
\subfloat[]{
\includegraphics[scale=0.5]{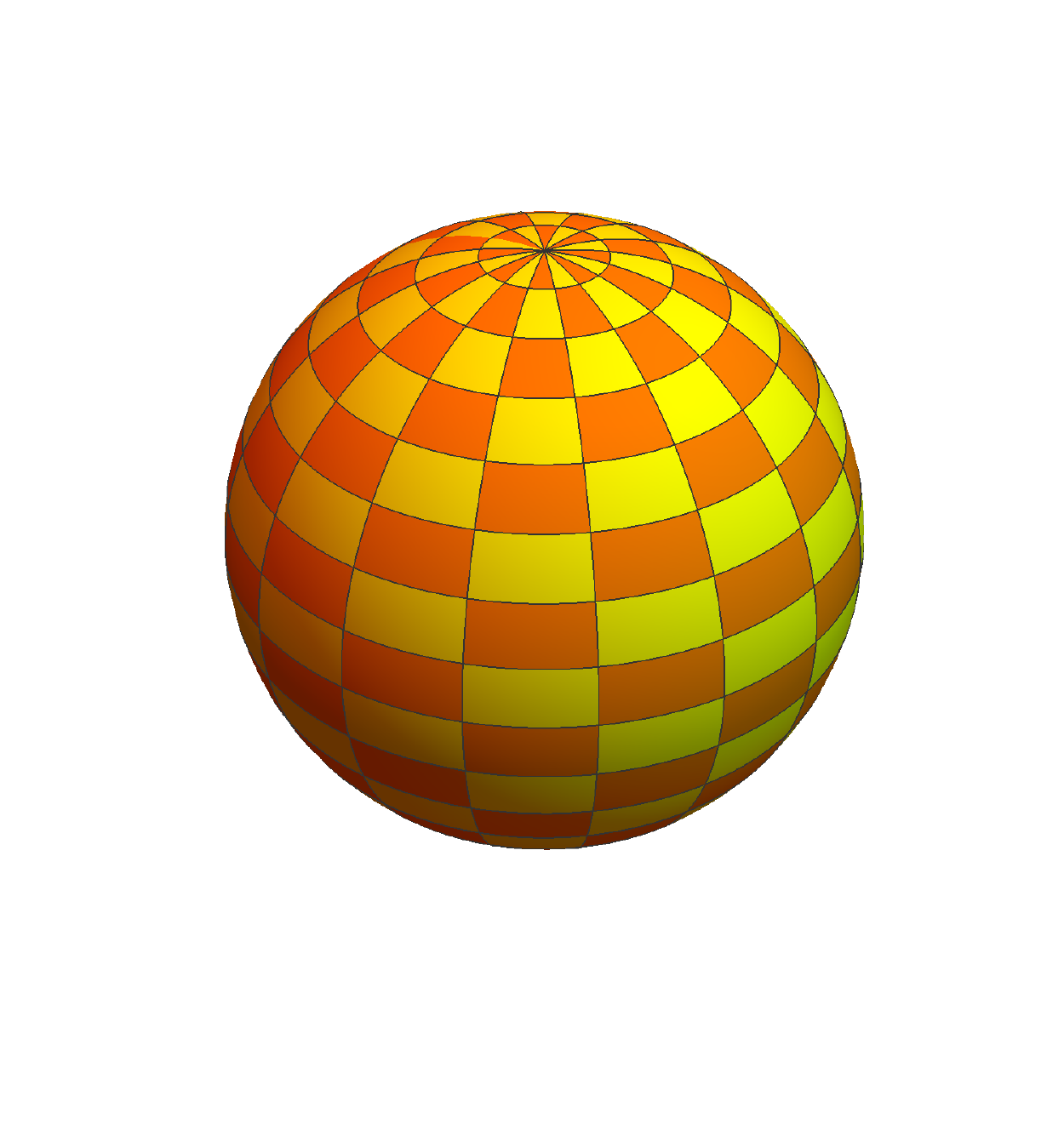}\label{fig:lorentzA}
}\qquad
\subfloat[]{
\includegraphics[scale=0.5]{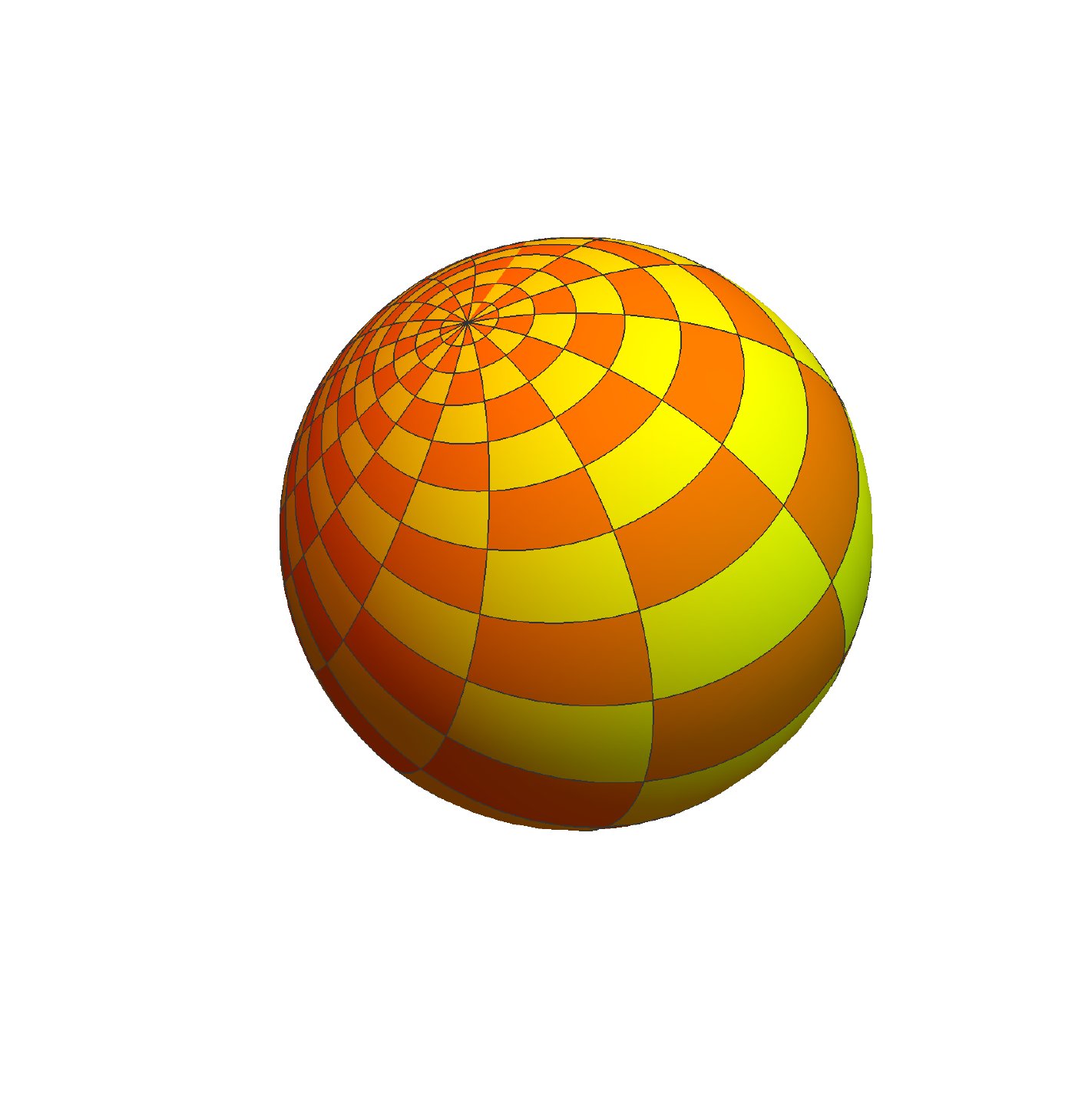}\label{fig:lorentzB}
}
%\captionsetup{font={footnotesize}}
\end{center}
\caption{An example of a Lorentz transformation on the celestial sphere. The operator $\exp{-i M^{+2}}$ transforms (a) to (b). }
\label{fig:lorentz_transf}
\end{figure}

For completeness, we present some finite Lorentz transformations of $n^i$ on the celestial sphere (see Figure \ref{fig:lorentz_transf}): 
\begin{align*}
&e^{-i \eta M^{03}}:%\textstyle 
\left(\frac{s_\theta c_\phi}{\sinh (\eta ) c_\theta+\cosh (\eta)},\frac{s_\theta s_\phi}{\sinh (\eta ) c_\theta+\cosh (\eta)},\frac{e^{2 \eta } (c_\theta+1)+c_\theta-1}{e^{2 \eta } (c_\theta+1)-c_\theta+1}\right),\\
&e^{-i \eta M^{12}}:  (s_\theta \cos (\eta +\phi ),s_\theta \sin (\eta +\phi ),c_\theta),\\
&e^{-i \eta M^{-1}}: %\textstyle
\left(\frac{2 (\eta -\eta  c_\theta +s_\theta c_\phi)}{\eta ^2-\eta ^2
   c_\theta+2 \eta  s_\theta c_\phi+2},\frac{2 s_\theta s_\phi}{\eta ^2-\eta ^2 c_\theta+2 \eta  s_\theta c_\phi+2},\frac{2 (c_\theta-1)}{\eta ^2-\eta ^2 c_\theta+2 \eta  s_\theta  c_\phi+2}+1\right),\\
&e^{-i \eta M^{-2}}: %\textstyle
\left(\frac{2 s_\theta c_\phi}{\eta ^2-\eta ^2 c_\theta+2 \eta 
   s_\theta s_\phi+2},\frac{2 (\eta-\eta  c_\theta +s_\theta
   s_\phi)}{\eta ^2-\eta ^2 c_\theta+2 \eta  s_\theta s_\phi+2},\frac{2 (c_\theta-1)}{\eta ^2 -\eta ^2 c_\theta+2 \eta  s_\theta s_\phi+2}+1\right),\\
&e^{-i \eta M^{+1}}: %\textstyle
   \left(\frac{2 (\eta  c_\theta+\eta +s_\theta c_\phi)}{\eta ^2 c_\theta +\eta ^2+2 \eta  s_\theta c_\phi+2},\frac{2 s_\theta s_\phi}{\eta ^2 c_\theta+\eta ^2+2 \eta  s_\theta c_\phi+2},\frac{2
   (c_\theta+1)}{\eta ^2 c_\theta+\eta ^2+2 \eta  s_\theta c_\phi+2}-1\right),\\
&e^{-i \eta M^{+2}}: %\textstyle
\left(\frac{2 s_\theta c_\phi}{\eta ^2 c_\theta+\eta ^2+2 \eta  s_\theta s_\phi+2},\frac{2 (\eta  c_\theta+\eta +s_\theta s_\phi)}{\eta ^2 c_\theta+\eta ^2+2 \eta  s_\theta s_\phi+2},\frac{2 (c_\theta+1)}{\eta ^2 c_\theta+\eta ^2+2 \eta  s_\theta
   s_\phi+2}-1\right).
\end{align*}
Here we have used the shorthand $c_\theta = \cos\theta$, $s_\theta=\sin\theta$, $s_\phi=\sin \phi$, and $c_\phi=\cos \phi$.

The well-known fact that Lorentz group acts on the celestial sphere as the conformal group is more transparent if we use the stereographic projection that maps $\vec{n}$ to a point $w=\tan \frac{\theta}{2} e^{i\phi}$ on the complex plane. Examples of the induced transformations on the complex plane are
\begin{align*}
&e^{-i \eta M^{03}}:\; w\to e^{-\eta} w,\quad\; e^{-i \eta M^{12}}:\; w\to e^{i\eta } w,\qquad e^{-i \eta M^{-1}}:\; w\to \frac{w}{1+\eta w}\,,\\
&e^{-i \eta M^{+1}}:\; w\to w+\eta,\quad e^{-i \eta M^{+2}}:\; w\to w+i\eta, \quad e^{-i \eta M^{-2}}:\; w\to \frac{i w}{i+\eta w},
\end{align*}
which can be summarized as a single Mobius transformation $w\to \frac{a w+ b}{c w+d}$ where $ad-bc=1$. From (\ref{eq: inf_transf_unit_vec}), we see there are 4 generators that fix a unit vector $n^i$ on the celestial sphere
\beq
n_a M^{0a}=-\frac{1}{2} M^{-+}, \quad M^{0 a}+n_b M^{b a}=M^{- a}, \quad \varepsilon_{abc}n^a M^{bc}.
\eeq
The first transformation corresponds to a boost in the null direction, the second to a special conformal transformation in the transverse plane, and the third to a transverse rotation.
This little group of transformations which fix the origin, is referred to as the parabolic subgroup of the Lorentz group (see e.g. \cite{Yamazaki:2016vqi,Karateev:2017jgd}). The representations of this subgroup can induce all admissible irreducible representations of the Lorentz group \cite{langlands1989irreducible}. In particular, as we will see, the light-ray operator $\mathbb{O}(n, \varepsilon)$ is an eigenstate of the boost generator $\vec{n}\cdot \vec{\mathbf{K}}=-n_a M^{0a}$ and the transverse rotation generator $\vec{n}\cdot \vec{\mathbf{J}}=\varepsilon_{abc}n^a M^{bc}$.

%%%%%%%%%%%%%%%%%%%%%%%%%%%%%%%%%%%%%%%%%%%%%%%%%%%%%%%%%%%%%%%%%%%%%%%%%%%%%%%%
\subsection{Symmetry Properties of Light-Ray Operators}
%%%%%%%%%%%%%%%%%%%%%%%%%%%%%%%%%%%%%%%%%%%%%%%%%%%%%%%%%%%%%%%%%%%%%%%%%%%%%%%%

%%%%%%%%%%%%%%%%%%%%
\begin{figure}
%[htp]
\begin{center}
\includegraphics[scale=0.25]{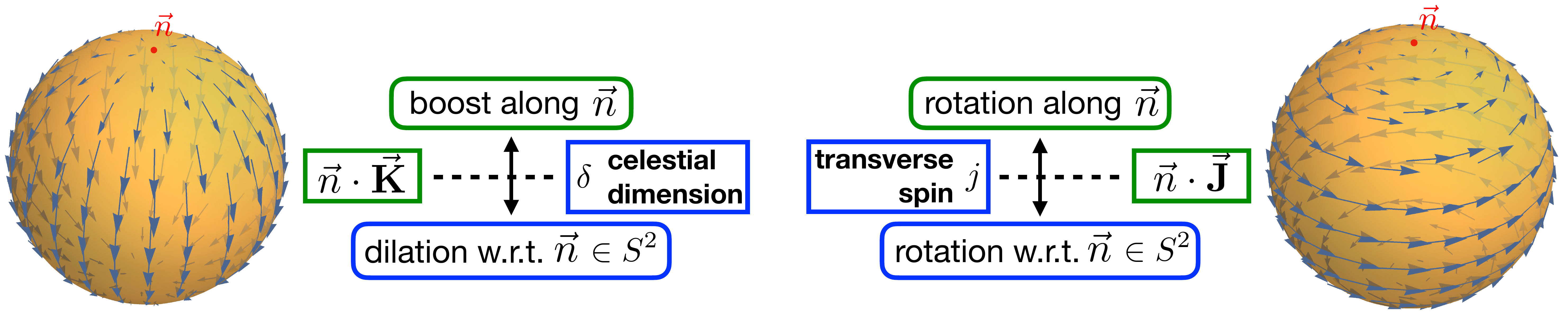}
\end{center}
\caption{Two important symmetry transformations for studying the behavior of light-ray operators are boosts along $\vec n$, and rotations about $\vec n$, whose action on the celestial sphere is illustrated. The quantum numbers under these transformations are the celestial dimension, $\delta$ and the transverse spin, $j$.}
\label{fig:summarize_symmetries}
\end{figure}
%%%%%%%%%%%%%%%%%%%%

We now review the transformation properties of the light-ray operators that lead to the nice features of their correlation functions. Recall again for convenience the definition of a light-ray operator $\mathbb{O}^{[J]}(n,\varepsilon)$ from local primary operator $\mathcal{O}^{\mu_1\dots\mu_J;\nu_1\dots\nu_j}$
\begin{equation*}
\mathbb{O}^{[J]}(n,\varepsilon)\equiv \mathbb{O}^{[J]}(\vec{n},\varepsilon)=\lim_{r\to \infty} r^{\Delta-J}\int_{0}^{\infty}\! dt\; \mathcal{O}^{\mu_1\dots\mu_J;\nu_1\dots\nu_j}(t,r\vec{n})\bar{n}_{\mu_1}\dots\bar{n}_{\mu_J}\varepsilon_{\nu_1}\dots\varepsilon_{\nu_j}.
\end{equation*}
First we note that the integration measure $\displaystyle \lim_{r\to \infty} r^{\Delta-J} \int_0^\infty dt$ shifts the dimension by $J-\Delta-1$, hence $\mathbb{O}^{[J]}(n,\varepsilon)$ has the dimension $J-1$.  More interesting is the behavior of this operator under Lorentz transformations. Our starting point is the transformation of the local operator
\beq
U_\Lambda \mathcal{O}^{\mu_1\dots\mu_J;\nu_1\dots\nu_j}(x) U_\Lambda^{-1} = {\left(\Lambda^{-1}\right)^{\mu_1}}_{\rho_1}\dots {\left(\Lambda^{-1}\right)^{\mu_J}}_{\rho_J} {\left(\Lambda^{-1}\right)^{\nu_1}}_{\sigma_1}\dots {\left(\Lambda^{-1}\right)^{\nu_j}}_{\sigma_j} \mathcal{O}^{\rho_1\dots\rho_J;\sigma_1\dots\sigma_j}(\Lambda x),
\eeq
or equivalently,
\beq
\begin{split}
U_\Lambda \mathcal{O}^{\mu_1\dots\mu_J;\nu_1\dots\nu_j}(x) U_\Lambda^{-1} \bar{n}_{\mu_1}\dots\bar{n}_{\mu_J}\varepsilon_{\nu_1}\dots\varepsilon_{\nu_j}&\\
&\hspace{-1cm}=(\Lambda\bar{n})_{\mu_1}\dots(\Lambda\bar{n})_{\mu_J}(\Lambda\varepsilon)_{\nu_1}\dots(\Lambda\varepsilon)_{\nu_j}  \mathcal{O}^{\mu_1\dots\mu_J;\nu_1\dots\nu_j}(\Lambda x) .
\end{split}
\eeq
Among all the Lorentz transformations, there are two special kinds generated by the collinear boost generator $\vec{n}\cdot \vec{\mathbf{K}}$ and the transverse rotation generator $\vec{n}\cdot \vec{\mathbf{J}}$ respectively, whose quantum numbers will play an important role. These are illustrated in \Fig{fig:summarize_symmetries}. Transverse rotations $\exp(-i \varphi\, \vec{n}\cdot \vec{\mathbf{J}})$ do not change $\bar{n}$ and $x=(t,r\vec{n})$ and can be easily diagonalized by choosing a polarization vector $\varepsilon_{\pm}$ such that $\exp(-i \varphi\, \vec{n}\cdot \vec{\mathbf{J}})\varepsilon =\exp(\mp i \varphi) \varepsilon_{\pm} $. This results in the transformation of $\mathbb{O}^{[J]}(n,\varepsilon_\lambda)$ under transverse rotations
\bea
e^{-i\varphi\, \vec{n}\cdot \vec{\mathbf{J}}} \mathbb{O}^{[J]}(n,\varepsilon_\lambda) e^{i\varphi\, \vec{n}\cdot \vec{\mathbf{J}}} &=& e^{-i \lambda j\varphi} \mathbb{O}^{[J]}(n,\varepsilon_\lambda),\\
\left[\vec{n}\cdot \vec{\mathbf{J}},\, \mathbb{O}^{[J]}(n,\varepsilon_\lambda) \right] &=& \lambda j \mathbb{O}^{[J]}(n,\varepsilon_\lambda),
\eea
where $\lambda=\pm$ is the helicity of the polarization vector $\varepsilon_\lambda$.

The collinear boost $\exp (-i \eta\,\vec{n}\cdot \vec{\mathbf{K}})$ does not change the polarization vector $\varepsilon$, but maps $n,\bar{n}, x=(t,r\vec{n})$ in the following way
\begin{gather}
n^\mu\to  e^\eta n^\mu\,,
\qquad
 \bar{n}^\mu\to 
 e^{-\eta} \bar{n}^\mu \,,
 \nn\\
  x^\mu=\left(\frac{t+r}{2}\right) n^\mu +\left(\frac{t-r}{2}\right)\bar{n}^\mu \to \left(\frac{t+r}{2}\right) e^{\eta} n^\mu +\left(\frac{t-r}{2}\right) e^{-\eta}\bar{n}^\mu.
\end{gather}
By rewriting the integration measure in terms of light-cone coordinates \cite{Belitsky:2013xxa}
\beq
\lim_{r\to \infty} r^{\Delta-J} \int_0^\infty dt = \lim_{(\bar{n}\cdot x)\to \infty} \left(\frac{\bar{n}\cdot x}{2}\right)^{\Delta-J} \int_{-\infty}^\infty d(n\cdot x)\,,
\eeq
we can obtain the transformation of $\mathbb{O}^{[J]}(n,\varepsilon)$
\beq
\begin{split}
&e^{-i\eta\, \vec{n}\cdot \vec{\mathbf{K}}} \mathbb{O}^{[J]}(n,\varepsilon) e^{i\eta\, \vec{n}\cdot \vec{\mathbf{K}}} \\
=\, & e^{-J \eta} \!\!\!\!\! \lim_{(\bar{n}\cdot x)\to \infty} \!\!\! \left(\frac{\bar{n}\cdot x}{2}\right)^{\Delta-J} \!\!\!\! \int_{-\infty}^\infty  \!\!\!\! d(n\cdot x)\,
\mathcal{O}^{\mu_1\dots\mu_J;\nu_1\dots\nu_j} \!\! \left( \frac{\bar{n}\cdot x}{2} e^{\eta} n^\mu \! +\! \frac{n\cdot x}{2} e^{-\eta}\bar{n}^\mu\right)\! \bar{n}_{\mu_1}\dots\bar{n}_{\mu_J}\varepsilon_{\nu_1}\dots\varepsilon_{\nu_j}\\
=\, & e^{-(\Delta-1) \eta}  \!\!\!\!\! \lim_{(\bar{n}\cdot x)\to \infty} \!\!\! \left(\frac{\bar{n}\cdot x}{2}\right)^{\Delta-J} \!\!\!\! \int_{-\infty}^\infty   \!\!\!\!  d(n\cdot x)\,
\mathcal{O}^{\mu_1\dots\mu_J;\nu_1\dots\nu_j}\!\! \left( \frac{\bar{n}\cdot x}{2}  n^\mu \! +\! \frac{n\cdot x}{2} \bar{n}^\mu\right)\! \bar{n}_{\mu_1}\dots\bar{n}_{\mu_J}\varepsilon_{\nu_1}\dots\varepsilon_{\nu_j}\\
=\, & e^{-(\Delta-1) \eta} \mathbb{O}^{[J]}(n,\varepsilon),
\end{split}
\eeq
where we rescaled $n\cdot x\to e^\eta n\cdot x,\,\bar{n}\cdot x\to e^{-\eta} \bar{n}\cdot x$ in the second line to obtain the third line. Since the collinear boost $\exp (-i \eta\,\vec{n}\cdot \vec{\mathbf{K}})$ is equivalent to a dilation centered at $\vec{n}$ on the celestial sphere,  the eigenvalues of $\vec{n}\cdot \vec{\mathbf{K}}$ are also called celestial dimensions. Now, we have
\beq
[ \vec{n}\cdot \vec{\mathbf{K}},\, \mathbb{O}^{[J]}(n,\varepsilon) ] = -i\delta\, \mathbb{O}^{[J]}(n,\varepsilon),
\eeq
where $\delta=\Delta-1$ is the celestial dimension of $\mathbb{O}^{[J]}(n,\varepsilon)$.

The operators $M^{+a}$ and $M^{-a}$ respectively play the role of translations and special conformal transformations on the celestial sphere. $M^{+a}$ can raise the celestial dimension while $M^{-a}$ lower the celestial dimension because of the commutation relations
\begin{align}
&[\vec{n}\cdot \vec{\mathbf{K}}, M^{+a}]=-i  M^{+a},\\
&[\vec{n}\cdot \vec{\mathbf{K}}, M^{-a}]=i  M^{-a}.
\end{align}
The action of $M^{+a}$ on the light-ray operator $\mathbb{O}^{[J]}(n,\varepsilon)$ is the transverse derivative. The action of $M^{\mu\nu}$ on the local operator $\mathcal{O}^{\mu_1\dots \mu_J;\nu_1\dots\nu_j}(x)$ is
\beq
\begin{split}
[M^{\mu\nu}, \mathcal{O}^{\mu_1\dots \mu_J;\nu_1\dots\nu_j}(x)] =& -i(x^\mu\partial^\nu-x^\nu\partial^\mu) \mathcal{O}^{\mu_1\dots \mu_J;\nu_1\dots\nu_j}(x)\\
&+ \sum_{k=1}^{J} \mathcal{O}^{\mu_1\dots \rho_k \dots \mu_J;\nu_1\dots\nu_j}(x)\, {\left(M^{\mu\nu}\right)_{\rho_k}}^{\mu_k}\\
&+ \sum_{k=1}^{j} \mathcal{O}^{\mu_1\dots \mu_J;\nu_1\dots\sigma_k\dots\nu_j}(x)\, {\left(M^{\mu\nu}\right)_{\sigma_k}}^{\nu_k}\,,
\end{split}
\eeq
where ${\left(M^{\mu\nu}\right)_\alpha}^\beta=i\left( \delta^\mu_\alpha g^{\nu\beta}- \delta^\nu_\alpha g^{\mu\beta}\right)$ is the vector representation matrix.
For the case of ${(M^{+a})_\alpha}^\beta=i( g^{a\beta} \bar{n}_\alpha-\delta^a_{\alpha} \bar{n}^\beta)$, we have
\begin{equation*}
{\left(M^{+ a}\right)_{\rho_k}}^{\mu_k}\, \bar{n}_{\mu_k}=0,\quad 
{\left(M^{+ a}\right)_{\sigma_k}}^{\nu_k}\, \varepsilon_{\nu_k}=i \varepsilon^a \bar{n}_{\sigma_k}.
\end{equation*}
Notice that for the irreducible representation $\mathcal{O}^{\mu_1\dots \mu_J;\nu_1\dots\nu_j}$, the indices $\mu_m$ and $\nu_k$ are anti-symmetrized, which leads to
\beq
\! \left(\sum_{k=1}^{J} \mathcal{O}^{\mu_1\dots \rho_k \dots \mu_J;\nu_1\dots\nu_j}\, {\left(M^{+ a}\right)_{\rho_k}}^{\mu_k}
\!+\! \sum_{k=1}^{j} \mathcal{O}^{\mu_1\dots \mu_J;\nu_1\dots\sigma_k\dots\nu_j}\, {\left(M^{+ a}\right)_{\sigma_k}}^{\nu_k} \!\right)\! \bar{n}_{\mu_1}\dots\bar{n}_{\mu_J}\varepsilon_{\nu_1}\dots\varepsilon_{\nu_j} =0.
\eeq
Along the integration contour, the transverse component of $x$ is $0$, so only transverse derivatives remain
\beq
[M^{+a}, \mathbb{O}^{[J]}(n,\varepsilon)]= \!\!\! \lim_{(\bar{n}\cdot x)\to \infty} \!\! \left(\frac{\bar{n}\cdot x}{2}\right)^{\! \Delta-J} \!\!\!\! \int_{-\infty}^\infty \!\!\! d(n\cdot x)\, \bar{n}_{\mu_1}\dots\bar{n}_{\mu_J} \varepsilon_{\nu_1}\dots\varepsilon_{\nu_j} \left(-i (\bar{n}\cdot x)\partial^a\mathcal{O}^{\mu_1\dots\mu_J;\nu_1\dots\nu_j}\right).
\eeq

As for $M^{-a}$, the action for general operators may be complicated, but we know it will lower the celestial dimension. Roughly speaking, its role is removing the transverse total derivative. If we impose the condition that celestial dimensions are bounded from below, which is reasonable because it relates to dimension through $\delta=\Delta-1$, there are a special class operator that cannot be lowered further and therefore should be annihilated by all of $M^{-a}$. We borrow the conventional name ''primary'' from CFT literature and call such operators the celestial primary light-ray operator. Since we start with the local primary operator $\mathcal{O}^{\mu_1\dots\mu_J;\nu_1\dots\nu_j}$, which in general does not contain total derivatives, we expect the corresponding light-ray operator $\mathbb{O}^{[J]}(n,\varepsilon)$ does not contain transverse total derivatives after the integration and satisfies the condition
\beq
[M^{-a}, \mathbb{O}^{[J]}(n,\varepsilon)]=0.
\eeq

We can organize the Lorentz transformation on the celestial primary light-ray operator in a simpler way using the embedding space formalism \cite{Dirac:1936fq,Mack:1969rr,Boulware:1970ty,Ferrara:1973yt,Cornalba:2009ax,Weinberg:2010fx,Costa:2011mg}, since the Minkowski space itself is the physical realization of the embedding space of the celestial sphere \cite{Dirac:1936fq}. By imposing the homogeneity condition on a light-ray operator $\mathbb{O}_\delta(n,\varepsilon)$ with celestial dimension $\delta$ \cite{Kravchuk:2018htv,Kologlu:2019mfz}
\beq
\mathbb{O}_\delta(\lambda n, \varepsilon) =\lambda^{-\delta} \mathbb{O}_\delta(n,\varepsilon),
\eeq
it transforms simply under Lorentz transformation
\beq
U_\Lambda \mathbb{O}_\delta( n, \varepsilon) U_\Lambda^{-1} = \mathbb{O}_\delta(\Lambda n, \Lambda\varepsilon).
\eeq

A more transparent way to understand and check these transformation properties is using the perturbative mode expansion, for example \Eq{eq: unpolarized_quark}, where we can explicitly see the covariant dependence on $n$ in $p^\mu=E n^\mu$. Using the action of Lorentz transformations on the creation and annihilation operators,
\beq
U_\Lambda b_{p,s} U_\Lambda^{-1} = b_{\Lambda p,s},\quad U_\Lambda d_{p,s} U_\Lambda^{-1} = d_{\Lambda p,s},
\eeq
we find for collinear boosts $\Lambda=e^{-i\eta\, \vec{n}\cdot \vec{\mathbf{K}}}$,
\beq
\begin{split}
e^{-i\eta\, \vec{n}\cdot \vec{\mathbf{K}}} \mathbb{O}_q^{[J]}(n) e^{i\eta\, \vec{n}\cdot \vec{\mathbf{K}}} 
&= \sum_{s}\int\!\! \frac{E^2 dE}{(2\pi)^3 2E}  E^{J-1} \left(b_{e^{\eta}p, s}^{\dagger}b_{e^{\eta}p,s}+(-1)^J d_{e^{\eta}p,s}^{\dagger} d_{e^{\eta}p,s} \right)\\
&=e^{-(J+1)\eta}\sum_{s}\int\!\! \frac{E^2 dE}{(2\pi)^3 2E}  E^{J-1} \left(b_{p, s}^{\dagger}b_{p,s}+(-1)^J d_{p,s}^{\dagger} d_{p,s} \right)\\
&= e^{-(J+1)\eta} \mathbb{O}_q^{[J]}(n),
\end{split}
\eeq
where $J+1=\Delta-1$ is the celestial dimension of twist-2 operator $\mathbb{O}_q^{[J]}(n)$.

%%%%%%%%%%%%%%%%%%%%%%%%%%%%%%%%%%%%%%%%%%%%%%%%%%%%%%%%%%%%%%%%%%%%%%%%%%%%%%%%
\subsection{Review of the Light-Ray OPE}
%%%%%%%%%%%%%%%%%%%%%%%%%%%%%%%%%%%%%%%%%%%%%%%%%%%%%%%%%%%%%%%%%%%%%%%%%%%%%%%%

Much like for correlators of local operators, multi-point correlators of light-ray operators can be reduced using the light-ray OPE \cite{Hofman:2008ar,Kologlu:2019mfz,1822249}. Intuitively, one expects that in the small angle limit light-ray operators should admit an OPE with the schematic form
\beq \label{eq: schematic_ope_ansatz}
\mathbb{O}^{[J_1]}_{\delta_1}(n_1) \mathbb{O}^{[J_2]}_{\delta_2}(n_2) \sim \sum \# \, (n_1\cdot n_2)^\kappa \mathbb{O}^{[J]}_{\delta,j}(n_2,\varepsilon) + \text{transverse derivatives}\,,
\eeq
where we choose external light-ray operators without transverse spin for simplicity. From the perturbative perspective, this can be viewed as replacing two particle detectors with a single more general particle detector, and is therefore quite natural. The scalar product simplifies $n_1\cdot n_2\sim \theta^2/4$ in the collinear limit giving rise to a scaling behavior in the angular separation. Since $n_1\cdot n_2$ is dimensionless, both sides of (\ref{eq: schematic_ope_ansatz}) should have the same dimension, a statement which is exact in a CFT. Recalling that the light-ray operator $\mathbb{O}^{[J]}$ has dimension $J-1$, we obtain the following constraint on $J$
\beq \label{eq: dimension_constraint}
(J_1-1)+(J_2-1)=J-1.
\eeq
Next, we consider constraints from the boost generator along the collinear direction. This corresponds to dilations on the celestial sphere, so we are simply doing dimensional analysis on the celestial sphere. The angle $\theta$ plays the role of length on the celestial sphere when $\theta\ll 1$ and thus we assign celestial dimension $-2$ to $(n_1\cdot n_2)$. In this way, we fix the exponent $\kappa$
\beq \label{eq: boost_constraint}
\delta_1+\delta_2=-2 \kappa +\delta \quad \Rightarrow \quad \kappa=\frac{\delta-\delta_1-\delta_2}{2}.
\eeq
More rigorously, we can act with a boost $\Lambda_{\vec{n}_2}$ along $\vec{n}_2$, which satisfies 
\beq
\Lambda_{\vec{n}_2} n_2 = \lambda_2 n_2,\quad \Lambda_{\vec{n}_2} n_1 = \lambda_1 n_1^\prime,
\eeq
on both sides of light-ray OPE:
\bea
\mathrm{l.h.s.: }& &  U_{\Lambda_{\vec{n}_2}} \mathbb{O}^{[J_1]}_{\delta_1}(n_1) \mathbb{O}^{[J_2]}_{\delta_2}(n_2) U_{\Lambda_{\vec{n}_2}}^{-1} = \lambda_1^{-\delta_1} \lambda_2^{-\delta_2} \mathbb{O}^{[J_1]}_{\delta_1}(n_1^\prime) \mathbb{O}^{[J_2]}_{\delta_2}(n_2) ,\nonumber\\
\mathrm{r.h.s.: }& & (n_1\cdot n_2)^\kappa U_{\Lambda_{\vec{n}_2}} \mathbb{O}^{[J]}_{\delta,j}(n_2,\varepsilon) U_{\Lambda_{\vec{n}_2}}^{-1} = ((\Lambda_{\vec{n}_2} n_1)\cdot (\Lambda_{\vec{n}_2}n_2))^\kappa \lambda_2^{-\delta}  \mathbb{O}^{[J]}_{\delta,j}(n_2,\varepsilon) \nonumber\\
&& \hspace{5 cm}= \lambda_1^{\kappa} \lambda_2^{\kappa-\delta} (n_1^\prime \cdot n_2)^\kappa \mathbb{O}^{[J]}_{\delta,j}(n_2,\varepsilon) . \nonumber
\eea
The scaling parameter is chosen such that the time component of null vectors $n_1,n_2,n_1^\prime$ are normalized to $1$. In the collinear limit, the scaling parameter $\lambda_1\to \lambda_2$, and we obtain the same constraint (\ref{eq: boost_constraint}).

We now apply the above conclusion to the $\mathcal{E}(n_1) \mathcal{E}(n_2)$ OPE. The celestial dimension of the energy flow operator is $\delta_1=\delta_2=\Delta_T-1=3$ and its dimension is $J_1-1=J_2-1=1$.  If the theory is conformal, the operators on the r.h.s. of the OPE should satisfy $J=J_1+J_2-1=3$. The value of $\kappa=(\delta-6)/2$, if written in terms of the dimension of the corresponding local operator, is 
\beq
\kappa =\frac{\Delta-7}{2}=\frac{\tau+J-7}{2}=\frac{\tau-4}{2},
\eeq
where $\tau$ is the twist, and we have used the constraint $J=3$.  Therefore, the schematic form of $\mathcal{E}(n_1) \mathcal{E}(n_2)$ OPE is:
\beq
\mathcal{E}(n_1)\mathcal{E}(n_2) \sim \sum_i \#\, (n_1\cdot n_2)^{\frac{\tau_i-4}{2}} \mathbb{O}_{i}^{[J=3]} (n_2) + \text{transverse derivatives}.
\eeq
From this formula, we see that the light-ray OPE is related to a twist expansion in a (nearly) conformal theory. As we will see in the example below, if the leading twist is $2$, the $\mathcal{E}(n_1) \mathcal{E}(n_2)$ OPE gives a leading scaling behavior $\sim 1/\theta^2$.

%%%%%%%%%%%%%%%%%%%%%%%%%%%%%%%%%%%%%%%%%%%%%%%%%%%%%%%%%%%%%%%%%%%%%%%%%%%%%%%%
\subsection{The $\mathcal{E}(n_1)\mathcal{E}(n_2)$ OPE in QCD}
%%%%%%%%%%%%%%%%%%%%%%%%%%%%%%%%%%%%%%%%%%%%%%%%%%%%%%%%%%%%%%%%%%%%%%%%%%%%%%%%

We can now show how to easily reproduce the perturbative calculation of the two-point correlator presented in \Sec{sec:pQCD} from the perspective of the light-ray OPE. In \cite{Chen:2021gdk} we derived the leading twist OPE coefficients for the $\mathcal{E}(n_1)\mathcal{E}(n_2)$ OPE in QCD. There we found that the leading twist $\mathcal{E}(n_1)\mathcal{E}(n_2)$ OPE takes the explicit form
\beq \label{eq: EE_OPE}
\begin{split}
&\mathcal{E}(n_1)\mathcal{E}(n_2)=\\
&-\frac{1}{2\pi} \frac{1}{2(n_1\cdot n_2)} \left\{ 
\left[ (\gamma_{qq}(2)-\gamma_{qq}(3))+(\gamma_{gq}(2)-\gamma_{gq}(3))\right]\mathbb{O}_q^{[3]}(n_2)\right.\\
& \qquad\qquad +\left[ (\gamma_{gg}(2)-\gamma_{gg}(3))+2 n_f(\gamma_{qg}(2)-\gamma_{qg}(3))\right]\mathbb{O}_g^{[3]}(n_2)\\
&\qquad\qquad \left. +\frac{1}{2} \left[(\gamma_{g\tilde{g}}(2)-\gamma_{g\tilde{g}}(3))+2 n_f (\gamma_{q\tilde{g}}(2)-\gamma_{q\tilde{g}}(3)) \right]
\left( e^{2i\phi_S} \mathbb{O}_{\tilde{g},-}^{[3]}(n_2) + e^{-2i\phi_S} \mathbb{O}_{\tilde{g},+}^{[3]}(n_2)\right)
\right\}\\
&+\mathcal{O}((n_1\cdot n_2)^0)\,.
\end{split}
\eeq
Expanding $\gamma_{ab}(J)$ perturbatively as $\gamma_{ab}(J)= \alpha_s/(4\pi) \gamma_{ab}^{(0)}(J)+\mathcal{O}\left(\alpha_s^2\right)$ we have
\beq
\begin{split} \label{eq: gamma_values}
&\gamma_{qq}^{(0)}(J)=C_F\left( 4\left(\psi^{(0)}(J+1)+\gamma_E\right)-\frac{2}{J(J+1)}-3\right),\quad 
\gamma_{qg}^{(0)}(J)=-T_F \frac{2(J^2+J+2)}{J(J+1)(J+2)},\\
&\gamma_{gq}^{(0)}(J)=-C_F \frac{2(J^2+J+2)}{(J-1)J(J+1)}, \\
&\gamma_{gg}^{(0)}(J)= 4 C_A \left( \psi^{(0)}(J+1)+\gamma_E -\frac{1}{(J-1)J}-\frac{1}{(J+1)(J+2)} \right)-\beta_0,\\
&\gamma^{(0)}_{q\tilde g}(J)= - T_F\frac{8}{(J+1)(J+2)}\,,  \quad
\gamma^{(0)}_{g\tilde g}(J)= C_A \left(\frac{8}{(J+1)(J+2)}+3\right) -\beta_0\,,
\end{split}
\eeq
where $\psi^{(0)}(z) = \Gamma'(z)/\Gamma(z)$ is the digamma function, and $\beta_0 = 11/3 C_A - 4/3 n_f T_F$ is the one-loop beta function in QCD. 

Focusing on the quark jet contribution, the light-ray OPE (\ref{eq: EE_OPE}) then gives the small angle limit of EEC
\beq
\langle \mathcal{E}(n_1)\mathcal{E}(n_2)  \rangle =\frac{\alpha_s}{4\pi} \frac{1}{2\pi} \frac{3}{2(n_1\cdot n_2)} \langle \mathbb{O}^{[3]}_q(n_2) \rangle + \mathcal{O}(\alpha_s^2, (n_1\cdot n_2)^0)\,.
\eeq
Before comparing with our earlier perturbative calculation of  $ J_{\mathrm{EEC}}^{q,(1)}$, we need to add a Jacobian factor. After gauge fixing $\vec{n}_2$ to a chosen point on the celestial sphere, we have
\beq
\langle \mathcal{E}(n_1)\mathcal{E}(n_2) \rangle\, d^2\vec{n}_1 = \langle \mathcal{E}(n_1)\mathcal{E}(n_2) \rangle\, \sin\theta d\theta d\phi =\langle \mathcal{E}(n_1)\mathcal{E}(n_2) \rangle\, 2d\zeta d\phi
\eeq
Integrating out the $2\pi$ rotation around the jet axis, we obtain the following relation between $\langle \mathcal{E}(n_1)\mathcal{E}(n_2) \rangle$ and $J_{\mathrm{EEC}}^{q,(1)}$
\beq
J_{\mathrm{EEC}}^{q,(1)} = 4\pi \langle \mathcal{E}(n_1)\mathcal{E}(n_2) \rangle,
\eeq
which agrees with the result $J_{\mathrm{EEC}}^{q,(1)}=\frac{\alpha_s}{4\pi} \frac{3}{2\zeta}$ in (\ref{eq:tree_qjet_res}). Therefore we see that we are able to reproduce a splitting function calculation from the OPE.

The power of the OPE is that it can be iterated to study higher point correlators, such as the three-point correlator. A schematic of this iterated OPE for the three-point correlator is shown in \Fig{fig: sequential_OPE}, and will be explained in the next section. While the calculation in terms of splitting functions is perhaps more familiar in the perturbative context, the operator based approach is particularly powerful for incorporating the constraints from the symmetries arising from the action of the Lorentz group on the celestial sphere.  To understand how to describe the shape dependence of the three-point correlator  from the perspective of the light-ray OPE, we will need to understand how to derive celestial blocks, which incorporate precisely these symmetries.

%%%%%%%%%%%%%%%%%%%%%%%%%%%%%%%%%%%%%%%%%%%%%%%%%%%%%%%%%%%%%%%%%%%%%%%%%%%%%%%%
\section{Three-Point Celestial Blocks for Jet Substructure}\label{sec:blocks}
%%%%%%%%%%%%%%%%%%%%%%%%%%%%%%%%%%%%%%%%%%%%%%%%%%%%%%%%%%%%%%%%%%%%%%%%%%%%%%%%

Having reviewed the structure of the three-point correlator from both the perturbative QCD perspective, and the light-ray perspective, and having shown that it exhibits interesting symmetry structures that are unexplained from the standard perspective of QCD splitting functions,  in this section we derive the celestial blocks describing the symmetry structure of OPE of the three-point function. 

We provide two separate derivations of these celestial blocks. First, we give a derivation by directly applying a Casimir operator to a lightlike source (i.e. to the jet functions discussed in \Sec{sec:pQCD}), which provides a simple derivation since it isolates the leading scaling behavior right from the beginning.  We then give a second derivation using the Casimir equation for the three-point correlator with generic angles sourced by a local operator, and then considering the collinear limit. These two derivations of course give the same result, but are designed to emphasize the relation to the factorization based approach discussed in \Sec{sec:pQCD}. 

Although the techniques we use here to derive the celestial blocks, in particular Casimir differential equations \cite{Dolan:2003hv,Dolan:2011dv}, are familiar to those from the CFT community, they have not been previously used in jet substructure, and therefore we aim to give a pedagogical presentation.

%%%%%%%%%%%%%%%%%%%%%%%%%%%%%%%%%%%%%%%%%%%%%%%%%%%%%%%%%%%%%%%%%%%%%%%%%%%%%%%%
\subsection{Casimir Equations with a Lightlike Source}\label{sec:cas_a}
%%%%%%%%%%%%%%%%%%%%%%%%%%%%%%%%%%%%%%%%%%%%%%%%%%%%%%%%%%%%%%%%%%%%%%%%%%%%%%%%

To illustrate the technique in its simplest form, we will begin by applying the technique of Casimir differential equations \cite{Dolan:2003hv} directly to the jet function for the three-point correlator. As described in \Sec{sec:pQCD}, this is designed to isolate the leading scaling behavior in the collinear limit, and can be viewed as taking the state in which the three-point correlator is measured to be sourced by a highly boosted (lightlike) quark or gluon state (In the SCET language, this is a collinear quark or gluon field.). 

For concreteness, we will choose the collinear anti-quark field $\chi$ as the source for the three-point jet function
\begin{equation}
\int dt\; e^{i t \bar{n}\cdot P}\mae{\Omega}{\bar{\chi}(t\bar{n}) \frac{\slashed{\bar{n}}}{2}  \mathcal{E}(n_1)\mathcal{E}(n_2)\mathcal{E}(n_3)\chi(0)}{\Omega}\equiv \vev{\mathcal{E}_1\mathcal{E}_2\mathcal{E}_3}_\chi\,.
\end{equation}
This jet function depends on one dimension-1 scalar product $\bar{n}\cdot P$ and six dimensionless scalar products
\begin{equation}
 \bar{n}\cdot n_1\,,\quad \bar{n}\cdot n_2\,,\quad \bar{n}\cdot n_3\,,\quad n_1\cdot n_2\,,\quad n_2\cdot n_3\,,\quad n_1\cdot n_3\,,
\end{equation}
and is a homogeneous function in $n_i$ and $\bar{n}$ with degree $-3$ and $0$ respectively. 
The homogeneity in $n_i$ comes from the property of energy flow operators $\mathcal{E}(\lambda n_i)=\lambda^{-3}\mathcal{E}(n_i)$. The homogeneity in $\bar{n}$ comes from the reparametrization invariance (RPI) in SCET \cite{Bauer:2000ew,Bauer:2000yr,Bauer:2001ct,Bauer:2001yt} arising from the arbitrariness in the choice of the lightcone basis $n,\bar{n}$. In particular, the RPI symmetry used here is  
\begin{equation}
n^\mu \to \lambda n^\mu,\quad \bar{n}^\mu\to \lambda^{-1} \bar{n}^\mu\,. 
\end{equation}
Since the quark field has dimension $\tfrac{3}{2}$, the dimension of $\vev{\mathcal{E}_1\mathcal{E}_2\mathcal{E}_3}_\chi$ is $5$. Therefore, $\vev{\mathcal{E}_1\mathcal{E}_2\mathcal{E}_3}_\chi$ takes the form
\begin{equation}
\vev{\mathcal{E}_1\mathcal{E}_2\mathcal{E}_3}_\chi =(\bar{n}\cdot P)^5 \frac{1}{(n_1\cdot n_2)^3}\frac{1}{(n_3\cdot\bar{n})^4} \left(\frac{n_1\cdot n_3}{n_1\cdot \bar{n}}\right) G(u,v) \label{eq: collnear_EEEC_ansatz}
\end{equation}
where the cross ratios are defined by
\begin{equation}
u=z\bar{z}=\frac{(n_1\cdot n_2)(n_3\cdot \bar{n})}{(n_1\cdot n_3)(n_2\cdot \bar{n})},\quad v=(1-z)(1-\bar{z})=\frac{(n_1\cdot\bar{n})(n_2\cdot n_3)}{(n_1\cdot n_3)(n_2\cdot \bar{n})}.
\end{equation}
The complex variable $z$ has the transparent physical meaning as the shape of the three energy detectors on the celestial sphere. 

Note that the dimensionless part of (\ref{eq: collnear_EEEC_ansatz}) has exactly the same structure as a 4-point  correlator of local scalar operators in a CFT
\begin{equation}
\vev{\mathcal{O}_1(x_1) \mathcal{O}_2(x_2) \mathcal{O}_3(x_3) \mathcal{O}_4(x_4)} =\frac{1}{(x_{12}^2)\frac{\Delta_1+\Delta_2}{2}} \frac{1}{(x_{34}^2)\frac{\Delta_3+\Delta_4}{2}}\left(\frac{x_{14}^2}{x_{24}^2}\right)^{\frac{\Delta_2-\Delta_1}{2}} \left(\frac{x_{14}^2}{x_{13}^2}\right)^{\frac{\Delta_3-\Delta_4}{2}} \mathcal{G}(u,v),
\end{equation}
after making the identification
\begin{equation}
x_{ij}^2 \to n_i\cdot n_j,\quad x_{i4}^2\to n_i\cdot \bar{n},\quad \Delta_1,\Delta_2,\Delta_3\to 3,\quad \Delta_4\to 5. 
\end{equation}
As discussed in \cite{Kologlu:2019mfz}, although Lorentz symmetry acts as conformal symmetry on the celestial sphere, the local source inside the Minkowski bulk plays the role of defect and breaks the conformal symmetry on the celestial sphere. By considering the collinear limit, the conformal symmetry emerges again, implying that the shape dependence for the leading term in the collinear limit is tightly constrained by symmetries.

Having understood the general structure of the three-point correlator in the collinear limit, we now wish to study the expansion about the squeezed limit $z\to 0$, which is governed by the light-ray OPE. When performing this expansion, we want to expand in functions with well defined quantum numbers under the symmetry group, in this case the Lorentz group acting on the celestial sphere. In the case of correlation functions of local operators in CFTs, these functions are referred to as conformal blocks, or conformal partial waves \cite{Ferrara:1972kab,Ferrara:1972uq,Ferrara:1974nf,Ferrara:1974ny}. The decomposition of a correlator into partial waves can be viewed as a generalization of Fourier analysis for the conformal group instead of the more familiar translation group.\footnote{While the particular case we will study here is quite simple, and so we will not need to fully develop the mathematical machinery, the study of harmonic analysis on the conformal group is well developed \cite{Dobrev:1977qv}, and has seen many recent applications and development in the context of the conformal bootstrap (see e.g. \cite{Karateev:2017jgd,Karateev:2018oml,Kravchuk:2018htv} and references therein for discussions most closely related to the present paper.) More generally, conformal blocks are well studied mathematically, due to their relations to the study of special functions and integrability (see e.g. \cite{Isachenkov:2016gim,Schomerus:2021ins}) } This same approach applies to correlation function of lightray operators defined by coordinates on the celestial sphere, and in this case the appropriate functions are referred to as celestial blocks \cite{Kologlu:2019mfz}. A fun aspect of the energy correlators is that since they can be directly measured, we will be able to directly see these ``harmonics of the Lorentz group" by eye in the energy distribution imprinted in the detector. This will be made extremely concrete in \Sec{sec:QCD_application} where we will plot the structure of these harmonics on the plane, and show how they build up the full result for the three-point correlator. The celestial blocks for the two-point correlator were derived in \cite{Kologlu:2019mfz}. Here we will derive the celestial blocks for the three-point correlator.

A powerful approach to deriving the structure of conformal blocks is the approach of Casimir differential equations \cite{Dolan:2003hv}. This approach can be illustrated with the elementary example of deriving the partial wave expansion of the $2 \to 2$ scattering amplitude $A(s,t)$ for massless, spinless particles in $d=4$ (For a detailed discussion in generic $d$, see e.g. \cite{Correia:2020xtr}). In the center of mass frame of the scattering, the symmetry is SO(3), and therefore one wants to expand in functions with definite quantum numbers under SO(3). To determine these functions, one can act with the quadratic Casimir of SO(3) on the outgoing particles, keeping the momentum of the incoming particles fixed. In this case the quadratic Casimir can be expressed as the differential operator
\beq
    \boldsymbol{L}^2= -\left[ \frac{1}{\sin\theta}\frac{\partial}{\partial \theta}\left(\sin\theta \frac{\partial}{\partial\theta}\right)
    +\frac{1}{\sin^2\theta}\frac{\partial^2}{\partial\phi^2}\right]\,,
\eeq
leading to the eigenvalue equation
\begin{align}
\boldsymbol{L}^2 P_{l}(\cos\theta) =  l (l + 1) P_{l} (\cos\theta) \,.
\end{align}
The solutions of this equation corresponding to the unitary representations of SO(3) are of course the familiar Legendre polynomials, giving rise to the standard $s$-channel partial wave expansion,
\beq
A(s,t) = \sum_{l = 0}^{\infty} (2 l + 1) a_l(s) P_l(\cos\theta) \,.
\eeq

We can now apply the same approach to derive the celestial blocks for the three-point energy correlator, which will provide a similar decomposition into partial waves on the celestial sphere. In this case, instead of considering SO(3), we must consider the full Lorentz group. The quadratic Casimir operator of the  Lorentz group is $C_2 = \frac{1}{2}{M}^{\mu\nu}{M}_{\mu\nu}$, where ${M}_{\mu\nu}$ is the angular momentum operator. Lorentz invariance implies that the energy correlators are annihilated by ${M}_{\mu\nu}$,
\begin{align}
M_{\mu\nu} \langle \cE_1 \cE_2 \cE_3 \rangle_\chi \equiv &\
-i\left(\cL_{\mu\nu}(P) + \cL_{ \mu\nu} (\bar n) + \sum_{i=1}^3 \cL_{\mu\nu} (n_i) \right) \langle \cE_1 \cE_2 \cE_3 \rangle_\chi
\nn
\\ 
= &\  0 \,,
\label{eq:lorentzinv}
\end{align}
where 
\beq
\cL_{\mu\nu}(p) = p_\mu \frac{\partial}{\partial p^\nu} - p_\nu \frac{\partial}{\partial p^\mu}  \,.
\eeq
This by itself is not particularly useful, as most event shape observables are also Lorentz invariant. What is special about energy correlators is the existence of light-ray OPE, 
\beq
\cE_1 (n_1) \cE_2(n_2) = \sum_{\delta, j}  C_{\delta, j} (n_1, n_2, \partial_{n_2}, \varepsilon) \mathbb{O}_{\delta, j}^{J=3} (n_2, \varepsilon) \,,
\eeq
where $\varepsilon^\mu$ is a null polarization vector for the light-ray operator, and the light-ray operator $\mathbb{O}_{\delta, j}^{J=3}$ is  labelled by the collinear boost quantum number $\delta$ and the transverse spin $j$ correpsonding to the generators $\vec{n}\cdot\vec{\mathbf{K}},\,\vec{n}\cdot\vec{\mathbf{J}}$ respectively. In a theory with conformal symmetry, such as QCD at leading order, the differential operator $C_{\delta, j} (n_1, n_2, \partial_{n_2}, \varepsilon)$ is completely fixed by symmetry. 
Inserting it into the collinear EEEC, we obtain
\beq
\langle \cE_1 \cE_2 \cE_3 \rangle_\chi =\sum_{\delta, j}  C_{\delta, j} (n_1, n_2, \partial_{n_2}, \varepsilon)    \langle   \mathbb{O}_{\delta, j}^{J=3} (n_2, \varepsilon) 
\cE_3(n_3)  \rangle_\chi  \,.
\eeq
where we have defined the partial waves of the Lorentz group, 
\beq
 C_{\delta, j} (n_1, n_2, \partial_{n_2}, \varepsilon)    \langle   \mathbb{O}_{\delta, j}^{J=3} (n_2, \epsilon) 
\cE_3(n_3)  \rangle_\chi  = 
(\bar{n}\cdot P)^5 \frac{1}{(n_1\cdot n_2)^3}\frac{1}{(n_3\cdot\bar{n})^4} \left(\frac{n_1\cdot n_3}{n_1\cdot \bar{n}}\right) c_{\delta, j}G_{\delta, j}(u,v) \,,
\eeq 
where $G_{\delta, j}(u, v)$ is the conformal block and $c_{\delta,j}$ is the block coefficient. As we shall see, $G_{\delta, j}$ is a pure kinematical function determined by symmetries, similar to the Legendre polynomial in the partial wave decomposition of scattering amplitudes, while $c_{\delta,j}$ encodes dynamical data, similar to the partial wave amplitudes $a_l(s)$. The celestial blocks provides a good basis for the expansion of the collinear EEEC. To proceed, we need to derive the Casimir equation for the conformal blocks. To that end, we use that $ \langle   \mathbb{O}_{\delta, j}^{J=3} (n_2, \varepsilon) 
\cE_3(n_3)  \rangle_\chi $ is also Lorentz invariant,
\beq
C_{\delta, j} (n_1, n_2, \partial_{n_2}, \varepsilon)    M_{\mu\nu} \Big[  \langle   \mathbb{O}_{\delta, j}^{J=3} (n_2, \epsilon) 
\cE_3(n_3)  \rangle_\chi  \Big] = 0 \,.
\label{eq:invariance1}
\eeq
Separating the action of $M_{\mu\nu}$ on $\mathbb{O}_{\delta, j}^{J=3}$ and the rest, we have
\begin{gather}
C_{\delta, j} (n_1, n_2, \partial_{n_2}, \epsilon)  \langle  \Big[ M_{\mu\nu},  \mathbb{O}_{\delta, j}^{J=3} (n_2, \epsilon) \Big]
\cE_3(n_3)  \rangle_\chi 
= 
\\
i \Big( \cL_{\mu\nu}(P)  +  \cL_{\mu\nu}(\bar n) +  \cL_{\mu\nu}(n_3) \Big) 
 C_{\delta, j} (n_1, \partial_{n_2}, \epsilon)    \langle   \mathbb{O}_{\delta, j}^{J=3} (n_2, \epsilon) 
\cE_3(n_3)  \rangle_\chi \,.
\end{gather}
Using \eqref{eq:lorentzinv}, we can also write
\beq
- \Big( \cL_{\mu\nu}(P)  +  \cL_{\mu\nu}(\bar n) +  \cL_{\mu\nu}(n_3) \Big)  =  \Big( \cL_{\mu\nu}(n_1)  +  \cL_{\mu\nu}(n_2) \Big) 
\eeq
when acting on  partial wave. We then find that the quadratic Casimir operator acting on the light ray operator gives
\begin{align}
&\ C_{\delta, j} (n_1, \partial_{n_2}, \epsilon)  \langle  \Big[ C_2,  \mathbb{O}_{\delta, j}^{J=3} (n_2, \epsilon) \Big]
\cE_3(n_3)  \rangle_\chi 
\nn\\
=&\ C_{\delta, j} (n_1, \partial_{n_2}, \epsilon)  \langle   \frac{1}{2} \Big[ M_{\mu\nu}, \Big[ M_{\mu\nu}, \mathbb{O}_{\delta, j}^{J=3} (n_2, \epsilon)  \Big] \Big]
\cE_3(n_3)  \rangle_\chi 
\label{eq:1}
\\
=&\ -\frac{1}{2}  \Big( \cL_{\mu\nu}(n_1)  +  \cL_{\mu\nu}(n_2) \Big)^2 (\bar{n}\cdot P)^5 \frac{1}{(n_1\cdot n_2)^3}\frac{1}{(n_3\cdot\bar{n})^4} \left(\frac{n_1\cdot n_3}{n_1\cdot \bar{n}}\right) c_{\delta, j} G_{\delta, j}(u,v) 
\\
=&\
(\bar{n}\cdot P)^5 \frac{1}{(n_1\cdot n_2)^3}\frac{1}{(n_3\cdot\bar{n})^4} \left(\frac{n_1\cdot n_3}{n_1\cdot \bar{n}}\right) c_{\delta, j}   \cC_2 G_{\delta, j}(u,v)  \,,
\label{eq:2}
\end{align}
%One possible choice of basis for (\ref{eq: collnear_EEEC_ansatz}) is to find the eigenfunctions of the Casimir differential operator
%\begin{equation}
%   \frac{1}{2}(M_{1}^{\mu \nu}+M_{2}^{\mu\nu})(M_{1\,\mu \nu}+M_{2\,\mu\nu})
%    =\frac{1}{2}\left[\left( n_1^\mu \frac{\partial}{\partial n_1^\nu}-n_1^\nu \frac{\partial}{\partial n_1^\mu}\right) 
%    +\left( n_2^\mu \frac{\partial}{\partial n_2^\nu}-n_2^\nu \frac{\partial}{\partial n_2^\mu}\right)\right]^2,
%\end{equation}
where $\cC_2$ is the action of Casimir operator acting on a function of cross ratios,
\begin{equation}
   \cC_2 = -2 u^2 (u-v-1)\partial_u^2 -4 u v (u-v+1)\partial_u\partial_v - 2 v (u(1+v)-(1-v)^2)\partial_v^2.
\end{equation}
On the other hand, the Casimir operator acting on the the light-ray operator $\mathbb{O}_{\delta, j}^{J=3} (n_2, \epsilon)$ gives
\beq
\Big[ C_2,  \mathbb{O}_{\delta, j}^{J=3} (n_2, \epsilon) \Big] =  \lambda_{\delta, j} \mathbb{O}_{\delta, j}^{J=3} (n_2, \epsilon)  \,.
\eeq
where $\lambda_{\delta,j}=\delta(\delta-2)+j^2$.
Combining \eqref{eq:1} and \eqref{eq:2}, we find the Casimir equation for the celestial block\footnote{Note that this is simply the eigenvalue equation for the conformal Laplacian.}
\begin{equation}
    \cC_2\, G_{\delta,j}(u,v) = \lambda_{\delta,j} G_{\delta,j}(u,v) \,.
\end{equation}
The differential operator factorizes in terms of $(z,\bar{z})$ variables\footnote{When interpreted as a quantum mechanical Hamiltonian, this is equivalent to a Poschl-Teller potential \cite{Poschl:1933zz}, relating it to the study of integrable Schrodinger problems \cite{Isachenkov:2016gim}.}
\begin{equation}\label{2dCasimir}
    2z^2(z-1)\partial_z^2 G_{\delta,j}+2\bar{z}^2(\bar{z}-1)\partial_{\bar{z}}^2 G_{\delta,j},
    =\lambda_{\delta,j} G_{\delta,j},
\end{equation}
which admits a closed form solution 
%\begin{equation}\label{2dBlock}
%    G_{\delta,j}(u,v)\equiv G_{\delta,j}(z,\bar{z})
%    =\frac{1}{1+\delta_{j,0}}\left[k_{\delta-j}(z)k_{\delta+j}(\bar{z})+k_{\delta+j}(z)k_{\delta-j}(\bar{z})\right],
%\end{equation}
\begin{equation}\label{2dBlock}
G_{\delta,j}(u,v)\equiv G_{\delta,j}(z,\bar{z})
=\frac{1}{1+\delta_{j,0}}\Big(k_{\frac{\delta-j}{2}}(z)k_{\frac{\delta+j}{2}}(\bar{z})+k_{\frac{\delta+j}{2}}(z)k_{\frac{\delta-j}{2}}(\bar{z})\Big)\,.
\end{equation}
These are simply the standard conformal blocks of a 2d CFT. Here the special function $k$ denotes
\begin{equation}
k_{h}(x)\equiv x^{h} \left._2 F_1 \right.\left(h+a,h+b,2h,x\right).
\end{equation}
%\begin{equation}
%    k_{\beta}(x)\equiv x^{\beta/2} \left._2 F_1 \right.\left(\frac{\beta}{2}+a,\frac{\beta}{2}+b,\beta,x\right).
%\end{equation}
In our specific example of EEEC with collinear quark source, we set $a=0, \, b=-1$. Note that the appearance of the standard 2d conformal blocks in the leading collinear limit is expected, since one has effectively expanded the celestial sphere to a plane, with boosts along the null direction $n$ act as dilatations in the plane.

To summarize, we have shown that the collinear celestial blocks for the EEEC coincide with standard 2d conformal blocks. The non-trivial function, $G(z,\bar z)$, describing the shape dependence of the EEEC therefore satisfies an expansion in celestial blocks,
\begin{equation}
    G(z,\bar{z})=\sum_{\delta,j} c_{\delta,j} G_{\delta,j}(z,\bar{z})\,,
\end{equation}
which cleanly separates kinematics $g_{\delta,j}$ from dynamics $c_{\delta,j}$. We will see that this re-organization of the small angle power corrections, as opposed to a naive Taylor series commonly used in perturbative QCD, is numerically beneficial.

%%%%%%%%%%%%%%%%%%%%%%%%%%%%%%%%%%%%%%%%%%%%%%%%%%%%%%%%%%%%%%%%%%%%%%%%%%%%%%%%
\subsection{Expansion of the Generic Angle Casimir Equation}\label{sec:cas_b}
%%%%%%%%%%%%%%%%%%%%%%%%%%%%%%%%%%%%%%%%%%%%%%%%%%%%%%%%%%%%%%%%%%%%%%%%%%%%%%%%

\begin{figure}
\begin{center}
\includegraphics[width=8 cm]{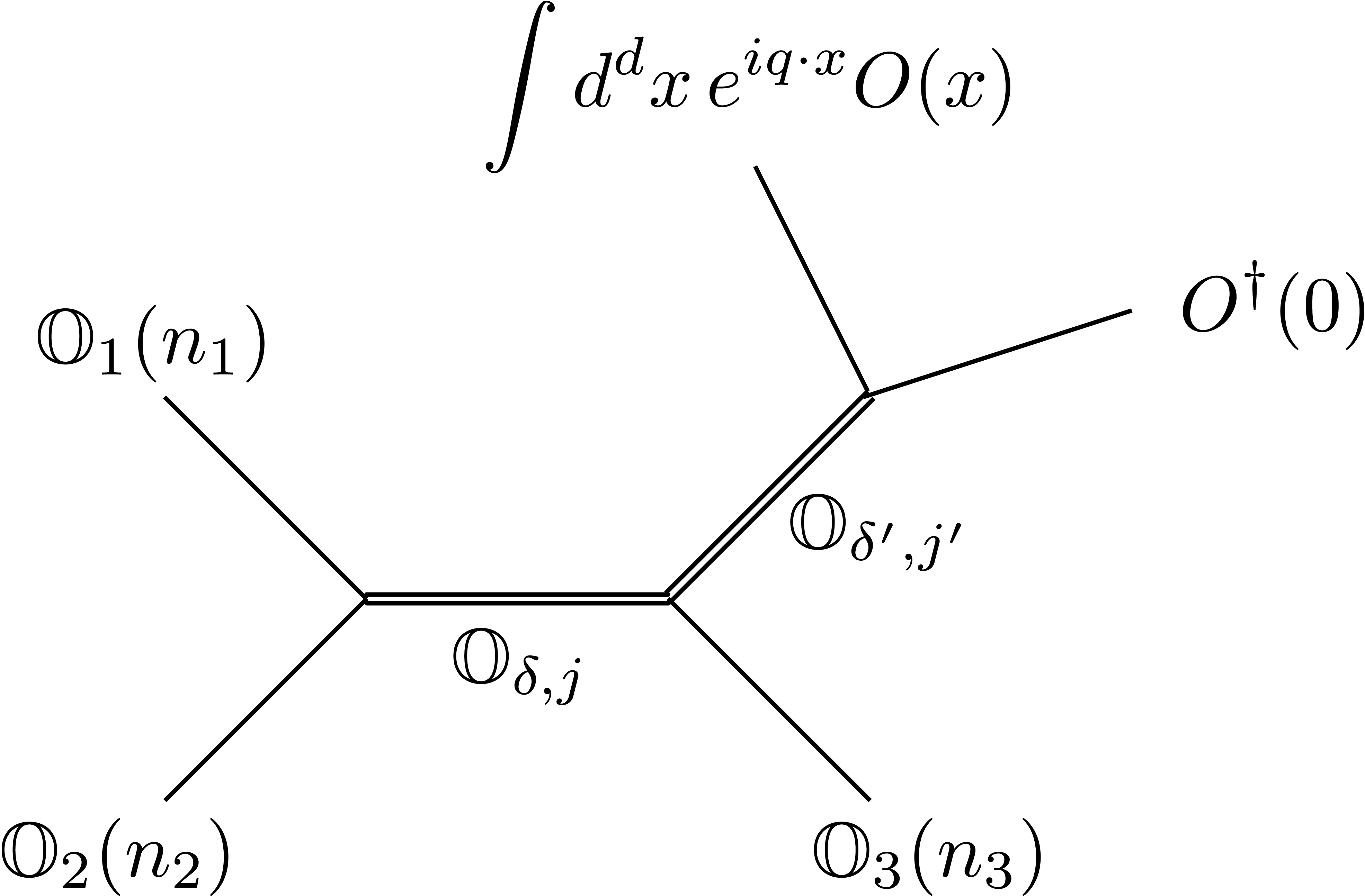}
\caption{A schematic illustration of the sequential light-ray OPE used to analyze the three-point energy correlator.}
\label{fig: sequential_OPE}
\end{center}
\end{figure}

To make contact with the CFT literature, we will now rederive the celestial blocks for general spacetime and scaling dimensions by considering the general angle three-point correlator, and then taking the collinear limit. This discussion follows the original derivation of the two-point celestial blocks in \cite{Kologlu:2019mfz}, but extends it to the three-point case. 
We consider the correlation of light-ray operators $\mathbb{O}_{\delta_i}(n_i)$, with transverse spin $j=0$ and celestial dimension $\delta_i$, in a source excited by a local operator $\mathcal{O}$ with time-like momentum $q^\mu$:
\begin{equation}
\langle \mathbb{O}_{\delta_1}(n_1) \mathbb{O}_{\delta_2}(n_2) \mathbb{O}_{\delta_3}(n_3)\rangle_{\mathcal{O}(q)} 
\equiv \int d^d x\, e^{i q\cdot x} \mae{\Omega}{\mathcal{O}(x) \mathbb{O}_{\delta_1}(n_1) \mathbb{O}_{\delta_2}(n_2) \mathbb{O}_{\delta_3}(n_3) \mathcal{O}^\dagger (0)}{\Omega}
\end{equation}
The correlator $\langle \mathbb{O}_{\delta_1}(n_1) \mathbb{O}_{\delta_2}(n_2) \mathbb{O}_{\delta_3}(n_3)\rangle_{\mathcal{O}(q)}$ is a function of Lorentz products and cross ratios
\beq
\zeta_{ij} = \frac{(2 n_i \cdot n_j)(q^2)}{(2 n_i\cdot q)(2 n_j \cdot q)}\,.
\eeq
Homogeneity in $n_i$ and dimensional analysis constrains the functional form to be 
\begin{equation}
\langle \mathbb{O}_{\delta_1}(n_1) \mathbb{O}_{\delta_2}(n_2) \mathbb{O}_{\delta_3}(n_3)\rangle_{\mathcal{O}(q)}
=\frac{(q^2)^{\frac{\delta_1+\delta_2+\delta_3+\Delta_{\mathrm{tot}}}{2}}}{(n_1\cdot q)^{\delta_1}(n_2\cdot q)^{\delta_2}(n_3\cdot q)^{\delta_3}} f(\zeta_{12}, \zeta_{23},\zeta_{13}),
\end{equation}
where $\Delta_{\mathrm{tot}}$ is the dimension of the whole correlator. The intuition of the sequential light-ray OPEs (see Figure \ref{fig: sequential_OPE})
\begin{equation*}
\mathbb{O}_{\delta_1}(n_1) \mathbb{O}_{\delta_2}(n_2) \mathbb{O}_{\delta_3}(n_3) 
\to \sum_{\delta,j}\mathbb{O}_{\delta,j}(n_2) \mathbb{O}_{\delta_3}(n_3) \to \sum_{\delta,j;\delta^\prime,j^\prime} \mathbb{O}_{\delta^\prime,j^\prime}(n_3)
\end{equation*}
guides us to decompose the correlator in a form
\begin{eqnarray}
\langle \mathbb{O}_{\delta_1}(n_1) \mathbb{O}_{\delta_2}(n_2) \mathbb{O}_{\delta_3}(n_3)\rangle_{\mathcal{O}(q)} =\sum_{\delta,j;\delta^\prime,j^\prime} \eta_{\delta,j;\delta^\prime,j^\prime} F_{\delta,j;\delta^\prime,j^\prime}, \label{eq: block_decomp_full}\\
F_{\delta,j;\delta^\prime,j^\prime} = \frac{(q^2)^{\frac{\delta_1+\delta_2+\delta_3+\Delta_{\mathrm{tot}}}{2}}}{(n_1\cdot q)^{\delta_1}(n_2\cdot q)^{\delta_2}(n_3\cdot q)^{\delta_3}} f_{\delta,j;\delta^\prime,j^\prime}(\zeta_{12}, \zeta_{23},\zeta_{13}).
\end{eqnarray}
The label $\{\delta,j\}$ represents the operator exchanged in the $1,2$-light-ray OPE channel and $\{\delta^\prime, j^\prime\}$ labels the operator appearing in the $1,2,3$-light-ray OPE. The reason why $\{\delta^\prime,j^\prime\}$ does not show up in the discussion of light-like source is that symmetry label $\{\delta^\prime,j^\prime\}$ is hidden as the symmetry properties (e.g. dimension, transverse spin) of the light-like source. As we will see, $F_{\delta,j;\delta^\prime,j^\prime}$ is purely kinematic whose functional form is completely determined by Lorentz symmetry. The numerical factor $\eta_{\delta,j;\delta^\prime,j^\prime}$ in (\ref{eq: block_decomp_full}) ensures certain normalization condition, which will be stated later in (\ref{eq: normalization_cond}), on $f_{\delta,j;\delta^\prime,j^\prime}$ and contains the dynamic information of the theory.

%%%%%%%%%%%%%%%%%%%%%%%%%%%%%%%%%%%
%[htbp]
%%%%%%%%%%%%%%%%%%%%%%%%%%%%%%%%%%%

Going through the same procedure as in last subsection, the function $F_{\delta,j;\delta^\prime,j^\prime}$ satisfies two differential equations:
\begin{eqnarray}
C_{12} F_{\delta, j;\delta^\prime,j^\prime} &=& \lambda_{\delta,j} F_{\delta, j;\delta^\prime,j^\prime}, \label{eq: F_casimir_eq_1}\\
C_{123} F_{\delta, j;\delta^\prime,j^\prime} &=& \lambda_{\delta^\prime ,j^\prime} F_{\delta, j;\delta^\prime, j^\prime}, \label{eq: F_casimir_eq_2}
\end{eqnarray}
where $C_{12}, \, C_{123}$ are quadratic Casimir operators acting on $n_1,n_2$ and $n_1,n_2,n_3$ respectively,
\begin{equation}
C_{12}=-\frac{1}{2} \left(\mathcal{L}_{\mu\nu}(n_1)+\mathcal{L}_{\mu\nu}(n_2)\right)^2,\quad
C_{123}=-\frac{1}{2} \left(\mathcal{L}_{\mu\nu}(n_1)+\mathcal{L}_{\mu\nu}(n_2)+\mathcal{L}_{\mu\nu}(n_3)\right)^2.
\end{equation}
In a general spacetime dimension $d$, the eigenvalues are 
\beq
\lambda_{\delta,j}=\delta(\delta-d+2) + j(j+d-4).
\eeq 
Differential equations (\ref{eq: F_casimir_eq_1}, \ref{eq: F_casimir_eq_2}) are equivalent to complicated differential equations of $f_{\delta,j;\delta^\prime,j^\prime}(\zeta)\equiv f_{\delta,j;\delta^\prime,j^\prime}(\zeta_{12},\zeta_{23},\zeta_{31})$:
\begin{multline} \label{eq: f_casimir_eq_1}
0=-2 \zeta _{12} (d+2 \zeta _{12}-4) f_{\delta,j;\delta^\prime,j^\prime}^{(1,0,0)}(\zeta)+(\delta _1+\delta _2) (-d+\delta _1+\delta _2+2) f_{\delta,j;\delta^\prime,j^\prime}(\zeta)\\
-4 (\delta _1+\delta _2) (\zeta _{12}-1) \zeta _{12} f_{\delta,j;\delta^\prime,j^\prime}^{(1,0,0)}(\zeta)+2 \delta _2 (-2 \zeta _{13} \zeta _{12}+\zeta _{12}+\zeta _{13}-\zeta _{23}) f_{\delta,j;\delta^\prime,j^\prime}^{(0,0,1)}(\zeta)\\
+2 \delta _1 (-2 \zeta _{23} \zeta _{12}+\zeta _{12}-\zeta _{13}+\zeta _{23}) f_{\delta,j;\delta^\prime,j^\prime}^{(0,1,0)}(\zeta)-4 (\zeta _{12}-1) \zeta _{12}^2 f_{\delta,j;\delta^\prime,j^\prime}^{(2,0,0)}(\zeta)\\
+2 (-2 \zeta _{13} \zeta _{12}+\zeta _{12}+\zeta _{13}-\zeta _{23}) \zeta _{12} f_{\delta,j;\delta^\prime,j^\prime}^{(1,0,1)}(\zeta)+2 (-2 \zeta _{23} \zeta _{12}+\zeta _{12}-\zeta _{13}+\zeta _{23}) \zeta _{12} f_{\delta,j;\delta^\prime,j^\prime}^{(1,1,0)}(\zeta)\\
-2 ((\zeta _{13}-\zeta _{23}){}^2-\zeta _{12} \zeta _{23}+\zeta _{12} \zeta _{13} (2 \zeta _{23}-1)) f_{\delta,j;\delta^\prime,j^\prime}^{(0,1,1)}(\zeta)-4 \delta _1 \delta _2 \zeta _{12} f_{\delta,j;\delta^\prime,j^\prime}(\zeta)-\lambda_{\delta,j}  f_{\delta,j;\delta^\prime,j^\prime}(\zeta)\,, 
\end{multline}
\begin{multline} \label{eq: f_casimir_eq_2}
0=-2 \zeta _{12} (d+2 \zeta _{12}-4) f_{\delta,j;\delta^\prime,j^\prime}^{(1,0,0)}(\zeta )-2 \zeta _{13} (d+2 \zeta _{13}-4) f_{\delta,j;\delta^\prime,j^\prime}^{(0,0,1)}(\zeta )-2 \zeta _{23} (d+2 \zeta _{23}-4) f_{\delta,j;\delta^\prime,j^\prime}^{(0,1,0)}(\zeta )
\\-4 \Big(\zeta _{13} \big(\delta _1 (\zeta _{13}-1)+\delta _3 (\zeta _{13}-1)+\delta _2 (\zeta _{12}+\zeta _{23}-1)\big) f_{\delta,j;\delta^\prime,j^\prime}^{(0,0,1)}(\zeta )+\zeta _{23} \big(\delta _1 (\zeta _{12}+\zeta _{13}-1)+(\delta _2+\delta _3) (\zeta _{23}-1)\big) \\
\times f_{\delta,j;\delta^\prime,j^\prime}^{(0,1,0)}(\zeta )+\zeta _{12} \big(\delta _1 (\zeta _{12}-1)+\delta _2 (\zeta _{12}-1)+\delta _3 (\zeta _{13}+\zeta _{23}-1)\big) f_{\delta,j;\delta^\prime,j^\prime}^{(1,0,0)}(\zeta )\Big)-4 (\zeta _{12}-1) \zeta _{12}^2 f_{\delta,j;\delta^\prime,j^\prime}^{(2,0,0)}(\zeta )\\
-4 \zeta _{13} (\zeta _{12}+\zeta _{13}+\zeta _{23}-2) \zeta _{12} f_{\delta,j;\delta^\prime,j^\prime}^{(1,0,1)}(\zeta )-4 \zeta _{23} (\zeta _{12}+\zeta _{13}+\zeta _{23}-2) \zeta _{12} f_{\delta,j;\delta^\prime,j^\prime}^{(1,1,0)}(\zeta )-4 (\zeta _{13}-1) \zeta _{13}^2 f_{\delta,j;\delta^\prime,j^\prime}^{(0,0,2)}(\zeta )\\
-4 \zeta _{13} \zeta _{23} (\zeta _{12}+\zeta _{13}+\zeta _{23}-2) f_{\delta,j;\delta^\prime,j^\prime}^{(0,1,1)}(\zeta )-4 (\zeta _{23}-1) \zeta _{23}^2 f_{\delta,j;\delta^\prime,j^\prime}^{(0,2,0)}(\zeta )+f_{\delta,j;\delta^\prime,j^\prime}(\zeta ) \Big(-\delta _1 \big(\delta _2 (4 \zeta _{12}-2)\\
+\delta _3 (4 \zeta _{13}-2)+d-2\big)+(\delta _2+\delta _3) (-d+\delta _2+\delta _3+2)-4 \delta _2 \delta _3 \zeta _{23}+\delta _1^2\Big)-\lambda_{\delta^\prime, j^\prime}  f_{\delta,j;\delta^\prime,j^\prime}(\zeta)\,.
\end{multline}

In the collinear limit, it's convenient to choose the longest side of the triangle as the variable that describes the degree of collinearity. Recall from eq. (\ref{crosssec}) that $\zeta_{ij}$ is related to $(x_L, u, v)$ by 
\begin{equation}\label{eq: cross_ratio_def}
x_L= \frac{2\zeta_{13}}{q^2}=\frac{n_1\cdot n_3}{(n_1\cdot q)( n_3\cdot q)},
\quad u = \frac{\zeta_{12}}{\zeta_{13}},
\quad v =\frac{\zeta_{23}}{\zeta_{13}}\,.
\end{equation}
When $x_L\ll 1$, we make the following ansatz
\begin{equation} \label{eq: sol_ansatz}
\begin{split}
 f_{\delta,j;\delta^\prime,j^\prime}(\zeta_{12},\zeta_{23},\zeta_{13}) &\equiv  f_{\delta,j;\delta^\prime,j^\prime}(x_L, u, v) \\
 &= x_L^{\alpha} \left(  f^{(0)}_{\delta,j;\delta^\prime,j^\prime}(u,v)+ x_L  f^{(1)}_{\delta,j;\delta^\prime,j^\prime}(u,v) + x_L^2 f^{(2)}_{\delta,j;\delta^\prime,j^\prime}(u,v) +\cdots \right),
\end{split}
\end{equation} 
which allows us to solve the differential equations order by order in $x_L$.

Before discussing the solution for the celestial blocks, note that when restricting to the scalar source, there is no angular dependence on the simultaneous rotation of all three light-ray operators, so only the $j^\prime=0$ contribution is of consideration in the present paper. In addition, the main focus is to analyze the LO perturbative results of leading power collinear EEEC in $\mathcal{N}=4$ SYM and QCD and the $\delta^\prime$ is actually fixed in this case. Therefore, we will save the notation by dropping the label $\{\delta^\prime, j^\prime\}$ in the main content of this paper, except the next subsection where we briefly discuss the subleading $x_L$ power corrections of celestial blocks in the example of collinear EEEC, which will be helpful for analysis once the full angle dependence of EEEC are calculated in the future.

For now, we only discuss the leading power $x_L^\alpha f^{(0)}_{\delta,j}(u,v)$ in the ansatz (\ref{eq: sol_ansatz})  for celestial blocks where we begin to hide the label $\{ \delta^\prime, j^\prime \}$ but always keep in mind it depends on $\delta^\prime$. The procedure is straightforward -- expanding the differential equations (\ref{eq: f_casimir_eq_1}, \ref{eq: f_casimir_eq_2}) to leading power in $x_L$ after changing the variables from $(\zeta_{12},\zeta_{23},\zeta_{13})$ to $(x_L, u, v)$. What we find is that the leading behavior $x_L^\alpha$ is constrained by the (\ref{eq: f_casimir_eq_2}) from quadratic Casimir $C_{123}$:  
\begin{equation}
\left(2 \alpha +\delta _1+\delta _2+\delta _3-\delta^\prime \right) \left(2 \alpha-d+\delta _1+\delta _2+\delta _3+\delta^\prime +2\right) f^{(0)}_{\delta,j}(u,v) = 0.
\end{equation}
In the light-ray OPE, higher twist operators correspond to higher powers in the small angle, so only one of the two solutions makes sense:
\begin{equation}
\alpha = \frac{\delta^\prime-\delta_1-\delta_2-\delta_3 }{2} = -\frac{\delta_1+\delta_2}{2} -b,
\end{equation}
where we define $b=\frac{\delta_3-\delta^\prime}{2}$. Then we can obtain the differential equation on $f^{(0)}_{\delta,j}(u,v)$ from (\ref{eq: f_casimir_eq_1}). To make contact with the CFT literature \cite{Dolan:2000ut, Dolan:2003hv}, we rewrite in cross-ratios $u=z\zb$, $v=(1-z)(1-\zb)$ and rescale $f^{(0)}_{\delta,j}(u,v) = u^{-\frac{\delta_1+\delta_2}{2}}G_{\delta,j}(z,\bar{z})$. With these redefinitions, we obtain the usual conformal block differential equation in $d-2$ dimensions,
\beq
\frac{1}{2}\Big[z^2(1-z) \frac{\partial^2}{\partial z^2} -(a+b+1) z^2\frac{\partial}{\partial z} - ab z + (z\to \zb) + (d-4)D_{z,\zb} \Big] G_{\delta,j}(z,\bar{z}) =  \lambda_{\delta,j}G_{\delta,j}(z,\bar{z})\,,
\eeq where we have defined $a = -\frac{\delta_1-\delta_2}{2}$. We have included differential operator
\beq
D_{z,\zb} = \frac{z \zb}{z-\zb}\Big((1-z)\frac{\partial}{\partial z}-(1-\zb)\frac{\partial}{\partial \zb}\Big)\,.
\eeq
For $d=4$ this differential equation is separable and we obtain the previous result from eqs. (\ref{2dCasimir}, \ref{2dBlock}). 

We end this part by clarifying the normalization condition we used for $F_{\delta,j;\delta^\prime,j^\prime}$ or $f_{\delta,j;\delta^\prime,j^\prime}$ in the case of scalar source. We have followed CFT convention on the normalization condition for $G_{\delta,j}(z,\bar{z})$:
\begin{equation*}
G_{\delta,j}(z,\bar{z}) \to z^{\frac{\delta-j}{2}} \bar{z}^{\frac{\delta+j}{2}} \quad \quad (0< z\ll \bar{z} \ll 1).
\end{equation*}
Therefore, our choice of the normalization condition for $f_{\delta,j;\delta^\prime,j^\prime}(\zeta_{ij})\equiv f_{\delta,j;\delta^\prime,j^\prime}(x_L, u,v)\equiv f_{\delta,j;\delta^\prime,j^\prime}(x_L,z,\bar{z})$ when $j^\prime =0 $ is
\begin{equation} \label{eq: normalization_cond}
f_{\delta,j;\delta^\prime,j^\prime =0 }(x_L,z,\bar{z}) \to x_L^{\frac{\delta^\prime-\delta_1-\delta_2-\delta_3 }{2}} z^{\frac{\delta-j-\delta_1-\delta_2}{2}} \bar{z}^{\frac{\delta+j-\delta_1-\delta_2}{2}}  \quad (0<x_L\ll 1,\, 0< z\ll \bar{z} \ll 1 ).
\end{equation}

\subsection{Three-Point Celestial Blocks at Subleading Powers \label{sec: eeec_subleading}}
%%%%%%%%%%%%%%%%%%%%%%%%%%%%%%%%%%%%%%%%%%%%%%%%%%%%%%%%%%%%%%%%%%%%%%%%%%%%%%%%

Although the primary focus of this paper is on the leading behavior in the collinear limit, the techniques are more general, and can be extended to a systematic expansion in $x_L$. In this section we derive the structure of the celestial blocks at the next few subleading powers in $x_L$. We believe that these will be useful when the full angle EEEC is calculated in $\mathcal{N}=4$ SYM and QCD in the near future. Although we can easily generalize the following procedure to general external celestial dimensions and general spacetime dimensions, for simplicity we restrict ourselves to the celestial blocks for   $\langle \mathcal{E}(n_1) \mathcal{E}(n_2) \mathcal{E}(n_3) \rangle$ in spacetime dimension $d=4$. Furthermore, we restrict ourselves to the case of an unpolarized source.
 
In $d=4$, the eigenvalues in (\ref{eq: F_casimir_eq_1}) and (\ref{eq: F_casimir_eq_2}) take the form
\begin{equation*}
\lambda_{\delta,j} = \delta(\delta-2) +j^2,\quad \lambda_{\delta^\prime, j^\prime=0} =\delta^\prime(\delta^\prime-2).
\end{equation*}
By changing variables from $(\zeta_{12},\zeta_{23},\zeta_{13})$ to $(x_L,u,v)$ and setting $\delta_1=\delta_2=\delta_3=3,\;d=4$ in (\ref{eq: f_casimir_eq_1}) and (\ref{eq: f_casimir_eq_2}), we obtain the following differential equations for $f_{\delta,j;\delta^\prime, j^\prime=0}(x_L,u,v)$:
\begin{eqnarray}
\mathcal{C}_{12} f_{\delta,j;\delta^\prime, 0}(x_L,u,v) &=& \lambda_{\delta,j} f_{\delta,j;\delta^\prime,0}(x_L,u,v), \label{eq: f_casimir1_uv}\\
\mathcal{C}_{123} f_{\delta,j;\delta^\prime, 0}(x_L,u,v) &=& \lambda_{\delta^\prime,0} f_{\delta,j;\delta^\prime, 0}(x_L,u,v),\label{eq: f_casimir2_uv}
\end{eqnarray}
where $\mathcal{C}_{12}, \mathcal{C}_{123}$ are the quadratic Casimir differential operators in variables $(x_L, u, v)$:
\begin{equation}
    \begin{split}
        \mathcal{C}_{12} = & -2 u \left(u x_L-u+v-1\right) x_L\partial_{x_L}\partial_u - \left(2 u v x_L+ 2 (v-1)^2-2 u (v+1)\right)x_L \partial_{x_L}\partial_v\\
        & -  6 \left(u x_L-u+v-1\right) x_L \partial_{x_L} - 2 u^2 (u-v-1) \partial_u^2 + 2 u v \left(u x_L-2 u+2 v-2\right) \partial_u \partial_v\\
        & +2 v \left(u v x_L-u v-u+v^2-2 v+1\right) \partial_v^2 - 2 u \left(3 u x_L+4 u-4 v-8\right) \partial_u \\
        & +2 \left(u v x_L-4 u v+2 u+4 v^2-2 v-2\right)\partial_v - 6 \left(3 u x_L-4\right) 
    \end{split},
\end{equation}
\begin{equation} \label{eq: C123_uv}
    \begin{split}
        \mathcal{C}_{123}=& -2 \left(x_L-2\right) x_L^2 \partial_{x_L}^2 - 2 u  (u+v-1) x_L^2 \partial_{x_L} \partial_u - 2 v  (u+v-1) x_L^2 \partial_{x_L}\partial_v\\
        & - 2 \left( (3 u +3 v+7) x_L-18\right)x_L\partial_{x_L} + u^2 v x_L \partial_u^2 - 2 u v x_L (u+v-1) \partial_u\partial_v +2 u v^2 x_L \partial_v^2\\
        & - 2 u x_L (3 u-v-3) \partial_u - 2 v x_L (-u+3 v-3) \partial_v - 18 x_L (u+v+1)+63.
    \end{split}
\end{equation}
After rewriting the cross ratios $u=z\bar{z},\,v=(1-z)(1-\bar{z})$, the ansatz (\ref{eq: sol_ansatz}) becomes
\begin{equation} 
\begin{split}
f_{\delta,j;\delta^\prime,0}(x_L, u, v) &\equiv f_{\delta,j;\delta^\prime,0}(x_L, z, \bar{z})\\
 &= x_L^{(\delta^\prime-9)/2} \left(  f^{(0)}_{\delta,j;\delta^\prime,0}(z, \bar{z})+ x_L  f^{(1)}_{\delta,j;\delta^\prime,0}(z, \bar{z}) + x_L^2 f^{(2)}_{\delta,j;\delta^\prime,0}(z, \bar{z}) +\cdots \right),
 \end{split}
\end{equation} 
where the leading power term $f^{(0)}_{\delta,j;\delta^\prime,0}(z, \bar{z})$ is the standard 2D conformal block,
\begin{equation}
f^{(0)}_{\delta,j;\delta^\prime,0}(z,\bar{z})=
    \frac{1}{1+\delta_{j,0}}\left[ z^{h-3} \left._2F_1\right.(h,h-\alpha-3,2h;z) \bar{z}^{\bar{h}-3} \left._2F_1\right.(\bar{h},\bar{h}-\alpha-3,2\bar{h};\bar{z})+\left(z\leftrightarrow \bar{z}\right)\right].
\end{equation}
Here we define the parameters $h\equiv\frac{\delta-j}{2},\bar{h}\equiv\frac{\delta+j}{2},\alpha\equiv \frac{\delta^\prime-9}{2}$.

Now, we proceed to find subleading powers $f^{(m\geq 1)}_{\delta,j;\delta^\prime,0}(z, \bar{z})$ in the celestial block. In principle, we can continue to solve the differential equation (\ref{eq: f_casimir1_uv}) order by order in $x_L$. However, this is non-trivial because it is a second order partial differential equation that contains lower order solutions (which are complicated hypergeometric functions) in its inhomogeneous term. 

To circumvent solving the full differential equations, we use a trick of using the full differential equation (\ref{eq: f_casimir2_uv}) which saves us from the labor of solving differential equation for subleading power corrections. The key observation is that in the Casimir operator $\mathcal{C}_{123}$, each term that contains $\partial_u,\,\partial_v$ has an additional suppression in $x_L$  (after replacing the homogeneous differential operators $x_L^k \partial_{x_L}^k$ with the corresponding constants at a given order in $x_L$). Hence, at $m$-th subleading power $(\mathrm{N}^m\mathrm{LP})$ in $x_L$, the equation takes the schematic form
\begin{equation*}
(\text{polynomials in } z,\bar{z}) f^{(m)}_{\delta,j;\delta^\prime,0}(z, \bar{z}) + (\text{polynomials in } z,\bar{z},\partial_z,\partial_{\bar{z}}) f^{(m-1)}_{\delta,j;\delta^\prime,0}(z, \bar{z}) = 0,
\end{equation*}
which is trivial to solve for $f^{(m)}_{\delta,j;\delta^\prime,0}(z, \bar{z}) $. Here we list the results for $m=1,2$:
\begin{eqnarray}\label{eq:sub_blocks}
    f^{(1)}_{\delta,j;\delta^\prime\!,0}\!\! =\!\!
    \left[ 
        \frac{(\alpha+3)(6 z \bar{z} -3z-3\bar{z}+\alpha+6)}{4\alpha+18}
         \!-\! \frac{z(z-1)(3z+6\bar{z}-6 z \bar{z}-3\bar{z}^2-\alpha(2z\bar{z}-z-\bar{z}))}{(4\alpha+18)(z-\bar{z})}\partial_z \right.\nonumber\\
     - \left. \frac{\bar{z}(\bar{z}-1)(3\bar{z}+6 z-6 z\bar{z}-3z^2-\alpha(2z\bar{z}-z-\bar{z}))}{(4\alpha+18)(\bar{z}-z)} \partial_{\bar{z}}
    -\frac{z\bar{z}(1-z)(1-\bar{z})}{4\alpha+18}\partial_{z}\partial_{\bar{z}}
    \right]\!\! f^{(0)}_{\delta,j;\delta^\prime\! ,0}\, , \qquad\\
    f^{(2)}_{\delta,j;\delta^\prime\!,0} \!\! =\!\! 
    \left[
        \frac{(\alpha+4)(6z\bar{z}-3z-3\bar{z}+\alpha+7)}{8(\alpha+5)}
        \!-\! \frac{z(z-1)(4z+7\bar{z}-8z\bar{z}-3\bar{z}^2-\alpha(2 z\bar{z}-z-\bar{z}))}{8(\alpha+5)(z-\bar{z})}\partial_{z}\right. \nonumber\\
    -\left. \frac{\bar{z}(\bar{z}-1)(4\bar{z}+7z-8z\bar{z}-3z^2-\alpha(2 z\bar{z}-z-\bar{z}))}{8(\alpha+5)(\bar{z}-z)}\partial_{\bar{z}}
    -\frac{z\bar{z}(1-z)(1-\bar{z})}{8(\alpha+5)}\partial_z\partial_{\bar{z}}
    \right]\!\! f^{(1)}_{\delta,j;\delta^\prime\! ,0}\, .\qquad
\end{eqnarray}

Since perturbative results for the full angle EEEC in $\mathcal{N}=4$ SYM or QCD at weak coupling are not currently available, we do an example of a celestial block expansion to the first few powers in the collinear limit for $\mathcal{N}=4$ SYM EEEC in the strong coupling limit in Appendix. \ref{secStrong}.

%%%%%%%%%%%%%%%%%%%%%%%%%%%%%%%%%%%%%%%%%%%%%%%%%%%%%%%%%%%%%%%%%%%%%%%%%%%%%%%%
\section{Relating Leading Twist Celestial Blocks and QCD Factorization}\label{sec:spin2}
%%%%%%%%%%%%%%%%%%%%%%%%%%%%%%%%%%%%%%%%%%%%%%%%%%%%%%%%%%%%%%%%%%%%%%%%%%%%%%%%

While the use of conformal blocks for organizing the OPE is familiar in the CFT community, it has not previously been used in the analysis of jet substructure observables. In this section we want to clearly illustrate how it relates to the standard factorization approach (perturbative splitting functions \cite{Altarelli:1977zs}) in the case that the OPE is performed onto a physical perturbative state (such as a quark or gluon).\footnote{Amusingly, (to our understanding) Parisi seems to have been involved both in the introduction of conformal partial waves \cite{Ferrara:1972kab,Ferrara:1972uq}, and in the introduction of perturbative splitting functions \cite{Altarelli:1977zs}. In this section we will see the interplay of these two ideas.} In this case the celestial blocks also have a transparent interpretation as incorporating ``kinematic" power corrections associated with the expansion of the perturbative propagator.

A convenient feature of the perturbative QCD result for the three-point energy correlator (as compared to the $\cN=4$ result), is that despite being more complicated, it enables us to look at particular partonic channels. In other words, we can tag particular partonic configurations to isolate specific operators in the OPE,  simplifying its interpretation. This will allow us to make a closer connection to the language of QCD splitting functions, where one factorizes onto (nearly on-shell) quark or gluon intermediate states. 

\begin{figure}
\begin{center}
\subfloat[]{
\includegraphics[scale=0.28]{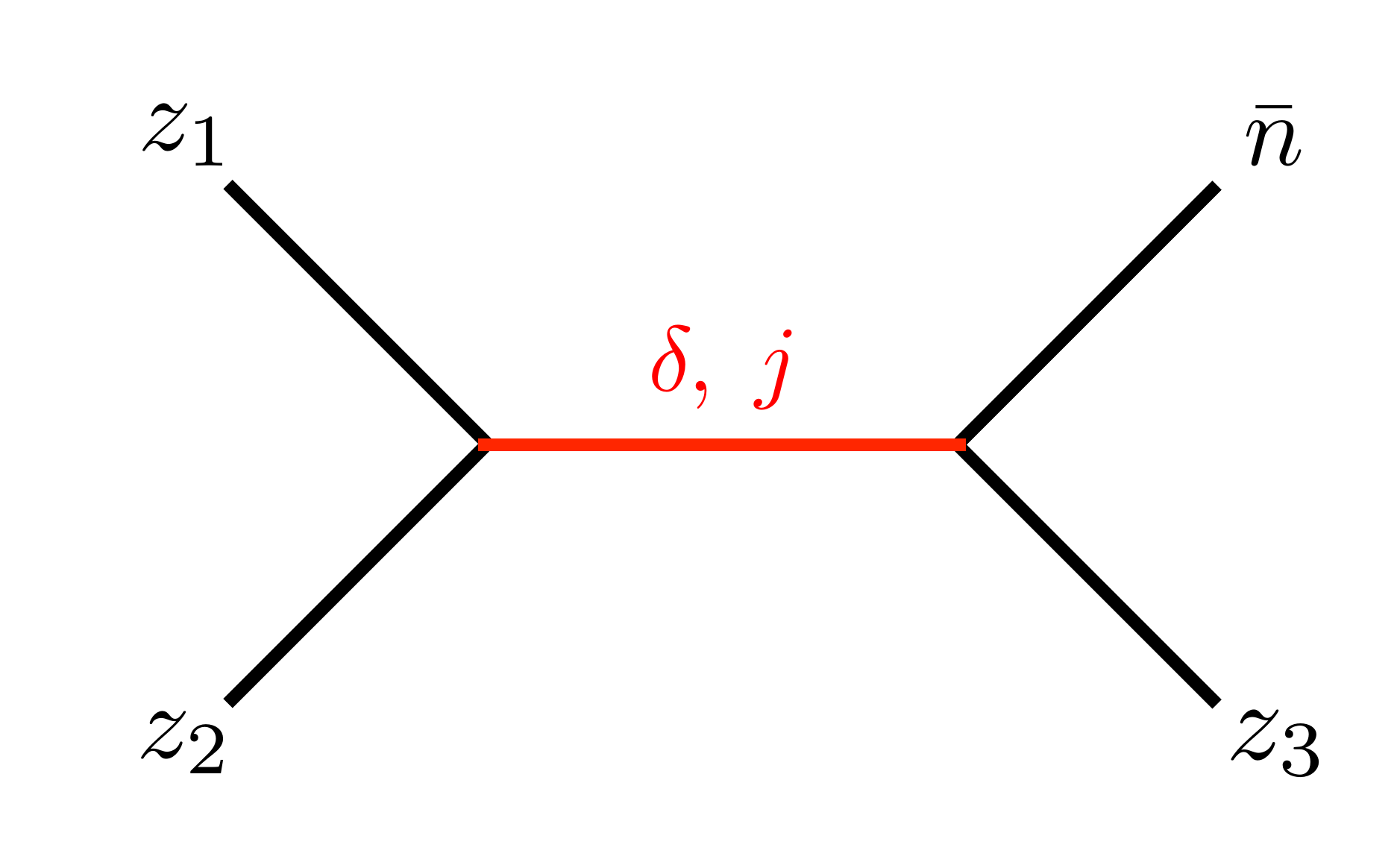}\label{fig:OPE_schematic_a}
}
\subfloat[]{
\includegraphics[scale=0.28]{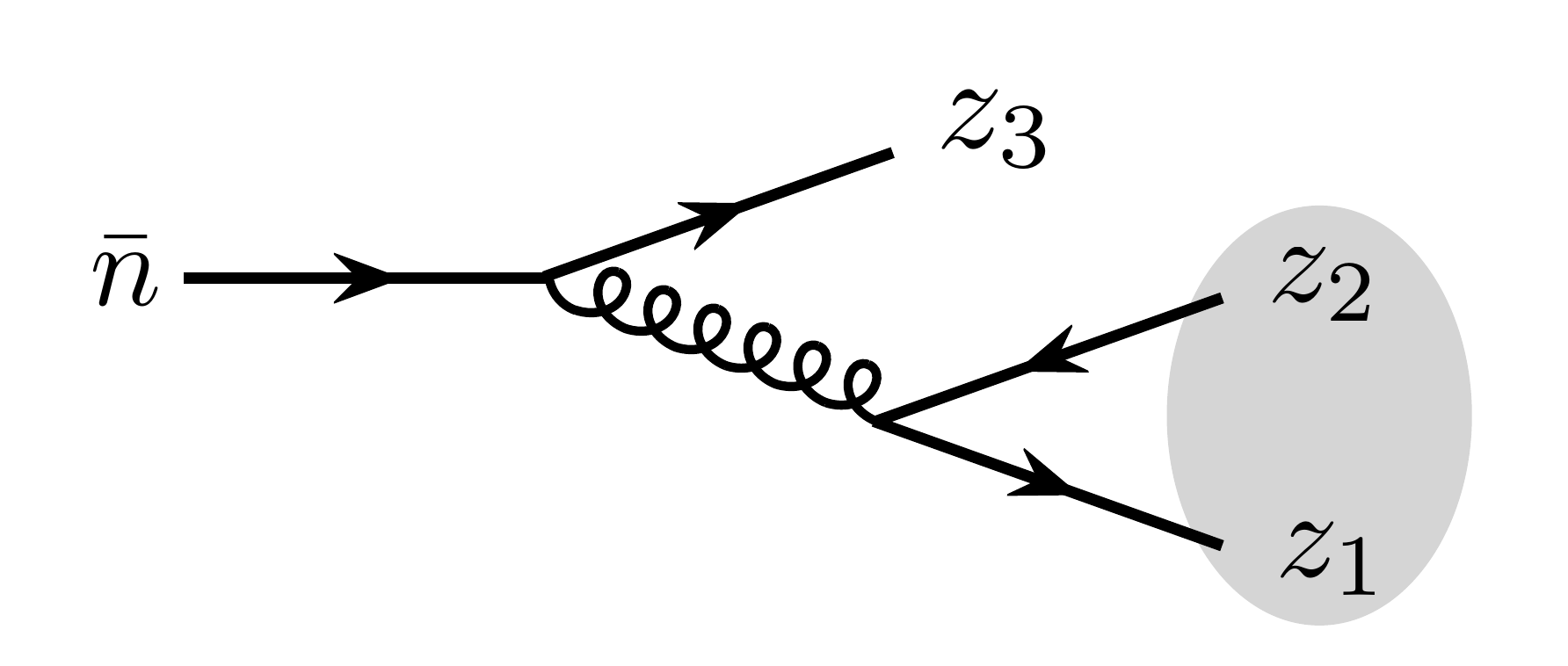}\label{fig:OPE_schematic_b}
}\qquad
%\captionsetup{font={footnotesize}}
\end{center}
\caption{For the simple case of the $q\to q^\prime \bar{q}^\prime q$ contribution to the three-point correlator, the OPE of the $q^\prime \bar{q}^\prime$ states isolates a perturbative gluon state with $j=0,2$. In this case the celestial block has the simple interpretation as kinematic power corrections to the leading splitting function. }
\label{fig:OPE_schematic}
\end{figure}

While the discussion of the light-ray OPE in terms of states of specific quantum numbers being pass between operators may seem quite abstract, it in fact has a transparent physical interpretation in perturbation theory. To see this in a particularly simple context,   we can consider the partonic channel where we squeeze a $q^\prime \bar{q}^\prime$ pair in the collinear splitting process $q\to q^\prime \bar{q}^\prime q$. This is illustrated in \Fig{fig:OPE_schematic}. To the order in perturbation theory at which we work, we have a perturbative gluon state in this channel,  and therefore we expect to see this from the perspective of the light-ray OPE. In particular, one expects that the light-ray states appearing in this OPE should only have transverse spin $j=0,2$.

The expansion of the squeezed limit for this particular configuration is given by
\begin{equation}
{\footnotesize
    \begin{split}
        &G(z,\bar{z}) =\\
        &\frac{1}{\pi^2}\left(\frac{\alpha_s}{4\pi}\right)^2\times\frac{1}{720}z^2\bar{z}^2\left[
    \left(-\frac{z}{\bar{z}}-\frac{\bar{z}}{z}+\frac{39}{10}\right)+\left(-\frac{z^2}{\bar{z}}-\frac{\bar{z}^2}{z}+\frac{39 z}{20}+\frac{39 \bar{z}}{20}\right) \right.\\
    & +\left(-\frac{6 z^3}{7 \bar{z}}+\frac{229 z^2}{140}-\frac{6 \bar{z}^3}{7 z}-\frac{211 z \bar{z}}{140}+\frac{229 \bar{z}^2}{140} \right)
    +\left(-\frac{5 z^4}{7 \bar{z}}+\frac{207 z^3}{140}-\frac{233 z^2 \bar{z}}{140}-\frac{5 \bar{z}^4}{7 z}-\frac{233 z \bar{z}^2}{140}+\frac{207 \bar{z}^3}{140}\right)\\
    & + \left(-\frac{25 z^5}{42 \bar{z}}+\frac{4 z^4}{3}-\frac{73 z^3 \bar{z}}{42}-\frac{25 \bar{z}^5}{42 z}-\frac{73 z \bar{z}^3}{42}+\frac{4 \bar{z}^4}{3}
    -\frac{152}{49} z^2 \bar{z}^2-\frac{15}{7} z^2 \bar{z}^2 \log (z \bar{z})\right)\\
    & + \left(-\frac{z^6}{2 \bar{z}}+\frac{1003 z^5}{840}-\frac{1517 z^4 \bar{z}}{840}-\frac{28739 z^3 \bar{z}^2}{5880}-\frac{28739 z^2 \bar{z}^3}{5880}-\frac{\bar{z}^6}{2 z}-\frac{1517 z \bar{z}^4}{840}+\frac{1003 \bar{z}^5}{840}\right. \\
    &+ \left.\left(-\frac{45}{14} z^3 \bar{z}^2-\frac{45 z^2 \bar{z}^3}{14}\right) \log (z \bar{z})\right)
    + \left(-\frac{14 z^7}{33 \bar{z}}+\frac{469 z^6}{440}-\frac{1541 z^5 \bar{z}}{840}-\frac{13172471 z^4 \bar{z}^2}{2134440}-\frac{14 \bar{z}^7}{33 z}+\frac{469 \bar{z}^6}{440}\right.\\
    & \left. \left.-\frac{5211499 z^3 \bar{z}^3}{711480}-\frac{13172471 z^2 \bar{z}^4}{2134440}-\frac{1541 z \bar{z}^5}{840}
    +\left(-\frac{295}{77} z^4 \bar{z}^2-\frac{1101 z^3 \bar{z}^3}{154}-\frac{295 z^2 \bar{z}^4}{77}\right) \log (z \bar{z}) \right)
    +\cdots \right].
    \end{split}
}
\end{equation}
Here we see more and more harmonics in $\arg z$ as we expand to higher powers. However this is an artifact of using a naive Taylor expansion, instead of performing an expansion in terms of celestial blocks with well defined quantum numbers under the Lorentz group. 

Using our basis of celestial blocks, we can re-write this expansion as
\begin{equation}\label{eq:sq_2}
    \begin{split}
        &G(z,\bar{z})= \\
        &\frac{1}{\pi^2}\! \left(\frac{\alpha_s}{4\pi}\right)^{\! 2}\!\! \left[  
            -\frac{1}{720}G_{4,2}(z,\bar{z}) \!+\! \frac{163}{252000}G_{6,2}(z,\bar{z})\! -\! \frac{2057}{4233600}G_{8,2}(z,\bar{z}) \!-\!\frac{82667}{768398400}G_{10,2}(z,\bar{z})\!+\!\cdots
        \right.\\
       & \left. \;\; + \frac{13}{2400} G_{4,0}(z,\bar{z}) - \frac{139}{40320} G_{6,0}(z,\bar{z})-\frac{10211}{5880000}G_{8,0}(z,\bar{z}) + \cdots \right.\\
       & \left. \;\; -\frac{1}{168}\partial_{\delta} G_{8,0}(z,\bar{z})-\frac{1}{1386}\partial_{\delta}G_{10,2}(z,\bar{z})+\cdots \right],
    \end{split}
\end{equation}
where we see only $j=0,2$ celestial blocks! While this is obvious from the light-ray OPE perspective, it is highly non-trivial from the naive expansion of the perturbative result, showing that symmetries play a large role. This provides an explanation for the ``hidden" structure observed in \Sec{sec:pQCD}.

One of the reasons we find this particularly fascinating is the interplay between the perturbative functions (logs, polylogs, etc), and the symmetry structure of the light-ray OPE. From the perspective of the functions appearing in the three-point function, the result of \Eq{eq:sq_2} is quite miraculous, while from the perspective of the OPE it is obvious. This suggests that these two different perspectives could be powerfully combined into some form of perturbative bootstrap.

%%%%%%%%%%%%%%%%%%%%%%%%%%%%%%%%%%%%%%%%%%%%%%%%%%%%%%%%%%%%%%%%%%%%%%%%%%%%%%%%
\subsection{Celestial Blocks and Kinematic Power Corrections}
%%%%%%%%%%%%%%%%%%%%%%%%%%%%%%%%%%%%%%%%%%%%%%%%%%%%%%%%%%%%%%%%%%%%%%%%%%%%%%%%

From the perspective of perturbative QCD factorization, it should be clear that the coefficients of the leading celestial blocks in \Eq{eq:sq_2} can be easily computed using iterated $1\to 2$ (polarized) splitting functions, which are based on the factorization onto intermediate quark and gluon states in the squeezed limit. This was explicitly shown in \cite{Chen:2020adz,Chen:2021gdk}. However, the symmetry structure of the light-ray OPE  tells us much more, namely that this coefficient should be associated with an entire celestial block. It is therefore interesting to understand the interpretation of the subleading power corrections associated with the celestial block from a perturbative QCD perspective.

Although much less developed than the conformal block expansion, when studying QCD at subleading power, there is an analogous characterization of power corrections as ``dynamic" or ``kinematic". Dynamic power corrections arise from genuinely new operators that are not related to leading power operators by symmetry, while ``kinematic" power corrections are entirely fixed by symmetries (and often arise through the expansion of some trivial kinematic factor, hence their name). For detailed discussions of dynamic vs. kinematic power corrections in the Sudakov regime, see  \cite{Moult:2016fqy,Moult:2017jsg,Ebert:2018gsn,Ebert:2018lzn,Moult:2019uhz}).  Since celestial blocks are simply kinematical functions,  one expects that they are related to a form of kinematical power corrections. Due to the simplicity of the energy correlator measurement function, there are no kinematic power corrections arising from its expansion, and so the kinematic power corrections should come from the expansion of the propagator. We will now make this precise using this simple example of the squeezed $q'\bar q'$ channel, where we will see how to reproduce the hypergeometric functions of the celestial blocks from the Taylor expansion of the perturbative propagators.

We consider the highest transverse spin series in the $q\to q^\prime \bar{q}^\prime q$ channel. The relevant $1 \to 3$ splitting function is given by
 \begin{equation}
    P_{q\to q^\prime_1 \bar{q}^\prime_2 q_3}\!\! = C_F T_F \frac{s_{123}}{2 s_{12}} \! \left[\frac{4\xi_3+(\xi_1-\xi_2)^2}{\xi_1+\xi_2}
    - \frac{[\xi_1(s_{12}+2s_{23})-\xi_2(s_{12}+2s_{13})]^2}{(\xi_1+\xi_2)^2 s_{12} s_{123}} 
    +\xi_1+\xi_2-\frac{s_{12}}{s_{123}}\right]
 \end{equation}
 where $\xi_i$ are the momentum fractions of the final state partons. Written in terms of the $z,\bar{z}$ variables, this gives 
 \begin{equation}\label{eq:qqsplit_z}
    \begin{split}
        P_{q\to q^\prime_1 \bar{q}^\prime_2 q_3} \!\! = C_F T_F \!\! \left[  -\frac{2 \xi _3^2}{\left(\xi _1+\xi _2\right){}^2} \frac{1}{z^2}
        +\frac{\left(\xi _1^2+\xi _2^2\right) \left(\xi _1^2+2 \xi _2 \xi _1+\xi _2^2+2 \xi _3\right) \xi _3}{\xi _1 \xi _2 \left(\xi _1+\xi _2\right){}^2 } \frac{1}{z \bar{z}}
        -\frac{2 \xi _3^2}{\left(\xi _1+\xi _2\right){}^2 }\frac{1}{\bar{z}^2} \right.\\
        + \frac{\xi _1^4+\left(\xi _2+\xi _3-1\right) \xi _1^3+\xi _2 \left(\xi _2+\xi _3\right) \xi _1^2+\xi _2 \left(\xi _2^2+\left(\xi _3-1\right) \xi _2+4 \xi _3\right) \xi _1+\xi _2 \xi _3 \left(\xi _2^2+2 \xi _3\right)}{\xi _1 \left(\xi _1+\xi _2\right){}^2}\\
       \left. -\frac{\xi _3 \left(\xi _1^3+\left(\xi _2-2\right) \xi _1^2+\left(\xi _2^2+2 \xi _2-2 \xi _3\right) \xi _1+\xi _2^3+2 \xi _2 \xi _3\right)}{\xi _1 \left(\xi _1+\xi _2\right){}^2} \left(\frac{1}{z}+\frac{1}{\bar{z}}\right) \right].
    \end{split}  
 \end{equation}
Recall from  \Sec{sec:pQCD} that to compute the three-point correlator, one must integrate over the energy fractions in the splitting function with a weight of $1/s_{123}^2$. From the form of \Eq{eq:qqsplit_z}, we see that  $P_{q\to q^\prime_1 \bar{q}^\prime_2 q_3}$ does not itself contribute to an infinite series in $z,\bar{z}$.  Therefore, for this particular channel, the infinite series of the celestial blocks comes only from the kinematic power corrections to  the propagator 
 \begin{equation}
     \frac{1}{s_{123}^2} = \frac{1}{E_J^4 x_L^2}\frac{1}{\left(\xi _1 \xi _3+\xi _1 \xi _2 z \bar{z}+\xi _3 \xi _2 (z-1) (\bar{z}-1)\right){}^2}.
 \end{equation}
 
To obtain the hypergeometric function appearing in the celestial blocks, we assume  that $z,\bar{z}$ are independent variables, which amounts to Wick rotating the Euclidean celestial sphere to the Lorentzian celestial sphere, and we Taylor expand in $\bar{z}$ to leading terms
 \begin{equation}
     \frac{1}{s_{123}^2}\to\frac{1}{E_J^4 x_L^2} \frac{1}{(\xi_1\xi_3+\xi_2\xi_3(1-z))^2}, \quad P_{q\to q^\prime_1 \bar{q}^\prime_2 q_3}\to -C_F T_F \frac{2 \xi _3^2}{\left(\xi _1+\xi _2\right){}^2 }\frac{1}{\bar{z}^2}.
 \end{equation}
 The phase space integral in this limit is proportional to 
 \begin{equation}
     \int \!\! d\xi_1 d\xi_2 d\xi_3 (\xi_1\xi_2\xi_3)^2\delta(1-\xi_1-\xi_2-\xi_3)\frac{P_{q\to q^\prime_1 \bar{q}^\prime_2 q_3}}{s_{123}^2}
     \!\to\! - \frac{2 C_F T_F}{E_J^4 x_L^2 \bar{z}^2}\!\!\! \int\!\! \frac{d\xi_1 d\xi_2 d\xi_3\, \xi_1^2\xi_2^2\xi_3^4\delta(1-\xi_1-\xi_2-\xi_3)}{(\xi_1+\xi_2)^2(\xi_1\xi_3+\xi_2\xi_3(1-z))^2}
 \end{equation}
in which the integral gives exactly the hypergeometic function obtained for highest transverse spin series
\begin{equation}
    -\frac{C_F T_F}{E_J^4 x_L^2 \bar{z}^2} \left[ \frac{z \left(z^2-12 z+12\right)+6 (z-2) (z-1) \log (1-z)}{18 z^5} \right] = -\frac{C_F T_F}{180 E_J^4 x_L^2 \bar{z}^2} \left._2 F_1\right.(2,3,6,z).
\end{equation}
This shows that (at least in this simple case), the celestial block expansion is indeed reproducing the expansion of the Lorentz invariant propagator.

On the other hand, if we consider the limit $z \to 0$ while $\bar{z}$ being finite, the leading term contains $\left._2 F_1\right.(2,3,6,\bar{z})$. 
However, the block $G_{\delta=4,j=2}(z,\bar{z})$ contains both $\left._2 F_1\right.(2,3,6,z)$ and $\left._2 F_1\right.(2,3,6,\bar{z})$, which is not clear how to achieve in a simple Taylor expansion when considering kinematic corrections. As we go to higher subleading powers, the relation between the block structures and the naive Taylor expansion of kinematic corrections is further obscured. This further illustrates the power of the systematic approach of the expansion in conformal blocks.

%%%%%%%%%%%%%%%%%%%%%%%%%%%%%%%%%%%%%%%%%%%%%%%%%%%%%%%%%%%%%%%%%%%%%%%%%%%%%%%%
\section{Analyticity in Transverse Spin}\label{sec:analyticity}
%%%%%%%%%%%%%%%%%%%%%%%%%%%%%%%%%%%%%%%%%%%%%%%%%%%%%%%%%%%%%%%%%%%%%%%%%%%%%%%%

\begin{figure}
\begin{center}
\subfloat[]{
\includegraphics[scale=0.28]{figures/hao_squeezed_gluon.pdf}\label{fig:inversion_a}
}
\subfloat[]{
\includegraphics[scale=1.88]{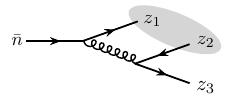}\label{fig:inversion_b}
}\qquad
%\captionsetup{font={footnotesize}}
\end{center}
\caption{Two distinct OPE channels. In (a) we squeeze onto an on-shell gluon state associated with a $1/\theta^2$ singularity in the OPE. This singularity must be reproduced  by the OPE in the crossed channel shown in (b). This requires an infinite number of operators of increasingly high transverse spin, and leads to analyticity in transverse spin of the light-ray OPE. }
\label{fig:inversion}
\end{figure}

Having derived the structure of the celestial blocks, and explored their relation with standard leading power factorization, in this section we investigate more systematically the higher twist structure of the OPE limit of light-ray operators. To do so, we will use the Lorentzian inversion formula \cite{Caron-Huot:2017vep}. While this formula is now widely used in the CFT community (see e.g. \cite{Albayrak:2019gnz,Caron-Huot:2020ouj,Liu:2020tpf,Atanasov:2022bpi} for applications related to the numerical conformal bootstrap \cite{Poland:2018epd}), it has not previously been applied in the context of jet substructure. Part of the goal of this section will therefore be to provide a pedagogical introduction to this technique. We will then use this approach to show that the OPE data in the light-ray OPE is analytic in transverse spin, much like the analyticity in spin for OPE data of correlation functions of local operators.   

While this story is by now familiar in the context of the OPE of local operators, we briefly review its physical origin. In \Fig{fig:inversion} we show two different OPE channels for a particular partonic configuration arising in the perturbative calculation of the three-point correlator. In (a), we perform the OPE onto a physical gluon state, giving rise to a $1/\theta^2$ singularity associated with the on-shell physical state. However, we can also perform the OPE in the cross channel, illustrated in (b). This OPE must also be able to reproduce the $1/\theta^2$ singularity. This is only possible from an infinite sum over operators with infinitely high transverse spin. Identical considerations apply to the OPE of local operators, but with spin instead of transverse spin. The study of the bootstrap equations in this high spin limit has seen significant attention \cite{Alday:2007mf,Komargodski:2012ek,Fitzpatrick:2012yx,Alday:2015ota,Alday:2015eya,Alday:2016njk}, and more recently has been efficiently codified in the Lorentzian inversion formula  \cite{Caron-Huot:2017vep,Simmons-Duffin:2017nub}. Here we will extend this to the study of transverse spin for light-ray operators, illustrating the essential role that transverse spin \cite{1822249} plays in understanding the structure of multi-point correlators of light-ray operators.

As shown above, the shape dependence of the three-point correlator can be expanded in terms of celestial blocks of definite quantum numbers, as
\begin{equation}
    G(z,\bar{z})=\sum_{\delta,j} c_{\delta,j} G_{\delta,j}(z,\bar{z})\,.
\end{equation}
The Lorentzian inversion formula \cite{Caron-Huot:2017vep,Simmons-Duffin:2017nub} provides an expression for the OPE coefficients $c_{\delta,j}$ in a manner that makes it manifest that they are analytic functions of $j$. This is a purely mathematical result, which arises due to the symmetry structure of the three-point correlator, combined with its analytic structure (the only singularities arise when light-rays are coincident), and behavior at infinity. However, this analyticity is quite remarkable, and shows that the correlators have an extremely rigid structure. Recall that at weak coupling higher transverse spin operators have higher twist. The Lorentzian inversion formula therefore says that the higher twist corrections contribute as an analytic function of the transverse spin. While any jet substructure observable can be expanded in a power expansion, naively one does not expect any nice features of the power expansion, and indeed this is part of the reason that little is known about higher power corrections to jet substructure observables. On the other hand, for the energy correlators not only do we have access to the complete structure of higher twist operators, but their contributions are analytic functions of the transverse spin! 

%%%%%%%%%%%%%%%%%%%%%%%%%%%%%%%%%%%%%%%%%%%%%%%%%%%
\subsection{Lorentzian Inversion Review and Tutorial}
%%%%%%%%%%%%%%%%%%%%%%%%%%%%%%%%%%%%%%%%%%%%%%%%%%%

%%%%%%%%%%%%%%%%%%%%
\begin{figure}
%[htp]
\begin{center}
\includegraphics[scale=0.5]{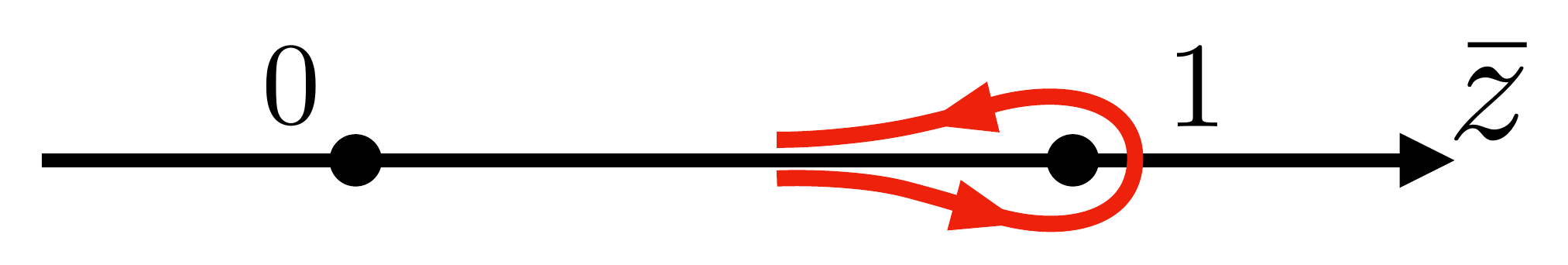}
\end{center}
\caption{The analytic continuation about $\bar z=1$, denoted by $\cG(z,\zb^\circlearrowleft)$, used in the Lorentzian inversion formula. $\cG(z,\zb^\circlearrowright)$ denotes the analytic continuation in the opposite direction.}
\label{fig:contour}
\end{figure}
%%%%%%%%%%%%%%%%%%%%

In this section we present a brief tutorial on Lorentzian inversion techniques. Let's start with a review of the main result of \cite{Caron-Huot:2017vep}; given a four-point function of scalar operators in a $d$-dimensional CFT, the OPE data density can be reconstructed from the Euclidean correlator minus its continuation onto the Lorentzian sheets. Concretely, we consider the correlator
\beq
\lb \cO_4(x_4)\cO_3(x_3)\cO_2(x_2)\cO_1(x_1)\rb = \frac{1}{(x_{12}^2)^{\frac{1}{2}(\Delta_1+\Delta_2)}(x_{34}^2)^{\frac{1}{2}(\Delta_3+\Delta_4)}}\Big(\frac{x_{14}^2}{x_{24}^2}\Big)^a\Big(\frac{x_{14}^2}{x_{13}^2}\Big)^b\cG(z,\zb)\,,
\eeq and represent it as an integral of the OPE data multiplying conformal blocks plus shadows over exchanged dimension $\Delta$,\footnote{See section 3.1 and equation (3.4) of \cite{Caron-Huot:2017vep} for a discussion of the partial waves.}
\beq\label{continuousexpansion}
\cG(z,\zb) = \sum_J \int_{\frac{d}{2} -i\infty}^{\frac{d}{2} +i\infty}\frac{d\Delta }{2\pi i} \,\,\Big[c^t(\Delta, J) +(-1)^J c^u(\Delta, J) \Big]G_{\Delta, J}^{(a,b)}(z,\zb) + \text{(non-norm)}\,.
\eeq
If we define the double-discontinuity as
\beq\label{dDisc}
\dDisc[\cG(z,\zb)]=\cos(\pi(a+b))\cG(z,\zb)-\frac{1}{2}e^{i\pi (a+b)}\cG(z,\zb^\circlearrowleft)-\frac{1}{2}e^{-i\pi (a+b)}\cG(z,\zb^\circlearrowright)\,,
\eeq where the arrow indicates the direction of the analytic continutation of $\zb$ around the branch point at $\zb=1$ to reach the Lorentzian sheets, as shown schematically in \Fig{fig:contour}. The inversion formula states \cite{Caron-Huot:2017vep, Simmons-Duffin:2017nub}
\beq\label{lorentzinversion}
c^t(\Delta,J) = \frac{\kappa_{J+\Delta}}{4}\int_0^1 dz d\zb \,\,\mu(z,\zb) G_{\Delta+1-d, J+d-1}(z,\zb) \dDisc[\cG(z,\zb)]\,,
\eeq where the measure is defined as
\beq
\mu(z,\zb)=\Big|\frac{z-\zb}{z\zb}\Big|^{d-2}\frac{((1-z)(1-\zb))^{a+b}}{(z\zb)^2}\,.
\eeq
The $u$-channel data is determined by a similar formula, with a few important differences. The first is the integral over $z$, $\zb$ runs from $-\infty$ to zero. Additionally, the conformal block (see eq. (\ref{2dBlock})) that is being integrated against is modified slightly via the replacement $k_h(x)= x^{h}\times(\cdots) \to (-x)^{h}\times(\cdots) $. As we'll see below, by change of variables and hypergeometric identities, the difference amounts to an exchange of operators $\mathcal{O}_1$ and $\mathcal{O}_2$ since $a\to-a$. Due to the symmetry of the OPE data and conformal blocks, we can neglect the second ``pure power" term of the conformal block, \textit{i.e.} we integrate against only one $k_{h}(z)k_{h'}(\zb)$.

The analogy between the conformal group and the rotation group was described in section \ref{sec:cas_a} when deriving the conformal Casimir. When expanding in spherical partial waves, we often simply integrate our function against the spherical harmonics and use orthogonality to read off the coefficients. A similar argument can be used for the conformal partial waves (the $F_{\Delta, J}$ above) and results in a ``Euclidean inversion formula" \cite{Caron-Huot:2017vep}. On the other hand, the ``Lorentzian inversion formula" in eq. (\ref{lorentzinversion}), obtained by analytic continuation, involves an integral over a simpler kinematical region which makes analyticity in spin manifest and computations simpler.

Based on the results from section \ref{sec:cas_b}, we expect to write our EEEC in terms of a sum of conformal blocks. To do this, we simply plug our expression in eq. (\ref{eq:three_point_N4}) into the inversion formula, use the values of $a=0$ and $b=-1$ described in section \ref{sec:cas_a} and replace $\Delta \to \delta$, $J\to j$ to indicate that we are working on the celestial sphere. The final step is to deform the $\delta$ contour to the positive real axis and pick up a discrete sum of residues. Note that there are additional poles in the complex $\delta$ plane which do not contribute due to symmetry constraints \cite{Kos:2013tga, Caron-Huot:2017vep}. We are left with
\beq\label{discretesum}
\mathcal{G}( z ,  \zb )= \sum_{j,\delta} \big(c^t_{j,\delta} +(-1)^j c^u_{j,\delta} \big)G_{\delta,j}(z,\zb) + \sum_{j,\delta} \big(\tilde{c}^t_{j,\delta} +(-1)^j \tilde{c}^u_{j,\delta} \big)\partial_\delta G_{\delta,j}(z,\zb)\,,
\eeq where we explicitly write the tilde coefficients from double poles in the $\delta$-plane in anticipation of the logarithms appearing in the EEEC correlators.

%%%%%%%%%%%%%%%%%%%%%%%%%%%%%%%%
\subsubsection{A Sample Inversion}
%%%%%%%%%%%%%%%%%%%%%%%%%%%%%%%%

To describe the inversion techniques, we'll work out a specific example calculation from $\mathcal{N}=4$ SYM \footnote{For additional examples of inversions and $\dDisc$ manipulations, see refs. \cite{Alday:2017vkk, Caron-Huot:2018kta, Alday:2019clp, Henriksson:2020jwk, Chicherin:2020azt, Caron-Huot:2020nem}}. Let's consider the weight one contribution to the EEEC,
\beq\label{weightone}
\mathcal{G}_1(u,v) = -u^2\Bigg(\frac{\log u}{2 v}(1+v) + \frac{\log v}{2v}(1+u)\Bigg)\,.
\eeq
We first rewrite in terms of the cross-ratios $z, \zb$ and expand near $\zb=1$ -- only the terms with brach points or singularities there will have non-vanishing double discontinuity. In the weight one case there are three of such terms, $(1-\zb)^{-1}$, $\log(1-\zb)$ and $(1-\zb)^{-1}\log(1-\zb)$.

There are two main strategies for computing the dDisc of a functions that we find in the $\mathcal{N}=4$ and QCD EEEC. The first involves a direct calculation of the singular terms in the $1-\zb$ expansion, which we focus on now, while the second uses plus distributions, which we discuss in section \ref{plusdist}. We start by using eq. (\ref{dDisc}), where we recall that $a=0$ and $b=-1$ in our case,
\beq
\dDisc\Big[\frac{\zb^p}{(1-\zb)^p }\Big] = \frac{\zb^p}{(1-\zb)^p}\Big(\cos \big(\pi  (a+b)\big)-\cos \big(\pi  (a+b-2 p)\big)\Big)\equiv \frac{\zb^p}{(1-\zb)^p} f^{(a,b)}(p) \,.
\eeq
While this term looks vanishing in the $p \to 1$ limit, we'll see that this term does indeed contribute and this will be made clear through the plus distribution techniques. Plugging our dDisc into the $\zb$ integral of the inversion formula gives
\begin{align}\label{zbintegration}
\inv_{\zb}\Big[\zb^{-b}\frac{\zb}{(1-\zb) }\Big]&=f^{(a,b)}(p)\int_0^1 d\zb\, \frac{(1-\zb)^{a+b}}{\zb^2} \zb^{-b}\Big(\frac{\zb}{1-\zb}\Big)^p \zb^{h}\,_2F_1\Big(h+a,h+b,2h,\zb\Big)\nonumber\\
&=f^{(a,b)}(p)\frac{\Gamma (2h ) \Gamma (1-p) \Gamma (a+b-p+1) \Gamma \left(-b+p+h-1\right)}{\Gamma \left(a+h\right) \Gamma \left(h-b\right) \Gamma \left(b-p+h+1\right)}\nonumber\\
&\xrightarrow[p\to 1]{} \frac{2\pi^2\Gamma(2h)}{\Gamma(h-1)\Gamma(h)}\,,
\end{align} where $\inv_{\zb}[x]$ denotes the $\zb$ integral over the measure, block and $\dDisc[x]$ and we made the $a$, $b$ substitution to obtain the last line. The integral above was computed by rewriting the hypergeometric function via the Euler identity
\beq
_2 F_1(a,b,c,z)=\frac{\Gamma(c)}{\Gamma(b)\Gamma(c-b)}\int_0^1dt\,\frac{u^{b-1}(1-u)^{c-b-1}}{(1-uz)^a}\,,
\eeq which leads to a separable integral after the replacement $z\to\frac{t}{1-u+tu}$. We also needed an extra factor of $\zb^{-b}$ to do this integral, which is factored into the correlator before doing the $\zb\sim1$ expansion. The $\zb$ integrals for the other terms are slightly more involved, although we can use
\begin{multline}
\dDisc\Big[\frac{\zb^p}{(1-\zb)^p }\log(1-\zb)\Big] \\ = -\partial_p\Big(\dDisc\Big[\frac{\zb^p}{(1-\zb)^p }\Big]\Big) + (\text{non-singular at $\zb=1$ and vanishing as $p\to0,1$}).
\end{multline} Using the integral in eq. (\ref{zbintegration}), we can compute (for $a=0$, $b=-1$)
\begin{align}\label{inversionsexplicit}
\inv_{\zb}\Big[\zb^{-b}\log(1-\zb)\Big] &=  -\frac{2 \pi ^2  \Gamma (2 h)}{\Gamma (h) \Gamma (h+1)}\,,\nonumber\\
\inv_{\zb}\Big[\zb^{-b}\frac{\zb}{(1-\zb) }\log(1-\zb)\Big] &=  \frac{2 \pi ^2 h \Gamma (2 h)}{\Gamma (h) \Gamma (h+1)}+\frac{2 \pi ^2 \Gamma (2 h) (-\psi(h)-\psi(h+1)+\psi(2)+\psi(1))}{\Gamma (h) \Gamma \left(h-1\right)}\,,
\end{align} 
where $\inv_{\zb}[x]$ denotes the $\zb$ integral over the measure, block and $\dDisc[x]$. Now only the $z$ integration remains in the inversion formula. Because we want the OPE data twist-by-twist, we can expand in powers of $z/(1-z)$, extract a factor of $z^{-b}$ and then integrate just as in eq. (\ref{zbintegration}) with the appropriate conformal block. As remarked in \cite{Alday:2017vkk} OPE coefficients corresponding to the blocks of order $z^k/(1-z)^k$ come from residues at the $\frac{\delta-j}{2}=1+ k + n$ poles  in the $\delta$-plane. Recall that the final step in rewriting the correlator in terms of a discrete block expansion involves a deformation of the principal series $\delta$ integral in eq. (\ref{continuousexpansion}) to lie along the real axis. In practice it is best to compute the residue before setting $k$ to be an integer.

It was shown in \cite{Chen:2021gdk} that there are logarithmic terms appearing in the QCD light-ray OPE expansion. These are a result of $\log(z/(1-z))$ terms in the expansion, which are integrated using a similar derivative trick as for the $\zb$ integration. The $\delta$ integration is slightly more subtle since these terms contain double poles -- we just have to remember to include the $\kappa_{j+\delta}$ functions along with block normalizations when computing the residue since the residue at a double pole involves $\delta$-derivatives of these coefficients, the OPE density and the blocks themselves. Hence we recover the logarithms in $z$ and $\zb$ due to these $\delta$-derivatives of blocks.

The $u$-channel OPE data is extracted in a similar manner. We first make the replacement $z\to \frac{-x}{1-x}$, $\zb\to\frac{-\xb}{1-\xb}$ so 
\beq
\int_{-\infty}^0 dz\, d\zb\, \frac{((1-z)(1-\zb))^{a+b}}{(z\zb)^2} \to \int_{0}^1 dx\, d\xb\,\frac{((1-x)(1-\xb))^{-(a+b)}}{(x\xb)^2}\,,
\eeq
and\footnote{Recall the $(-z)^h$ comes from the definition of the $u$-channel inversion.}
\beq
(-z)^h \, _2F_1(h+a,h+b,2h,z) \to (1-x)^b x^h\, _2F_1(h-a,h+b,2h,x)\,,
\eeq and likewise for $\zb$, where we used standard hypergeometric identitites. Comparing to the $t$-channel inversion, we simply have to extract a factor of $((1-x)(1-\xb))^{b+2a}$ from the correlator before expanding in $\xb$ and $x$, at least for $a=0$ (this is justified using the plus distribution techniques described below). Similar tricks can be applied for $a\neq0$ and after these subsitutions the $t$ and $u$-channel calculations can be done simultaneously. For instance, in the MHV case, the $t$ and $u$-channel contributions are identical after these manipulations (up to the $(-1)^J$ factor)

From these calculations we obtain the OPE data for the weight one contribution (\ref{weightone}) to sub-subleading order. We find, for $c_{\delta, j} = c^t_{\delta, j} + (-1)^j c^u_{\delta, j} $, 
\begin{align}
c_{j+4,j} &= \frac{1}{2 \Gamma (2 j+3)}\Big((-1)^j \big(-4 (j+1) H_j+2 (j+1) H_{2 j+2}+j-4\big) \Gamma (j+1) \Gamma (j+3)\nonumber\\&\quad\quad\quad-\Gamma (j+2) \big((j+2)^2 \Gamma (j+1)-2 H_{2 j+2} \Gamma (j+3)\big)\Big)\,,\nonumber\\
c_{j+6,j} &= \frac{1}{4 \Gamma (2 j+5)}\Big(\Gamma (j+3) \big(-4 (2 (-1)^j-1\big) H_{j+1} \Gamma (j+4)+2 \big((-1)^j+1\big) H_{2 j+4} \Gamma (j+4)\nonumber\\&\quad\quad\quad+\big(2 (-1)^j j^2+((-1)^j+3) j-7 (-1)^j+5) \Gamma (j+2)\big)\Big)\,,\nonumber\\
c_{j+8,j}&=\frac{\Gamma (j+4) }{{50 \Gamma (2 j+7)}}\Big(\big(5 (H_{j+2}-H_{j+3}+2 H_{2 j+6}\big) \Gamma (j+5)+(-1)^{j+1} \big(5 (H_{j+2}+H_{j+3}\nonumber\\&\quad\quad\quad+2 H_{j+4}-2 H_{2 j+6}) \Gamma (j+5)+10 \Gamma (j+3)+28 \Gamma (j+5))+37 \Gamma (j+5)\big)\Big)
\end{align}
where $H_n = \sum_{k=1}^n 1/k$ is the $n^{th}$ harmonic number. The OPE coefficients for derivatives of the conformal blocks are (as in eq. (\ref{discretesum}))
\begin{align}
\tilde{c}_{j+4,j}&=-\frac{\left((-1)^j+1\right) \Gamma (j+2) \Gamma (j+3)}{\Gamma (2 j+3)}\,,\nonumber\\
\tilde{c}_{j+6,j}&=-\frac{\left((-1)^j+1\right) \Gamma (j+3) \Gamma (j+4)}{2 \Gamma (2 j+5)}\,,\nonumber\\
\tilde{c}_{j+8,j}&=-\frac{\left((-1)^j+1\right) \Gamma (j+4) \Gamma (j+5)}{5 \Gamma (2 j+7)}\,.
\end{align}
This data will combine with the weight zero and weight two data in eq. (\ref{N4BlockExp}) to give the full block expansion of eq. (\ref{eq:three_point_N4}).

%%%%%%%%%%%%%%%%%%%%%%%%%%%%%%%%
\subsubsection{Inversions with Plus Distributions \label{plusdist}}
%%%%%%%%%%%%%%%%%%%%%%%%%%%%%%%%

It is significantly easier and more general to use plus distributions\footnote{For a review of plus distributions see  \cite{Ebert:2016gcn, Ebert:2018gsn}, and also \cite{Henn:2019gkr} for applications to Lorentzian inversion.} to extract the singular part of the $\zb$ integrands. We start with the distribution identity

\beq
\frac{1}{(1-\zb)^{1+\eps}} = -\frac{\delta(1-\zb)}{\eps} + \bigg[\frac{1}{(1-\zb)^{1+\eps}}\bigg]_+ =  -\frac{\delta(1-\zb)}{\eps} +  \bigg[\frac{1}{(1-\zb)}\bigg]_+ -\eps  \bigg[\frac{\log(1-\zb)}{(1-\zb)}\bigg]_+ +\cdots\,,
\eeq which we can differentiate to obtain expressions for higher-order singularitites,
\begin{align}\label{plussubs}
\frac{\log(1-\zb)}{(1-\zb)^{1+\eps}} &=  -\frac{\delta(1-\zb)}{\eps^2}  -  \bigg[\frac{\log(1-\zb)}{(1-\zb)}\bigg]_+ +\cdots\,, \nonumber\\\frac{\log^2(1-\zb)}{(1-\zb)^{1+\eps}} &= - 2\frac{\delta(1-\zb)}{\eps^3}  + \bigg[\frac{\log^2(1-\zb)}{(1-\zb)}\bigg]_+ +\cdots\,,\nonumber\\
\frac{1}{(1-\zb)^{2+\eps}} &=  \frac{\delta'(1-\zb)}{\eps}- \frac{\delta(1-\zb)}{1+\eps}+\bigg[\frac{1}{(1-\zb)^{2+\eps}}\bigg]_+ +\cdots \,,\nonumber\\
\frac{\log(1-\zb)}{(1-\zb)^{2+\eps}} &=  \frac{\delta'(1-\zb)}{\eps^2}- \frac{\delta(1-\zb)}{(1+\eps)^2} +\bigg[\frac{\log(1-\zb)}{(1-\zb)^{2+\eps}}\bigg]_+ +\cdots\,.
\end{align} 
Note that for higher power laws one often uses a notation of multiple plus distributions (see e.g. \cite{Ebert:2018gsn}), however, for simplicity of notation we will suppress this.
This expression will be sufficient for our example, the reproduction of eq. (\ref{inversionsexplicit}). We compute (with $a=0$ and $b=-1$)
\begin{equation}\label{explicitdDisc}
\dDisc\bigg[\zb^{-b}\frac{\zb\log(1-\zb)}{(1-\zb)^{1+\eps}}\bigg] = \frac{1}{(1-\zb)^{1+\eps}}\Big(4 \pi^2\zb^2 \,\eps -2\pi^2\zb^2 \log(1-\zb)\eps^2 + \cdots\Big)\,.
\end{equation} For our values of $a$ and $b$ we were able to use the following expansion of the hypergeometric around $\zb=1$ and for $n$ a positive integer ($n=1$ in our case):\footnote{See e.g. http://functions.wolfram.com/07.23.06.0015.01}
\begin{multline}\label{hyperexpansion}
_2F_1(h+a,h+b,2h+a+b+n,\zb) = \frac{(n-1)! \Gamma(2h+a+b+n)}{\Gamma(h+a+n)\Gamma(h+b+n)}\sum_{k=0}^{n-1}\frac{(h+a)_k(h+b)_k (1-\zb)^k}{k!(1-n)_k} \\ + \frac{\Gamma(2h+a+b+n)}{\Gamma(h+a)\Gamma(h+b)}(\zb-1)^n\sum_{k=0}^\infty\frac{(h+a+n)_k(h+b+n)_k}{k!(k+n)!}(-\log(1-\zb) + \psi(k+1)+\psi(k+n+1)\\-\psi(h+a+k+n)-\psi(h+b+k+n))(1-\zb)^k\,.
\end{multline}
With this we can compute
\begin{multline}
\inv_{\zb}\Big[\frac{\zb^2}{(1-\zb) }\log(1-\zb)\Big] = \int_0^1 \frac{d\zb}{\zb^{2-h}}\frac{1}{1-\zb}\Bigg(\frac{\Gamma(2h)}{\Gamma(h+1)\Gamma(h)} + \frac{\Gamma(2h)}{\Gamma(h)\Gamma(h-1)}(1-\zb)\\\times\big(\log(1-\zb) +\psi(h+1)+\psi(h)- \psi(1) - \psi(2)\big) + \mathcal{O}\big((1-\zb)^2\big)\Bigg)\dDisc\bigg[\frac{\zb^2\log(1-\zb)}{(1-\zb)^{1+\eps}}\bigg]\,.
\end{multline}
From here we substitute in the expression for the dDisc, eq. (\ref{explicitdDisc}), and rewrite the divergent pieces in terms of plus distributions using eq. (\ref{plussubs}). Keeping only the terms that are non-zero as $\eps\to0$ and integrating the remaining delta functions we find perfect agreement with eq. (\ref{inversionsexplicit}).

%%%%%%%%%%%%%%%%%%%%%%%%%%%%%%%%%%%%%%%%
\subsection{OPE Data in $\mathcal{N}=4$ SYM}
%%%%%%%%%%%%%%%%%%%%%%%%%%%%%%%%%%%%%%%%

\begin{figure}
%[htbp]
	\begin{center}
		\includegraphics[width=12 cm]{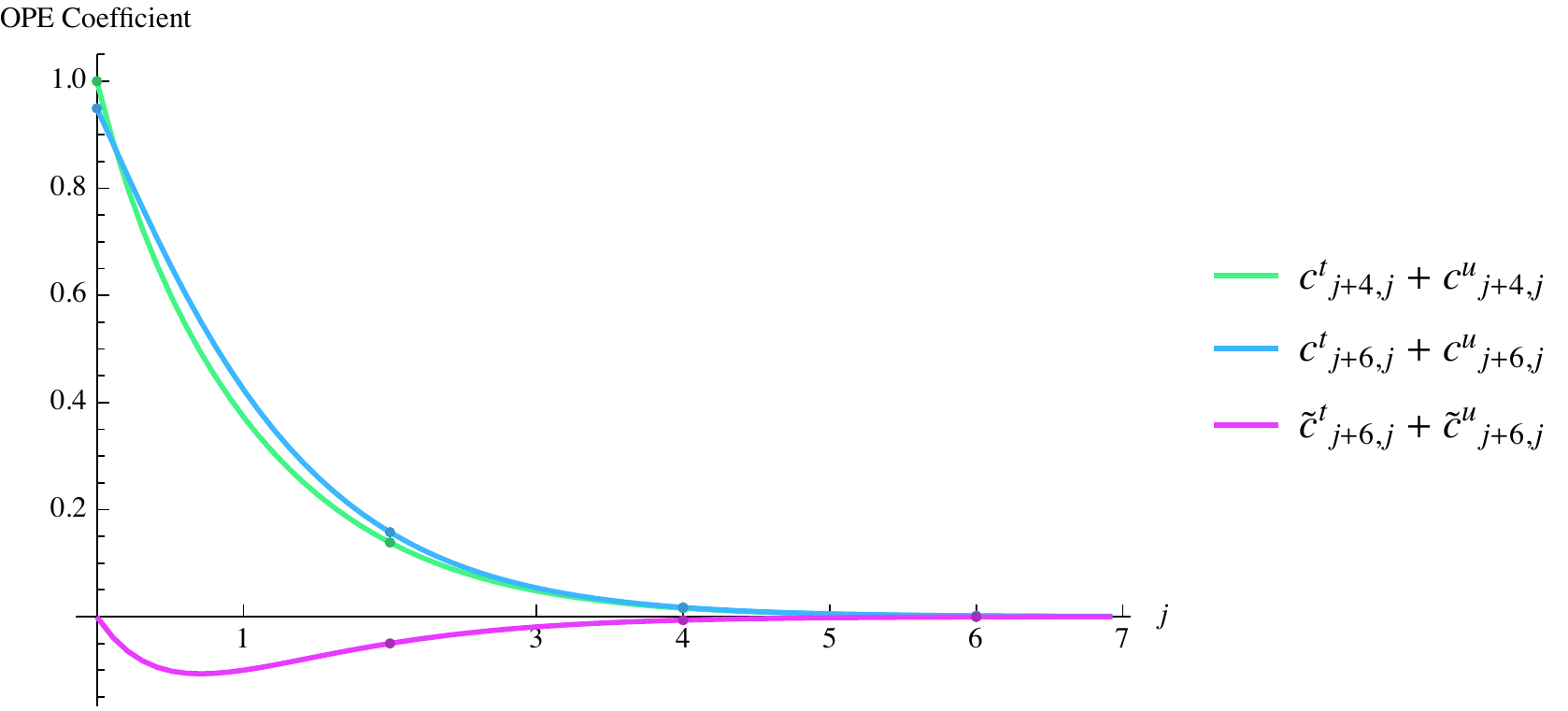}
		\caption{The even-spin inversion data plotted as a function of spin. The dots along the curve correspond to the coefficients appearing in the discrete OPE decomposition (eq. (\ref{discretesum})). In $\cN=4$ SYM, the coefficients agree with the analytic formula for all $j\geq 0$. The odd-spin coefficients are zero.}
		\label{fig: N4OPE}
	\end{center}
\end{figure}

Using the methods decribed above, we inverted the EEEC appearing in eq. (\ref{eq:three_point_N4}) to leading and subeading order, \textit{i.e.} for $\delta = j+4, j+6$. The data is plotted in \Fig{fig: N4OPE}. Using the notation in eq. (\ref{discretesum}) we find
%\begin{multline}
%c^t_{j+4,j}=c^u_{j+4,j} =\frac{\Gamma (j+1) \Gamma (j+2)}{12 (j+3) \Gamma (2 j+3)} \Big((j+3) \Big(\left(\pi ^2-6\right) j (j+3)+2 \pi ^2-9\Big)\\+6 \Gamma (j+2) \Gamma (j+3) \Gamma (j+4) \, _3\tilde{F}_2\big(j+1,j+2,j+3;j+4,2 j+4;1\big) \Big)\,,
%\end{multline} 
\begin{multline}
c^t_{j+4,j}=c^u_{j+4,j} =\frac{1}{12 \Gamma (2 j+3)}\Bigg(\frac{\Gamma (j+2)}{(j+3) \Gamma (2 j+4)} \bigg(6 \Gamma (j+1) \Gamma (j+2) \Gamma (j+3) \\\times\, _3F_2(j+1,j+2,j+3;j+4,2 j+4;1)+\Gamma (2 j+4) \Big(12 (2 H_j-H_{2 j+2}) \Gamma (j+4)+\\(j+3) (\pi ^2 (j+1) (j+2)-3 (6 j^2+8 j-9)) \Gamma (j+1)\Big)\bigg)+6 \Gamma (j+2) \Gamma (j+3)\\+6 \Big(-4 (j+1) H_j+2 (j+1) H_{2 j+2}+j-4\Big) \Gamma (j+1) \Gamma (j+3)\Bigg)\,,
\end{multline}
where $H_n$ is the $n^{th}$ harmonic number and $F$ is the generalized hypergeometric function.
%\begin{multline}
%c^u_{j+4,j} =\frac{\sqrt{\pi } 4^{-j-2} \Gamma (j+1)}{3 \Gamma \left(j+\frac{3}{2}\right)} \Big(\left(\pi ^2-6\right) j (j+3)+2 \pi ^2-9\\+6 \Gamma (j+2) \Gamma (j+3)^2 \, _3\tilde{F}_2(j+1,j+2,j+3;j+4,2 j+4;1)\Big)\,.
%\end{multline}
The logarithmic blocks appear at subleading twist. The data is
\begin{multline}
c^t_{j+6,j} =c^u_{j+6,j} =\frac{\Gamma (j+3) }{{8 \Gamma (2 j+5)}}\Big(12 \Gamma (j+2) \psi ^{(0)}(j+2)+12 \Gamma (j+2) \psi ^{(0)}(j+3)\\-\frac{6 \Gamma (j+3) \Gamma (j+4) \Gamma (j+2) \, _3F_2(j+2,j+3,j+4;j+5,2 j+6;1)}{(j+4) \Gamma (2 j+6)}\\\quad\quad\quad+2 H_{j+1} \Gamma (j+4)-6 H_{j+3} \Gamma (j+4)+4 H_{2 j+4} \Gamma (j+4)+6 j \Gamma (j+2)+29 \Gamma (j+2)\\-3 \Gamma (j+4)+\pi ^2 \Gamma (j+4)-24 \Gamma (j+2) \psi ^{(0)}(2 j+5)\Big)
\end{multline}
%\begin{multline}
%c^t_{j+6,j} =c^u_{j+6,j} =\frac{2^{-4 j-11} }{(j+2) (j+4) \Gamma (j+\frac{5}{2}) \Gamma (j+\frac{7}{2})}\bigg(-3 \pi  (j+2) \Gamma (j+2) \Gamma (j+4) \\\times _3F_2(j+2,j+3,j+4;j+5,2 j+6;1)-\sqrt{\pi } 4^{j+2} (j+4) \Gamma \Big(j+\frac{7}{2}\Big)\\\times \Big(\Gamma (j+2) \big(6 j^2+4 (j^2+7 j+10) j \psi ^{(0)}(j+2)-4 (j^2+7 j+10) j \psi ^{(0)}(2 j+5)\\+13 j-10\big)-(\pi ^2-3) (j+2) \Gamma (j+4)\Big)\bigg)\,,
%\end{multline}\josh{make data more clearly analytic in $j$.}
%and
%\begin{multline}
%c^u_{j+6,j} =\frac{2^{-4 j-11}}{(j+2) (j+4) \Gamma (j+\frac{5}{2}) \Gamma (j+\frac{7}{2})} \bigg(-3 \pi  (j+2) \Gamma (j+2) \Gamma (j+4) \\\times _3F_2(j+2,j+3,j+4;j+5,2 j+6;1)-\sqrt{\pi } 4^{j+2} (j+4) \Gamma \Big(j+\frac{7}{2}\Big) \\\times \Big(\Gamma (j+2) \big(-4 j^3-22 j^2+4 (j^2+7 j+10) j \psi ^{(0)}(j+2)-4 (j^2+7 j+10) j \psi ^{(0)}(2 j+5)\\-51 j-58\big)-(\pi ^2-7) (j+2) \Gamma (j+4)\Big)\bigg)\,.
%\end{multline} 
Finally, the logarithmic OPE coefficients which multiply the $\partial_\delta G_{\delta,j}|_{\delta=j+6}$ blocks are
\beq
\tilde{c}_{j+6,j}^t=\tilde{c}_{j+6,j}^u=-\frac{j (j+5) \Gamma (j+2) \Gamma (j+3)}{2 \Gamma (2 j+5)}\,.
\eeq
We see that no odd-spin blocks appear in the OPE decomposition. Combining this data and expanding to order $z^3$ gives
\begin{multline}\label{N4BlockExp}
u^3\mathcal{G}_{\mathcal{N}=4}(z)  =  G_{4,0}(z,\zb)+\left(\frac{107}{60}-\frac{\pi ^2}{6}\right) G_{6,2}(z,\zb)+\left(\frac{2843}{5040}-\frac{\pi ^2}{18}\right) G_{8,4}(z,\bar{z})\\
+\frac{1}{12} \left(41-3 \pi ^2\right) G_{6,0}(z,\bar{z})+\left(\frac{4883}{2800}-\frac{9 \pi ^2}{56}\right) G_{8,2}(z,\bar{z})\\+\left(\frac{33394601}{85377600}-\frac{5 \pi ^2}{132}\right) G_{10,4}(z,\bar{z})-\frac{1}{20}\partial_\delta G_{8,2}(z,\zb)\\-\frac{1}{154}\partial_\delta G_{10,4}(z,\zb) + \mathcal{O}(\zb^8, z^4)\,.
\end{multline}

%%%%%%%%%%%%%%%%%%%%%%%%%%%%%%%%%%%%%%%%
\subsubsection{Discussion and Open Questions}
%%%%%%%%%%%%%%%%%%%%%%%%%%%%%%%%%%%%%%%%

Throughout this section we have used the Lorentzian inversion formula as a purely mathematical device to extract the OPE coefficients in the celestial block expansion of the three-point energy correlator. This was facilitated by the fact that it has the mathematical form of a four point correlator of local operators, and shares the same conformal blocks as a 2d CFT. The Lorentzian inversion formula allows us to show that (at least to this order in perturbation theory), the OPE data is an analytic function of the transverse spin. This is quite interesting, since in the light-ray OPE the spin, $J$ is fixed (depending on the number of initial light-ray operators), and it is the transverse spin, $j$, that instead plays a role analogous to spin in the OPE of local operators. Since the study of analyticity in spin for correlators of local operators led to the study of light-ray operators \cite{Kravchuk:2018htv}, it is natural to ask if the analyticity in transverse spin observed in correlators of light-ray operators will lead to some more general structure that makes this manifest.

There are many aspects of the transverse spin that would be interesting to understand better, and from a more physical perspective. First, for correlators of local operators, the Lorentzian inversion formula relies on boundedness in the Regge limit \cite{Maldacena:2015waa}. While this can be checked in our perturbative results, it would be interesting to understand if it can be proven generically for correlators of light-ray operators. Furthermore, in our results for $\cN=4$, we find that the Lorentzian inversion works for all $j\geq 0$. For the case of normal spin, $J$, there is a clear physical interpretation of the lowest spin at which the Lorentzian inversion works, related to the Regge intercept of the theory (see e.g. \cite{Caron-Huot:2020ouj,Liu:2020tpf} for detailed discussions). It would be interesting to understand the physical interpretation of the Regge intercept in transverse spin. 

In this section we have also used the high transverse spin behavior of the energy correlators. For local operators, the high spin behavior is physically well understood, and exhibits many universal properties, such as the logarithmic scaling of anomalous dimensions with spin, and the relation to the cusp anomalous dimension and Wilson lines \cite{Korchemsky:1988si,Korchemsky:1992xv,Alday:2007mf}. It would be interesting to develop a similar understanding of the high transverse spin limit of light-ray operators, either from the perspective of light-ray operators, or from the perspective of operators living on the sphere. Perhaps the high transverse spin limit can be viewed as some form of Wilson line on the celestial sphere.  Therefore, despite our ability to apply the Lorentzian inversion formula, and study the high transverse spin limit,  many physical questions about the understanding of transverse spin in light-ray operators remain open, and would be interesting to understand better. 

Finally, we note that the Lorentzian inversion formula with transverse spin has also been considered for the case of correlators involving Wilson line defects \cite{Barrat:2021yvp,Barrat:2020vch}. Here the Wilson line behaves quite similarly to the light-like direction $n$, giving rise to similar kinematics as for the collinear limit of the three-point correlator. To our knowledge, an understanding of the Regge intercept in transverse spin in that situation is also not known.

%%%%%%%%%%%%%%%%%%%%%%%%%%%%%%%%%%%%%%%%
%%%%%%%%%%%%%%%%%%%%%%%%%%%%%%%%%%%%%%%%

%%%%%%%%%%%%%%%%%%%%%%%%%%%%%%%%%%%%%%%%
\subsection{A QCD Example}
%%%%%%%%%%%%%%%%%%%%%%%%%%%%%%%%%%%%%%%%

Although the collinear EEEC in QCD is much more complicated than that in $\mathcal{N}=4$ SYM, they share the same set of transcendental functions. Therefore the procedure of applying the Lorentzian inversion formula is essentially the same as has been discussed above. To keep the discussion short, we use the simplest $q\to q q^\prime \bar{q}^\prime$ channel to illustrate that the Lorentzian inversion formula is also useful in perturbative QCD. Results for the full expansion of the three-point correlator up to twist-8 will then be given in the next section.

We again use the $q\to q q^\prime \bar{q}^\prime$ partonic channel, and we use $\mathcal{E}_q$ to indicate flavor selection in addition to the ordinary energy weighting.  Instead of $\langle \mathcal{E}_{q^\prime}(\vec{n}_1) \mathcal{E}_{\bar{q}^\prime}(\vec{n}_2) \mathcal{E}_{q}(\vec{n}_3) \rangle_q$ whose transverse spins in the 1,2-OPE channel truncate at $j=2$, we are going to consider its crossed channel $\langle \mathcal{E}_q(\vec{n}_1) \mathcal{E}_{\bar{q}^\prime}(\vec{n}_2) \mathcal{E}_{q^\prime}(\vec{n}_3) \rangle_q$, see \Fig{fig:inversion}. By expanding the result and matching to the expressions for the conformal blocks, we find
\begin{equation}
\begin{split}
\langle \mathcal{E}_q(\vec{n}_1) \mathcal{E}_{\bar{q}^\prime}(\vec{n}_2) & \mathcal{E}_{q^\prime}(\vec{n}_3) \rangle_q \\
= \frac{1}{\pi^2}\left(\frac{\alpha_s}{4\pi}\right)^2 &\left[
\left(\frac{11939}{1200}-\pi ^2\right) G_{6,0} 
+\left( \frac{2953}{100}-3 \pi ^2 \right) G_{7,1}
+\left(\frac{233603}{4200}-\frac{79 \pi ^2}{14}\right) G_{8,2} \right.\\
&
+\left(\frac{662863}{8400}-8 \pi ^2\right) G_{9,3} 
+\left(\frac{9863251}{110880}-\frac{595 \pi ^2}{66}\right) G_{10,4}\\
&
+\left(\frac{99805933}{1201200}-\frac{1204 \pi ^2}{143}\right) G_{11,5} 
+\left(\frac{96097693}{1441440}-\frac{966 \pi ^2}{143}\right) G_{12,6}\\
&\left. 
+\left(\frac{3380047087}{71471400}-\frac{1059 \pi ^2}{221}\right) G_{13,7}
\right]+\text{higher celestial twist contributions},
\end{split}
\end{equation}
where we only show celestial twist $\delta-j=6$ blocks. As expected, we find a tower of infinite transverse spin states contributing in this channel. 

The $t$-channel analytic formula extracted from the Lorentzian inversion formula is
\begin{equation}
c_{j+6,j}^{t}= - \frac{8-25j-5 j^2}{60}\frac{\Gamma(j+3)\Gamma(j+4)}{\Gamma(2j+5)},
\end{equation}
while the formula for the $u$-channel is
%\begin{equation}
\begin{align}
&c_{j+6,j}^u = -\frac{\Gamma(j+2)\Gamma(j+3)^2 \Gamma(2j+7)}{1920\Gamma(j+5)\Gamma(2j+5)\Gamma(2j+6)} \times\\
&\Big[
 (j+1) (j+2)^2 (j+3)^2 (j+4)^2 \left(j^4+10 j^3+39 j^2+70 j+240\right) \!\left(\!\psi^{(1)}\!\left(\frac{j+3}{2}\right)\!-\psi ^{(1)}\!\left(\frac{j+4}{2}\right)\!\right) \nn\\
& -2 (j+4) \left(j^8+20 j^7+173 j^6+845 j^5+2738 j^4+6755 j^3+12156 j^2+12280 j+5256\right)
\Big]\,. \nn
\end{align}
%\end{equation}

%%%%%%%%%%%%%%%%%%%%%%%%%%%%%%%%%%%%%%%%%%
\begin{figure}
%[htbp]
\begin{center}
\includegraphics[width=12 cm]{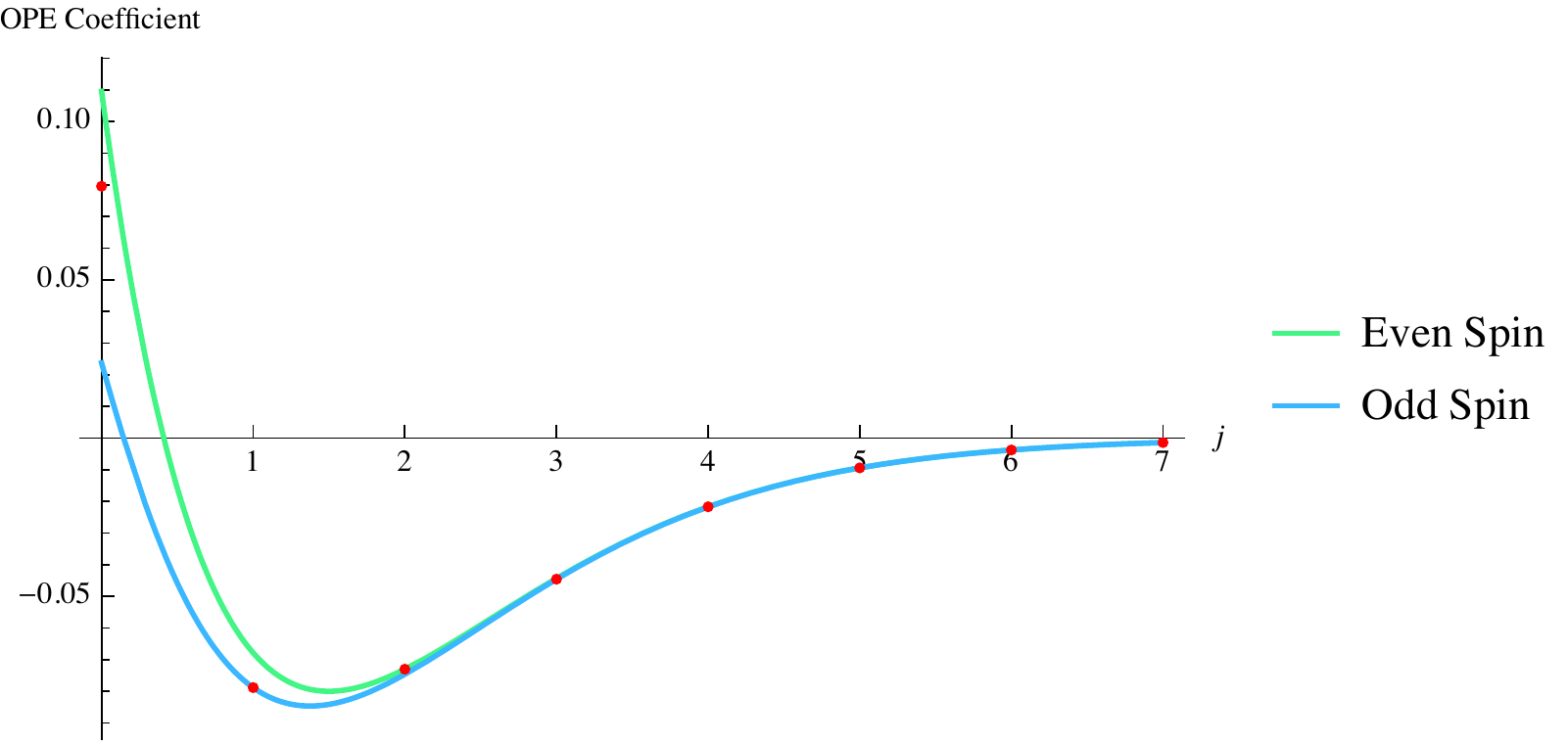}
\caption{The leading celestial twist data for the crossed channel. The red dots are the block coefficients $c_{j+6,j}$ found by expanding our perturbative result. The green and blue curves are obtained from the Lorentzian inversion formula, representing the even spin branch $c^t_{j+6,j} + c^u_{j+6,j}$ and  the odd spin branch $c^t_{j+6,j} - c^u_{j+6,j}$ respectively. In this example, the block coefficients agree with the analytic formula for $j\geq 1$.}
\label{fig: CrossChannelOPE}
\end{center}
\end{figure}
%%%%%%%%%%%%%%%%%%%%%%%%%%%%%%%%%%%%%%%%%%

For this particular channel, we find perfect agreement between the brute force block expansion and the analytic expression from the Lorentzian inversion formula when the transverse spin $j\geq 1$, as shown in \Fig{fig: CrossChannelOPE}. As for the case of the $\cN=4$ result, it would be interesting to understand the physical nature of this ``transverse spin Regge intercept".

From the analytic expression in this example, we see that the coefficients of both the even spin branch $\left(c^{t}_{j+6,j}+c^{u}_{j+6,j}\right)$ and the odd spin  branch $\left(c^{t}_{j+6,j}-c^{u}_{j+6,j}\right)$ are exponentially suppressed at large transverse spin
\begin{equation}
c^{t}_{j+6,j}+c^{u}_{j+6,j}, \;c^{t}_{j+6,j}-c^{u}_{j+6,j} \sim -\frac{\sqrt{\pi}}{3} \frac{j^{7/2}}{2^{2j+6}}.
\end{equation}
This is one of the reasons in favor of organizing the EEEC expansion using celestial blocks when study the power corrections. Transverse spin is dual to the angular distribution in the observable so that the celestial block expansion is better than a naive Taylor expansion in this aspect. This will be illustrated numerically in the next section. 

Again, we emphasize that it is quite remarkable that we are able to understand the high twist structure of jet substructure observables in QCD, and that they are analytic functions of the transverse spin! This illustrates the beautiful underlying structure of the energy correlators, made even more exciting by the fact that they can be measured in real LHC data \cite{Komiske:2022enw}.

%%%%%%%%%%%%%%%%%%%%%%%%%%%%%%%%%%%%%%%%%%%%%%%%%%%%%%%%%%%%%%%%%%%%%%%%%%%%%%%%
\section{Applications of Celestial Blocks to Jet Substructure}\label{sec:QCD_application}
%%%%%%%%%%%%%%%%%%%%%%%%%%%%%%%%%%%%%%%%%%%%%%%%%%%%%%%%%%%%%%%%%%%%%%%%%%%%%%%%

One of the practical interests in understanding the light-ray OPE for multi-point correlators, is that if it converges rapidly, it can be used to efficiently approximate complicated correlators using only a few terms. This could have significant phenomenological applications, enabling correlators that cannot be calculated exactly to be well approximated. 
While we have extracted the OPE data in previous sections, and discussed some of its theoretical properties,  in this section we take a more phenomenological approach, and numerically study the expansion of the three-point correlator for quark and gluon jets in QCD in terms of conformal blocks.

%%%%%%%%%%%%%%%%%%%%%%%%%%%%%%%%%%%%%%%%%%%%%%%%%%%%%%%%%%%%%%%%%%%%%%%%%%%%%%%%
\subsection{QCD Celestial Data to Twist-8}
%%%%%%%%%%%%%%%%%%%%%%%%%%%%%%%%%%%%%%%%%%%%%%%%%%%%%%%%%%%%%%%%%%%%%%%%%%%%%%%%

For completeness we begin by giving explicit results for the celestial block expansion of the EEEC for quark and gluon jets in QCD up to twist-8 (celestial dimension $\delta=10$). We write the EEEC jet functions at tree level in QCD as
\beq
\langle \mathcal{E}(\vec{n}_1)\mathcal{E}(\vec{n}_2)\mathcal{E}(\vec{n}_3) \rangle_i 
\equiv \frac{1}{32\pi} \frac{1}{x_L} \frac{d^3\sigma_i}{dx_L d\mathrm{Re} z \, d\mathrm{Im} z}
 = \frac{1}{\pi^2} \left( \frac{\alpha_s}{4\pi}\right)^2 \frac{1}{x_L^2} \frac{1}{(z\bar{z})^3}g_i(z),
\eeq
where $i=q,g$ refers to quark or gluon jet, and $z$ is the complex variable encoding the shape information of EEEC. Note that $\frac{1}{x_L} \frac{d^3\sigma_i}{dx_L d\mathrm{Re} z \, d\mathrm{Im} z}$ is related to $\frac{d^3\sigma_i}{d \zeta_{12} d\zeta_{23} d\zeta_{31}}$ by
\beq
\frac{1}{x_L}  \frac{d^3\sigma_i}{dx_L d\mathrm{Re} z \, d\mathrm{Im} z} = 2\sqrt{2\zeta_{12}
\zeta_{23}+2\zeta_{23}\zeta_{31}+2\zeta_{31}\zeta_{12}-\zeta_{12}^2 -\zeta_{23}^2-\zeta_{31}^2}  \frac{d^3\sigma_i}{d \zeta_{12} d\zeta_{23} d\zeta_{31}}.
\eeq
Since we chose to normalize to the jet energy $Q/2$, $g_i(z)$ here equals $4$ times $G_{i}(z)$ in \cite{Chen:2019bpb}.

For quark jets, the expansion up to celestial dimension $\delta=10$ is
\beq
\begin{split}
g_{q}(z) &=  C_F n_f T_F \left[
-\frac{1}{360}G_{4,2} + \frac{13}{1200}  G_{4,0}
+ \frac{163}{126000} G_{6,2} + \left(\frac{111199}{33600}-\frac{\pi ^2}{3}\right) G_{6,0} \right.\\
&\qquad -\frac{67}{420} \partial_{\delta}G_{8,0} + \left(\frac{39243247}{2116800}-\frac{79 \pi ^2}{42}\right) G_{8,2}
+ \left(\frac{201264317}{8820000}-\frac{7 \pi ^2}{3}\right) G_{8,0}\\
&\qquad -\frac{751}{4620} \partial_{\delta}G_{10,2} -\frac{12317}{18480} \partial_{\delta}G_{10,0}
+\left(\frac{9863251}{332640}-\frac{595 \pi ^2}{198}\right) G_{10,4}\\
&\qquad \left. +\left(\frac{2801569019}{64033200}-\frac{40 \pi ^2}{9}\right) G_{10,2} 
+\left(\frac{168438023821}{3585859200}-\frac{937 \pi ^2}{196}\right) G_{10,0} +\cdots
\right]\\
&+ C_F\left(C_F-\frac{C_A}{2}\right)
\left[
\left(\frac{38483}{3600}-\frac{13 \pi ^2}{12}\right) G_{6,2} + \left(\frac{50339}{3600}-\frac{17 \pi ^2}{12}\right) G_{6,0} 
-\frac{1}{105}\partial_{\delta}G_{8,0}
\right.\\
&\qquad  + \left(\frac{5057099}{302400}-\frac{61 \pi ^2}{36}\right) G_{8,4}
+ \left(\frac{2502343}{100800}-\frac{845 \pi ^2}{336}\right) G_{8,2} + \left(\frac{51053323}{1764000}-\frac{44 \pi ^2}{15}\right) G_{8,0}\\
&\qquad
+\frac{1}{120}\partial_{\delta}G_{10,2} -\frac{1}{60} \partial_{\delta}G_{10,0}
+ \left(\frac{77243299}{6177600}-\frac{1087 \pi ^2}{858}\right) G_{10,6} 
+ \left(\frac{28650319}{1900800}-\frac{2419 \pi ^2}{1584}\right) G_{10,4}\\
&\qquad \left. 
+ \left(\frac{30317549}{1512000}-\frac{5119 \pi ^2}{2520}\right) G_{10,2}
+\left(\frac{2564731}{120960}-\frac{421 \pi ^2}{196}\right) G_{10,0} +\cdots
\right]\\
&+ C_F^2 \left[
\frac{1}{5} G_{4,0}+ \frac{47}{1800}G_{6,2}+\frac{59}{360}G_{6,0} -\frac{1}{42}\partial_{\delta}G_{8,2}
+\frac{211}{75600}G_{8,4}+\frac{6863}{352800}G_{8,2}\right.\\
&\qquad 
-\frac{341}{63000}G_{8,0}-\frac{23}{6930}\partial_{\delta}G_{10,4} -\frac{13}{630} \partial_{\delta}G_{10,2} -\frac{17}{1470}  \partial_{\delta}G_{10,0}
+\frac{19}{72072} G_{10,6}\\
&\qquad \left.
 + \frac{1516849}{768398400}G_{10,4}+\frac{15193}{3969000}G_{10,2}+\frac{46537}{7408800} G_{10,0} +\cdots
\right]\\
&+C_A C_F \left[
\frac{1}{720} G_{4,2} + \frac{91}{1200} G_{4,0} + \left(\frac{\pi ^2}{12}-\frac{11211}{14000}\right) G_{6,2}+
\left(\frac{3 \pi ^2}{8}-\frac{41443}{11200}\right) G_{6,0}\right.\\
&\qquad
+\frac{13}{1680} \partial_{\delta}G_{8,2}-\frac{13}{210} \partial_{\delta}G_{8,0} + \left(\frac{7 \pi ^2}{36}-\frac{1159027}{604800}\right) G_{8,4}+ \left(\frac{389 \pi ^2}{336}-\frac{12072217}{1058400}\right) G_{8,2}\\
&\qquad
+\left(\frac{17 \pi ^2}{10}-\frac{294973877}{17640000}\right) G_{8,0}
+\frac{113}{110880} \partial_{\delta}G_{10,4} + \frac{316}{3465} \partial_{\delta}G_{10,2} + \frac{22721}{129360}\partial_{\delta}G_{10,0}\\
&\qquad
+\left(\frac{427 \pi ^2}{1716}-\frac{212378659}{86486400}\right) G_{10,6}
+\left(\frac{63 \pi ^2}{44}-\frac{86840432003}{6147187200}\right) G_{10,4}\\
&\qquad \left.
+\left(\frac{727 \pi ^2}{280}-\frac{196430295877}{7683984000}\right) G_{10,2}
+\left(\frac{537 \pi ^2}{196}-\frac{192973595057}{7171718400}\right) G_{10,0}
+\cdots
\right],
\end{split}
\eeq
while for gluon jets, it is
\beq
\begin{split}
g_g(z) & = C_F n_f T_F \left[
\frac{3}{80} G_{4,0} + \frac{1}{100}G_{6,2}-\frac{13}{1600}G_{6,0} -\frac{3}{28} \partial_{\delta}G_{8,0}+\frac{1}{840}G_{8,4}-\frac{307}{7840} G_{8,0}\right. \\
&\qquad \left.
-\frac{1}{140}\partial_{\delta}G_{10,2}-\frac{67}{560} \partial_{\delta}G_{10,0}+\frac{1}{8580}G_{10,6}-\frac{389 G_{10,2}}{352800}+\frac{30829}{1411200}G_{10,0}+\cdots
\right]\\
&+C_A n_f T_F \left[
-\frac{1}{360}G_{4,2} + \frac{7}{200}G_{4,0}+ \left(\frac{5 \pi ^2}{12}-\frac{129587}{31500}\right) G_{6,2}
+\left(\frac{\pi ^2}{4}-\frac{120899}{50400}\right) G_{6,0}\right.\\
&\qquad
-\frac{1}{315}\partial_{\delta}G_{8,2} -\frac{11}{126} \partial_{\delta}G_{8,0} 
+ \left(\frac{77 \pi ^2}{36}-\frac{1063963}{50400}\right) G_{8,4} 
+ \left(\frac{593 \pi ^2}{336}-\frac{110698537}{6350400}\right) G_{8,2}\\
&\qquad
+ \left(\pi ^2-\frac{399089339}{39690000}\right) G_{8,0}
-\frac{1}{2640} \partial_{\delta}G_{10,4} -\frac{11981}{83160}\partial_{\delta}G_{10,2} 
-\frac{337}{588}\partial_{\delta}G_{10,0}\\
&\qquad
+\left(\frac{3031 \pi ^2}{858}-\frac{753853073}{21621600}\right) G_{10,6}
+\left(\frac{2429 \pi ^2}{528}-\frac{3323123011}{73180800}\right) G_{10,4}\\
&\qquad \left.
\left(\frac{667 \pi ^2}{280}-\frac{136102341667}{5762988000}\right) G_{10,2}
+\left(\frac{151 \pi ^2}{98}-\frac{2512589293}{162993600}\right) G_{10,0} + \cdots
\right]\\
&+ C_A^2 \left[
\frac{1}{720}G_{4,2}+\frac{49}{200}G_{4,0}+\left(\frac{834469}{126000}-\frac{2 \pi ^2}{3}\right) G_{6,2}+\left(\frac{318193}{50400}-\frac{5 \pi ^2}{8}\right) G_{6,0}\right.\\
&\qquad
-\frac{73}{5040}  \partial_{\delta}G_{8,2} -\frac{31}{630} \partial_{\delta}G_{8,0}
+\left(\frac{1713863}{100800}-\frac{31 \pi ^2}{18}\right) G_{8,4}
+\left(\frac{241456351}{12700800}-\frac{323 \pi ^2}{168}\right) G_{8,2}\\
&\qquad
+\left(\frac{563610307}{39690000}-\frac{43 \pi ^2}{30}\right) G_{8,0}
-\frac{13}{6160}\partial_{\delta}G_{10,4} + \frac{5743}{83160} \partial_{\delta}G_{10,2}+ \frac{331}{1960} \partial_{\delta}G_{10,0}\\
&\qquad
+ \left(\frac{13114631}{617760}-\frac{3691 \pi ^2}{1716}\right) G_{10,6}
+\left(\frac{10567048573}{341510400}-\frac{1655 \pi ^2}{528}\right) G_{10,4}\\
&\qquad \left.
+\left(\frac{104601961181}{5762988000}-\frac{2309 \pi ^2}{1260}\right) G_{10,2}
+\left(\frac{1212949489}{81496800}-\frac{293 \pi ^2}{196}\right) G_{10,0} +\cdots
\right].
\end{split}
\eeq

%%%%%%%%%%%%%%%%%%%%%%%%%%%%%%%%%%%%%%%%%%%%%%%%%%%%%%%%%%%%%%%%%%%%%%%%%%%%%%%%
\subsection{Numerical Results}
%%%%%%%%%%%%%%%%%%%%%%%%%%%%%%%%%%%%%%%%%%%%%%%%%%%%%%%%%%%%%%%%%%%%%%%%%%%%%%%%

We now use these results to study the numerical convergence of the celestial block expansion. This will also allow us to make the idea of the decomposition into celestial blocks extremely concrete. The shape dependence of the three-point correlator was recently analyzed  with the LHC Open Data \cite{Komiske:2022enw}. To measure it in data, it is convenient to map the domain in \Fig{fig:region} into a square in which data can be easily binned. For consistency with \cite{Komiske:2022enw}, we will use these same coordinates in this section. 

We define the new coordinates 
\begin{align}
\xi=\sqrt{\frac{\zeta_{12}}{\zeta_{23}}} \,, \qquad \phi&=\arcsin \sqrt{1 - \frac{(\sqrt{\zeta_{31}}-\sqrt{\zeta_{23}})^2}{\zeta_{12}}}
\,,
\label{eq:transf}
\end{align}
which induces the following relation
\beq
 \frac{d^3\sigma_i}{dx_L d\xi d\phi} = \frac{|z|^3|1-z|}{|\sqrt{z}+\sqrt{z^*}|} \frac{d^3\sigma_i}{dx_L d\mathrm{Re} z \, d\mathrm{Im} z}.
\eeq
These coordinates are chosen to blow up the OPE region into a line to study its azimuthal structure. The variables $\xi$ and $\phi$ can be viewed as radial and angular coordinates about the OPE limit, respectively. A more detailed discussion of this parameterization can be found in \cite{Komiske:2022enw}.

\begin{figure}%[htbp]
\begin{center}
\subfloat[]{
\includegraphics[scale=0.44]{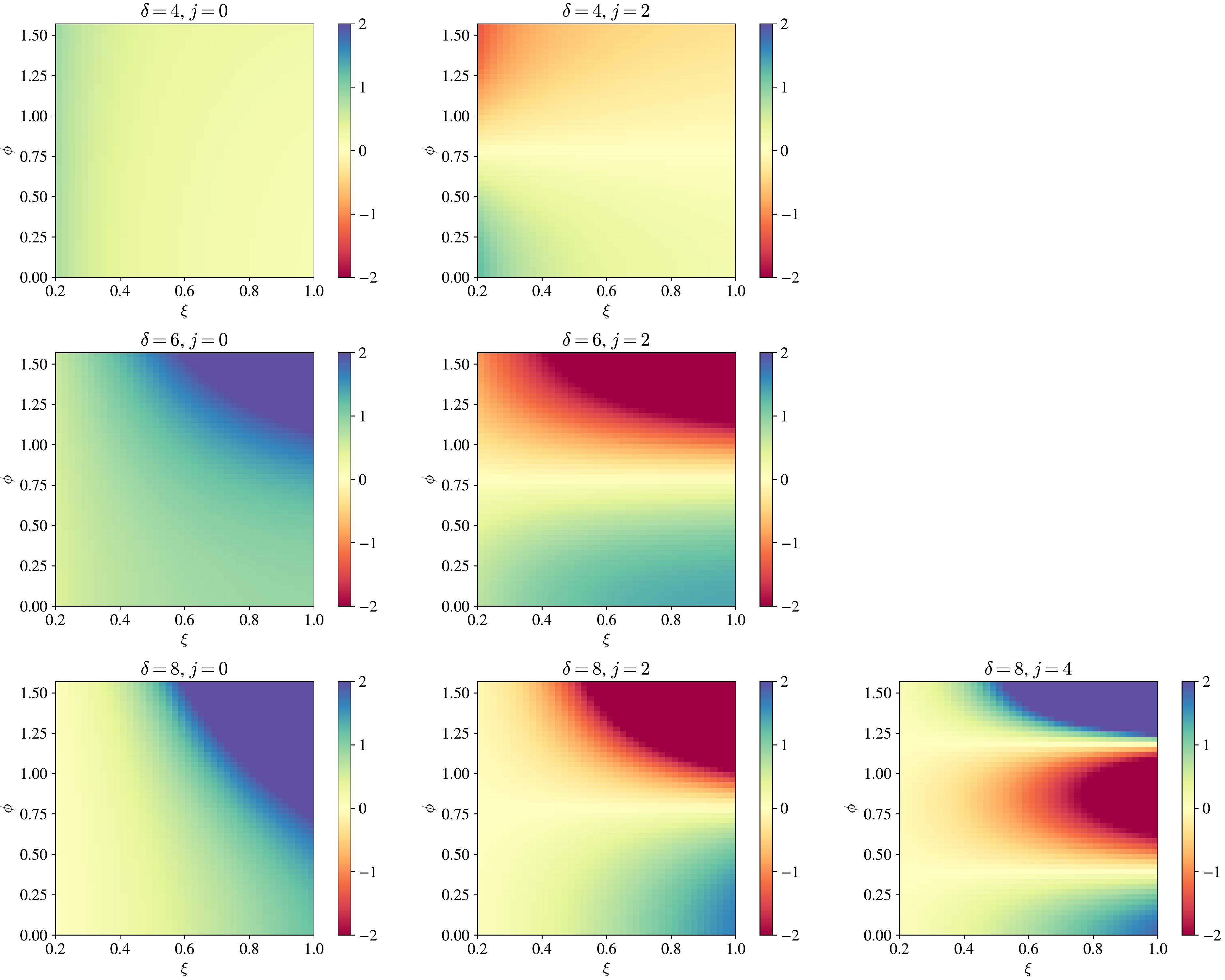}
}
%\captionsetup{font={footnotesize}}
\end{center}
\caption{Plots of the celestial blocks for several low lying values of the celestial dimensions, $\delta$, and transverse spins, $j$. In the higher transverse spin blocks, we see clearly the oscillatory behavior in $\phi$, which is the azimuthal coordinate about the OPE limit. The full three-point correlator is decomposed as a sum over these blocks, much in analogy with a standard Fourier series.}
\label{fig:2d_blocks}
\end{figure}

Before plotting the full distribution for the three-point energy correlator, and its expansion into blocks, it is useful to visualize the celestial blocks themselves. These are shown in \Fig{fig:2d_blocks} for several low lying celestial dimensions, $\delta$, and for several values of the transverse spin, $j$. In the higher transverse spin blocks, one can clearly see the oscillatory structure in  $\phi$, which is the azimuthal variable about the squeezed limit. As discussed above, the decomposition  of the three-point correlator into celestial blocks can just be viewed as a form of Fourier analysis, where now the partial waves are labelled by $\delta$ and $j$ and can be thought of as the harmonics of the Lorentz group. The full result is then just a sum over these partial waves. Here we see that this is made extremely concrete.

\begin{figure}%[htbp]
\begin{center}
\subfloat[]{
\includegraphics[scale=0.24]{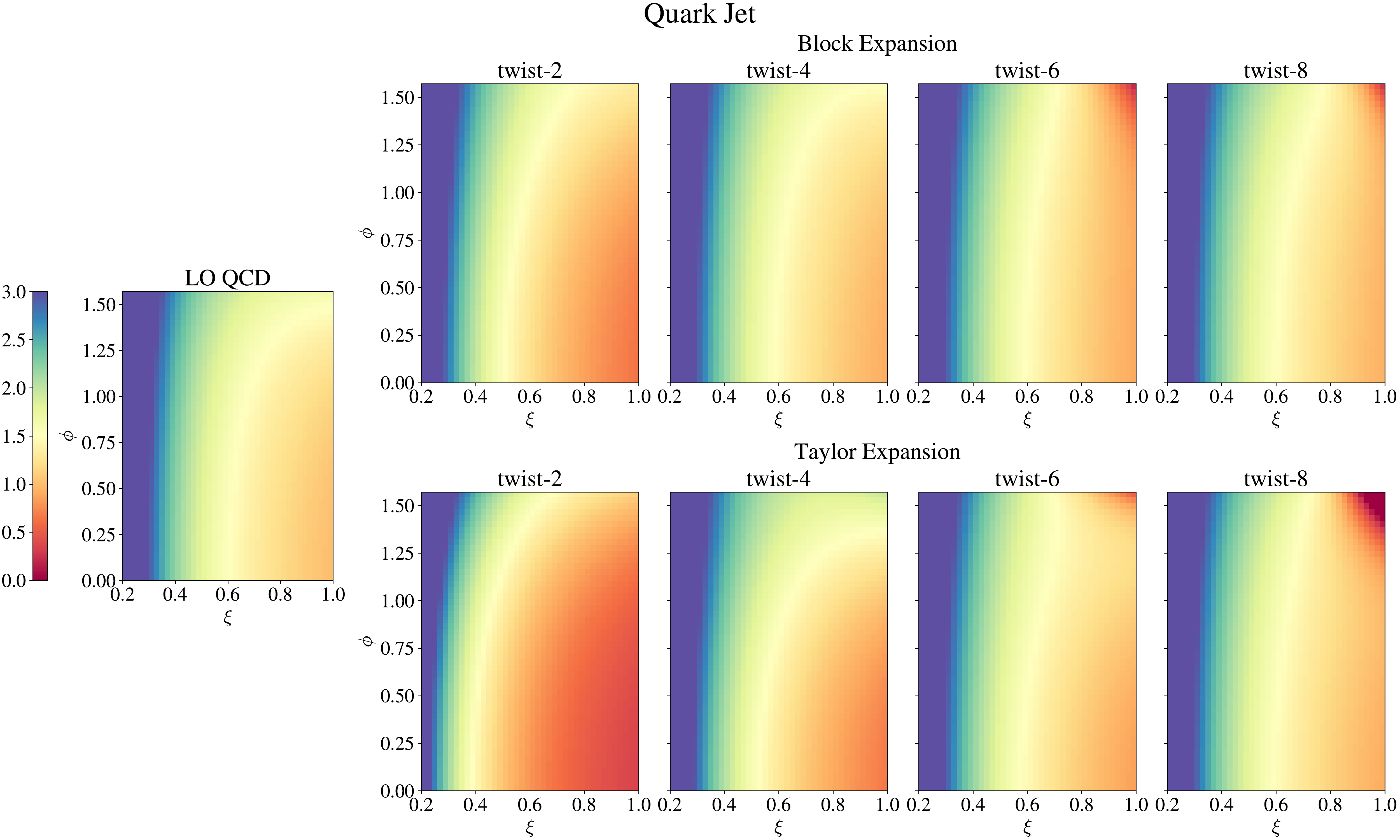}\label{fig:quark_2d}
}\\
\subfloat[]{
\includegraphics[scale=0.24]{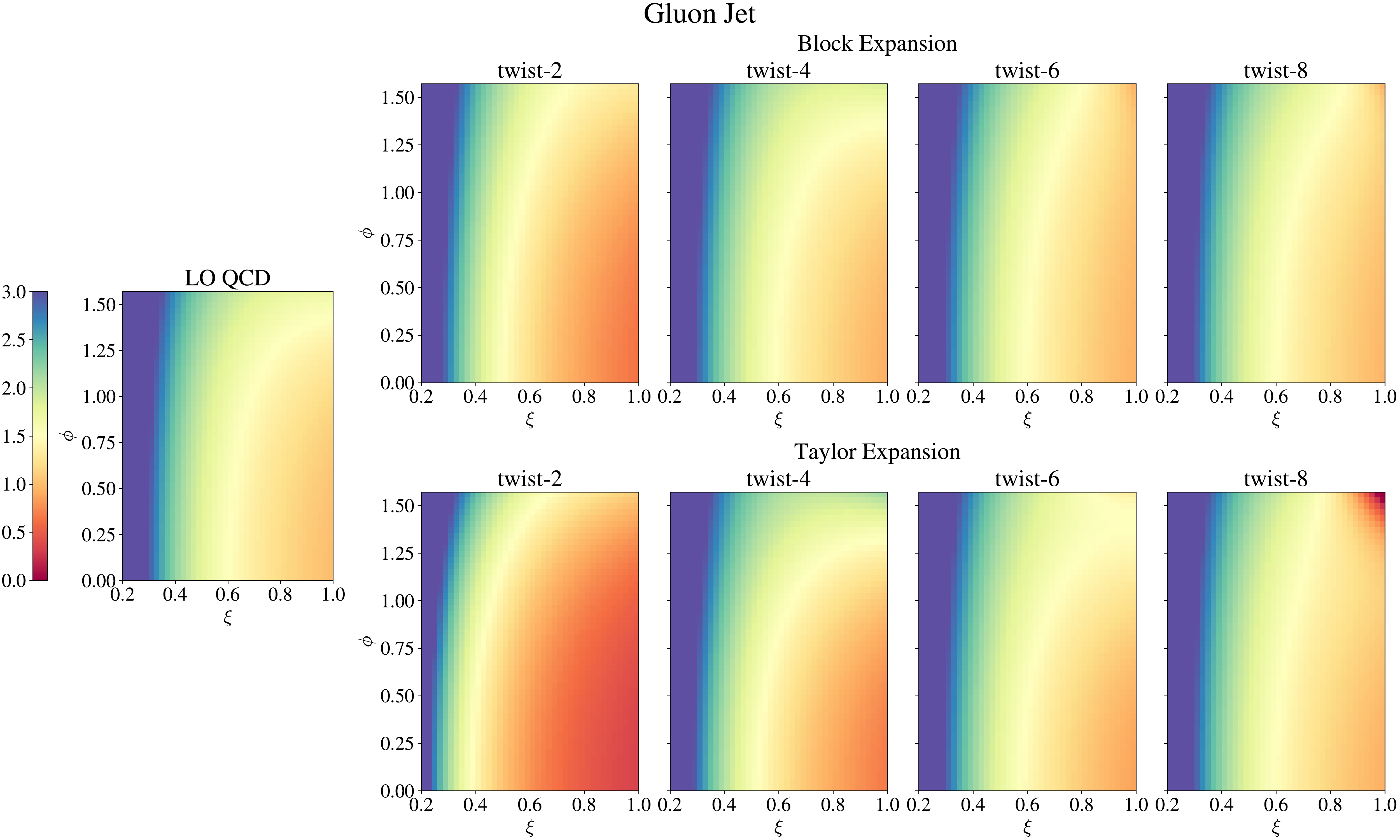}\label{fig:gluon_2d}
}\qquad
%\captionsetup{font={footnotesize}}
\end{center}
\caption{The leading order shape dependence of the three-point correlator for quark jets (a) and gluon jets (b), compared with the expansion in terms of celestial blocks and the naive Taylor expansion about the OPE limit.}
\label{fig:2d_plots}
\end{figure}

In \Fig{fig:2d_plots}, we show two-dimensional plots of the shape dependence of the three-point correlator for both quark and gluon jets. We show both the leading order result on the left, as well as a comparison of an expansion in celestial blocks, compared with a naive Taylor expansion about the squeezed limit on the right. We see overall that the block expansion generates a good approximation to the overall shape. As emphasized above, this decomposition into celestial blocks can be viewed as a decomposition into the harmonics of the Lorentz group, which can be directly seen by eye in these two-dimensional distributions, as well as in the measurement of the three-point function in real LHC data \cite{Komiske:2022enw}. 

Since it is hard to visualize the convergence in a heat-map plot, it is useful to consider slices through the two-dimensional shape. We consider slices at fixed values of $\xi$, which allows us to control the approach to the squeezed limit. At smaller values of $\xi$, we expect to have good convergence of the light-ray OPE, and that the celestial blocks should correctly describe the $\phi$ dependence.

\begin{figure}%[htbp]
\begin{center}
\subfloat[]{
\includegraphics[scale=0.24]{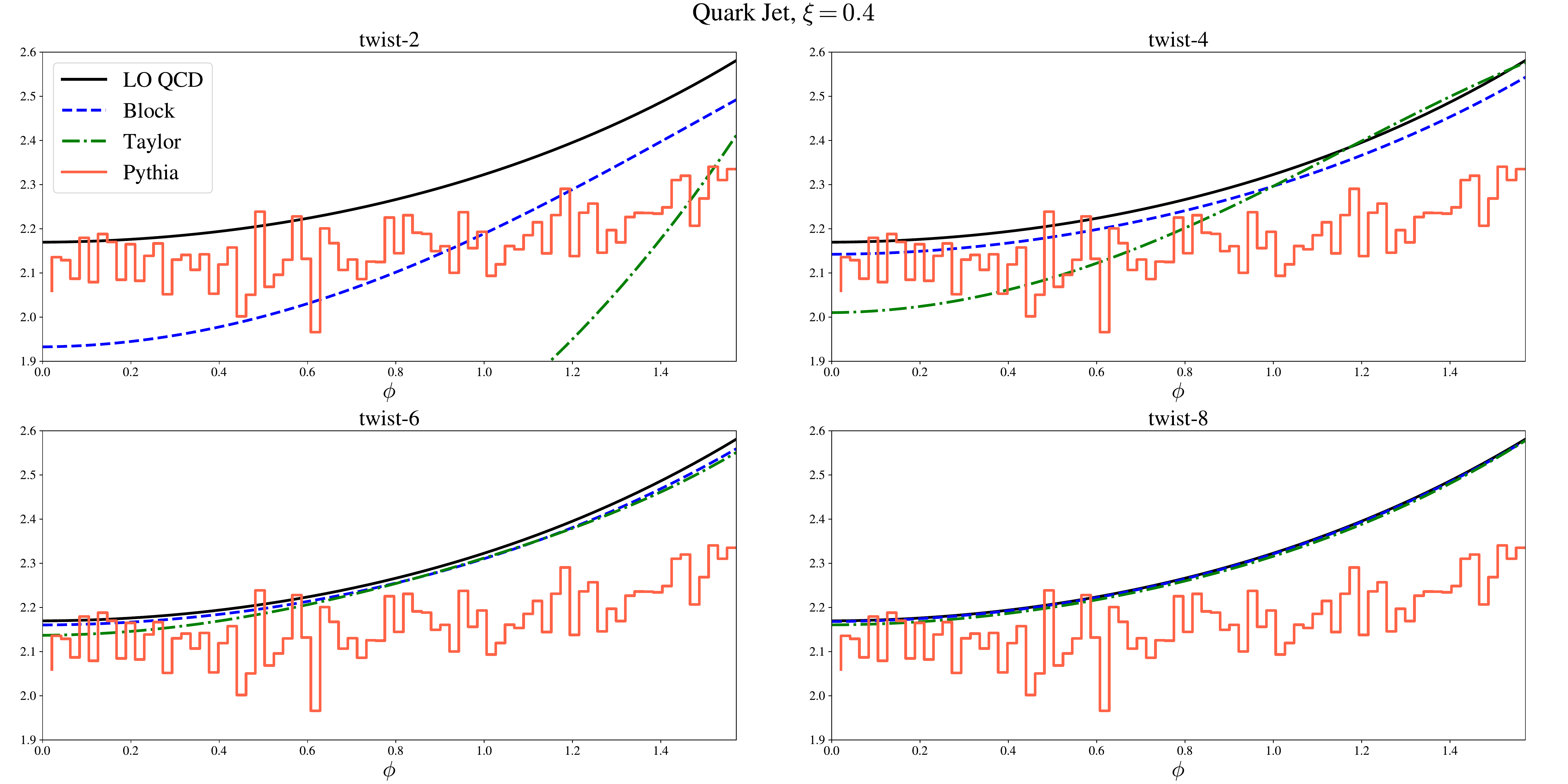}\label{fig:quark_2d}
}\\
\subfloat[]{
\includegraphics[scale=0.24]{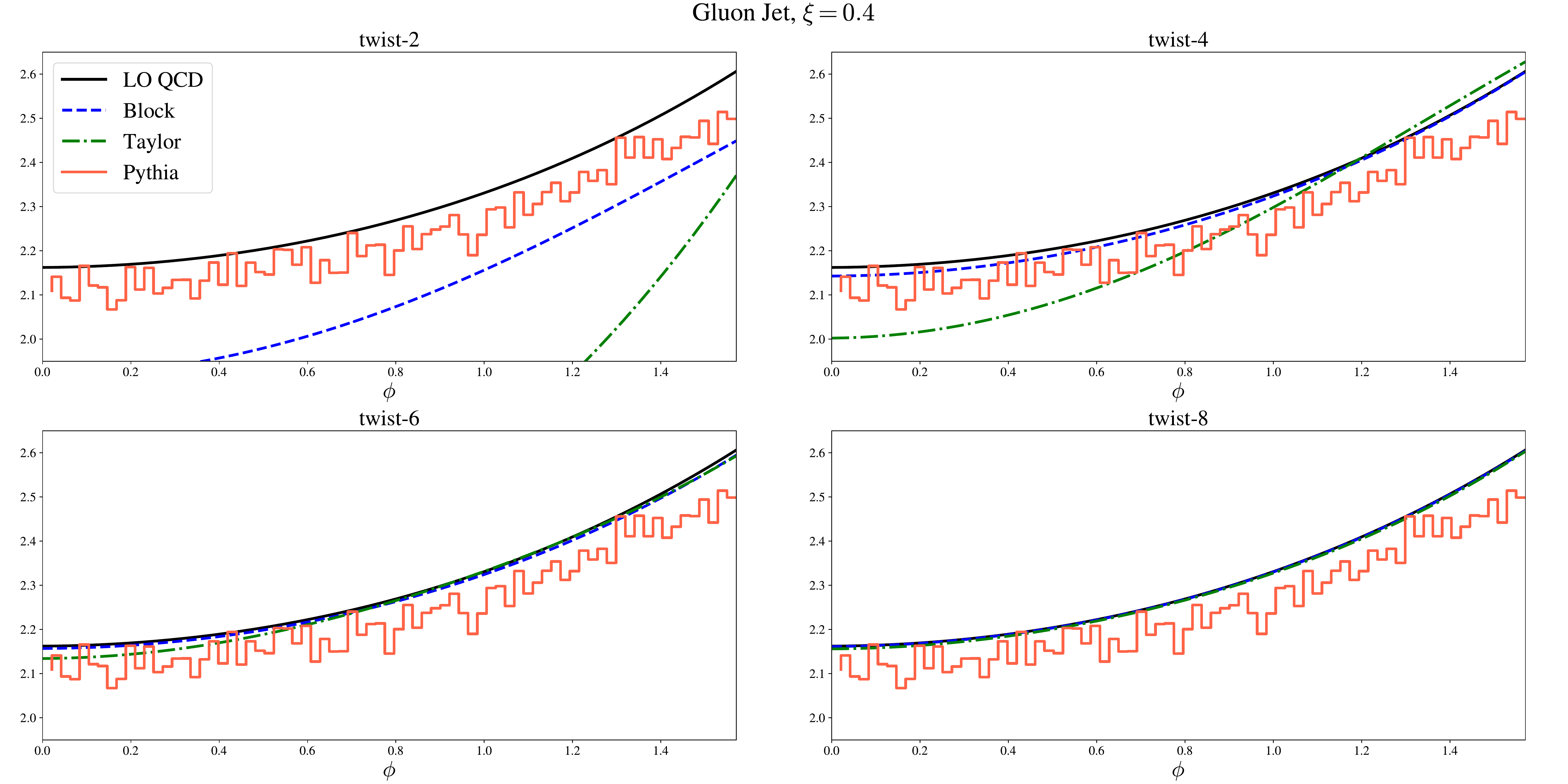}\label{fig:gluon_2d}
}\qquad
%\captionsetup{font={footnotesize}}
\end{center}
\caption{A slice of the leading order three-point corelator at $\xi=0.4$ for quark jets in (a) and gluon jets in (b). We compare the full QCD result, the expansion in celestial blocks, the naive Taylor expansion, and the result from Pythia. The expansion in celestial blocks converges rapidly. }
\label{fig:slice_04}
\end{figure}

\begin{figure}%[htbp]
\begin{center}
\subfloat[]{
\includegraphics[scale=0.24]{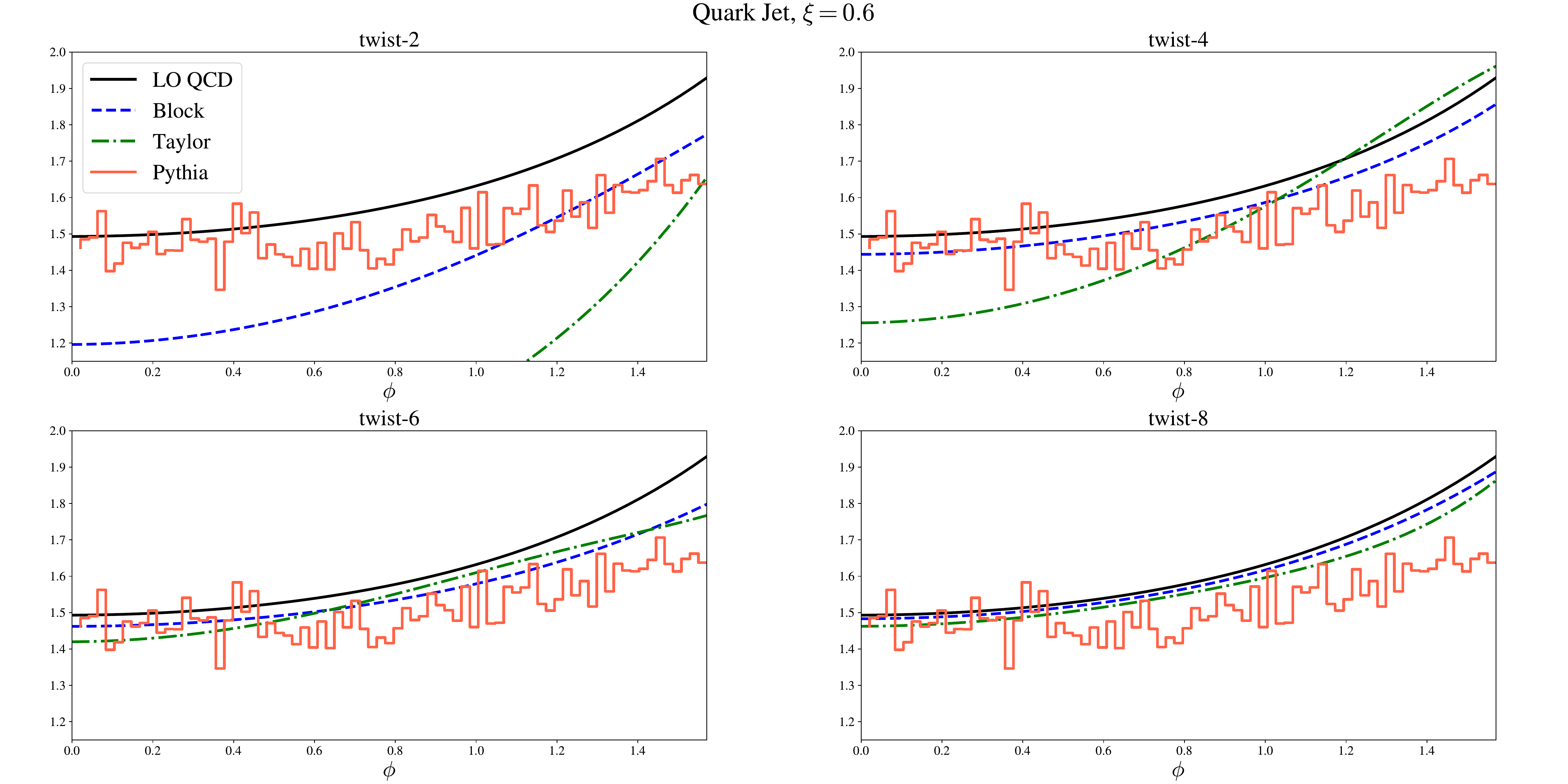}\label{fig:quark_2d}
}\\
\subfloat[]{
\includegraphics[scale=0.24]{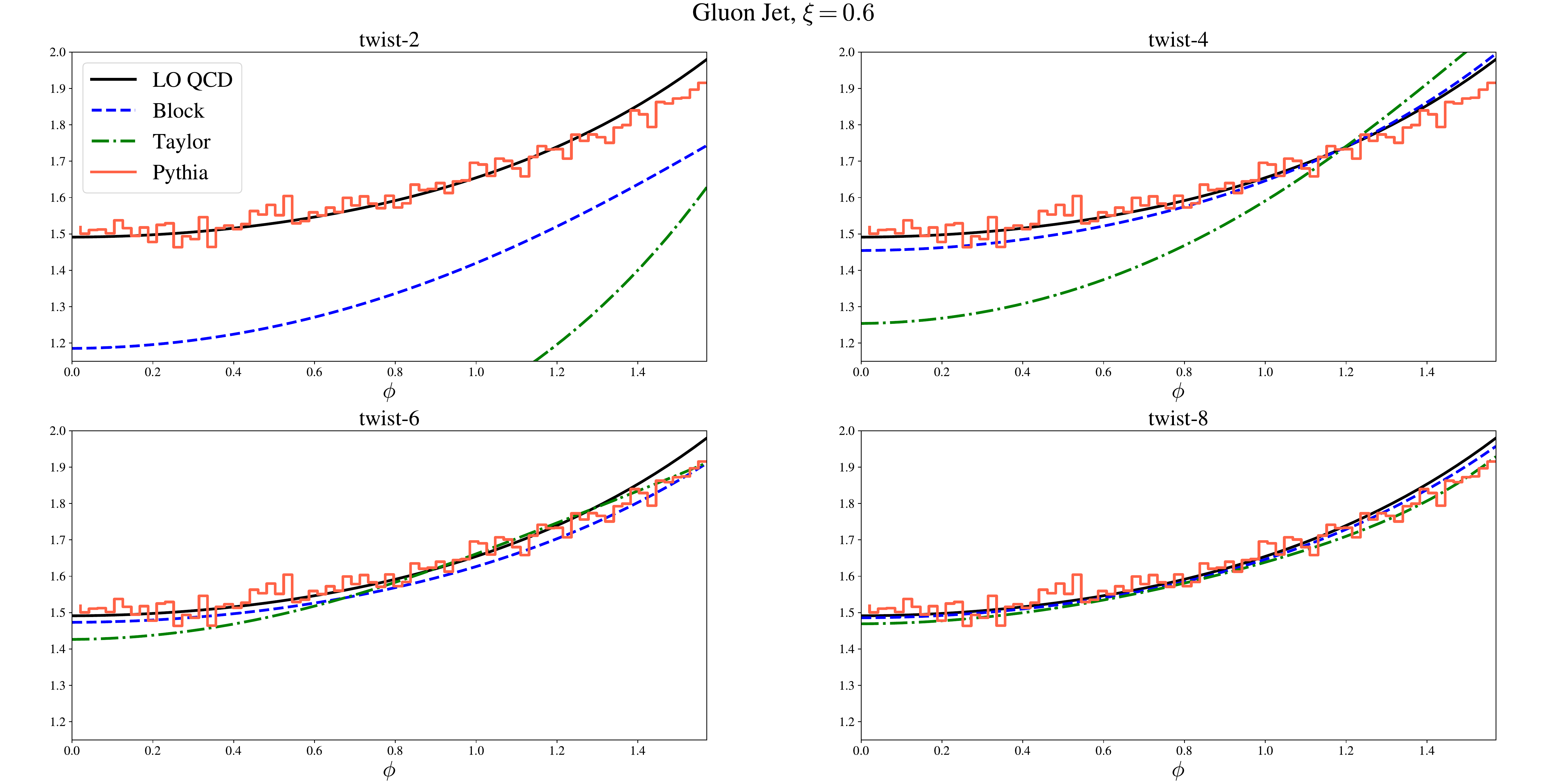}\label{fig:gluon_2d}
}\qquad
%\captionsetup{font={footnotesize}}
\end{center}
\caption{A slice of the leading order three-point corelator at $\xi=0.6$ for quark jets in (a) and gluon jets in (b). We compare the full QCD result, the expansion in celestial blocks, the naive Taylor expansion, and the result from Pythia. The expansion in celestial blocks converges rapidly. }
\label{fig:slice_06}
\end{figure}

\begin{figure}%[htbp]
\begin{center}
\subfloat[]{
\includegraphics[scale=0.24]{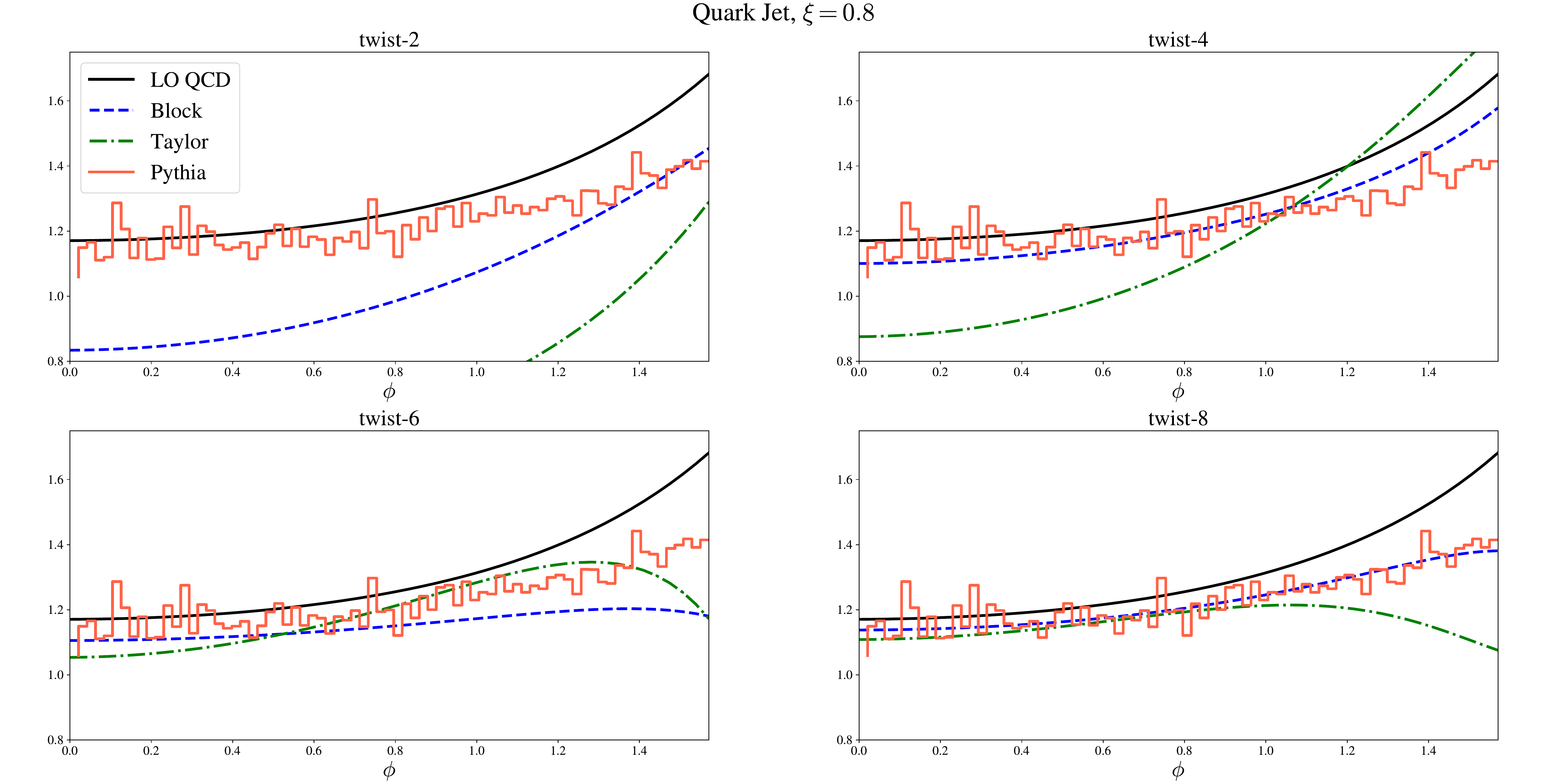}\label{fig:quark_2d}
}\\
\subfloat[]{
\includegraphics[scale=0.24]{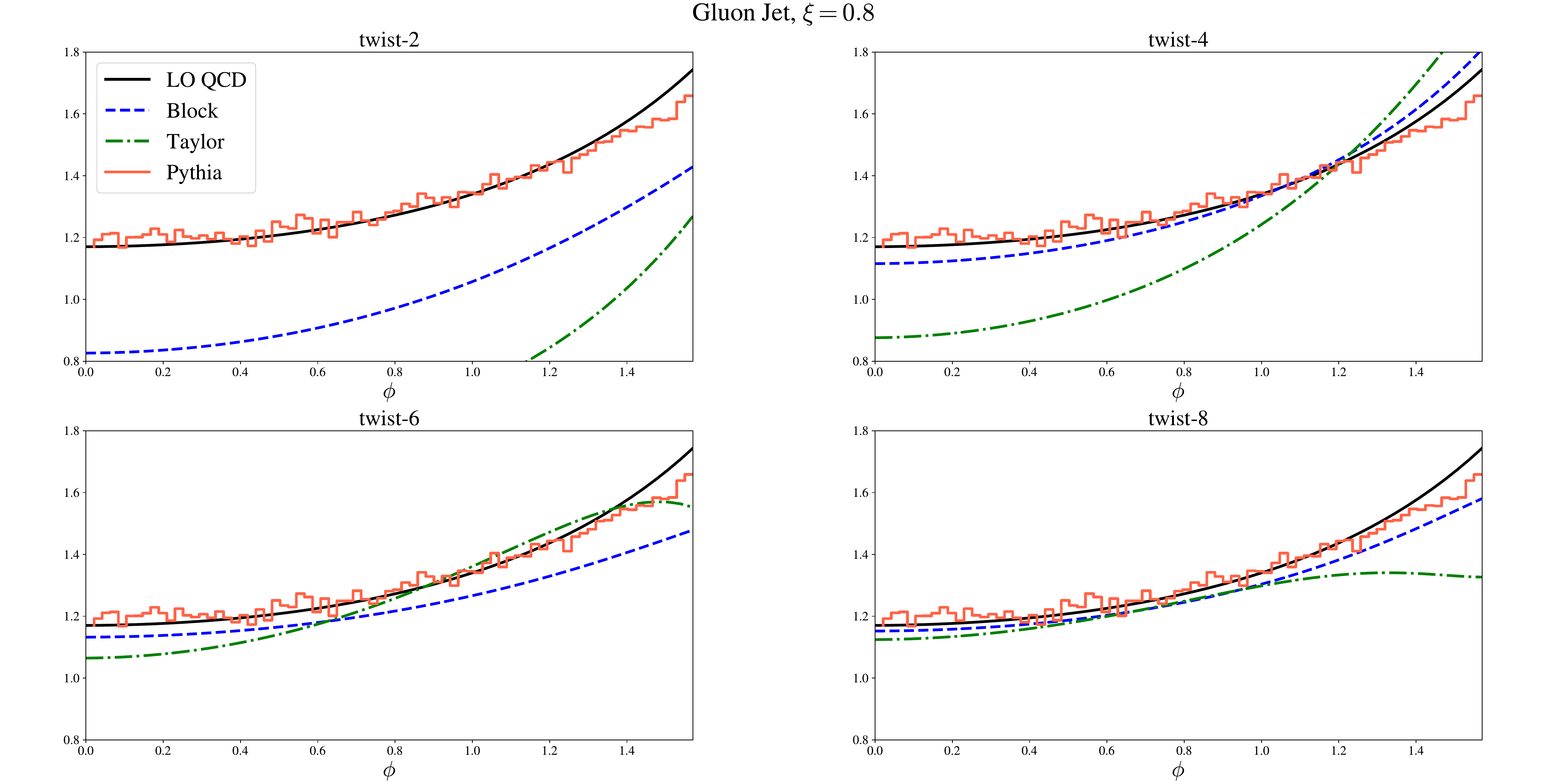}\label{fig:gluon_2d}
}\qquad
%\captionsetup{font={footnotesize}}
\end{center}
\caption{A slice of the leading order three-point corelator at $\xi=0.8$ for quark jets in (a) and gluon jets in (b). We compare the full QCD result, the expansion in celestial blocks, the naive Taylor expansion, and the result from Pythia. The expansion in celestial blocks converges rapidly. }
\label{fig:slice_08}
\end{figure}

\begin{figure}%[htbp]
\begin{center}
\subfloat[]{
\includegraphics[scale=0.24]{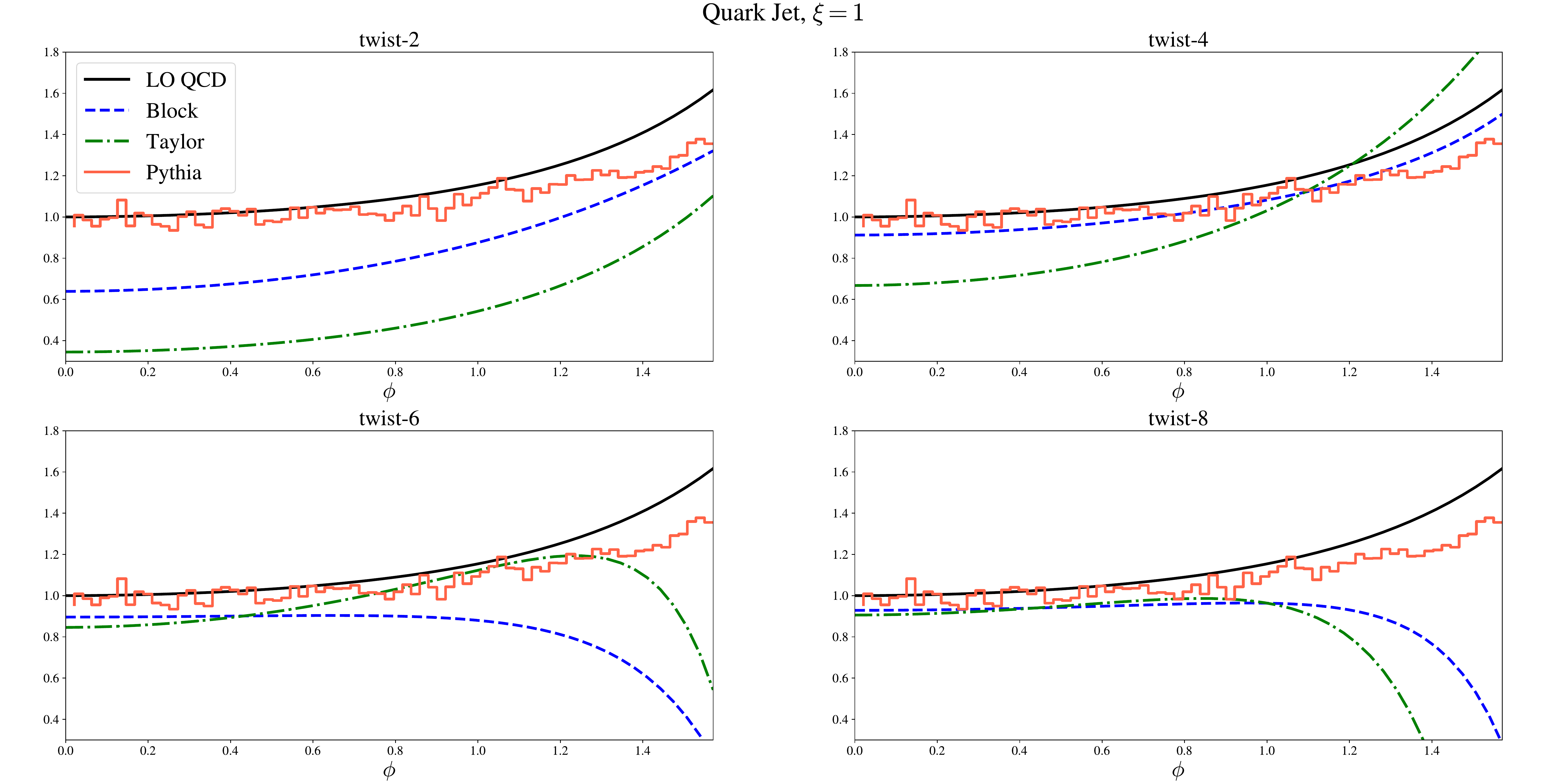}\label{fig:quark_2d}
}\\
\subfloat[]{
\includegraphics[scale=0.24]{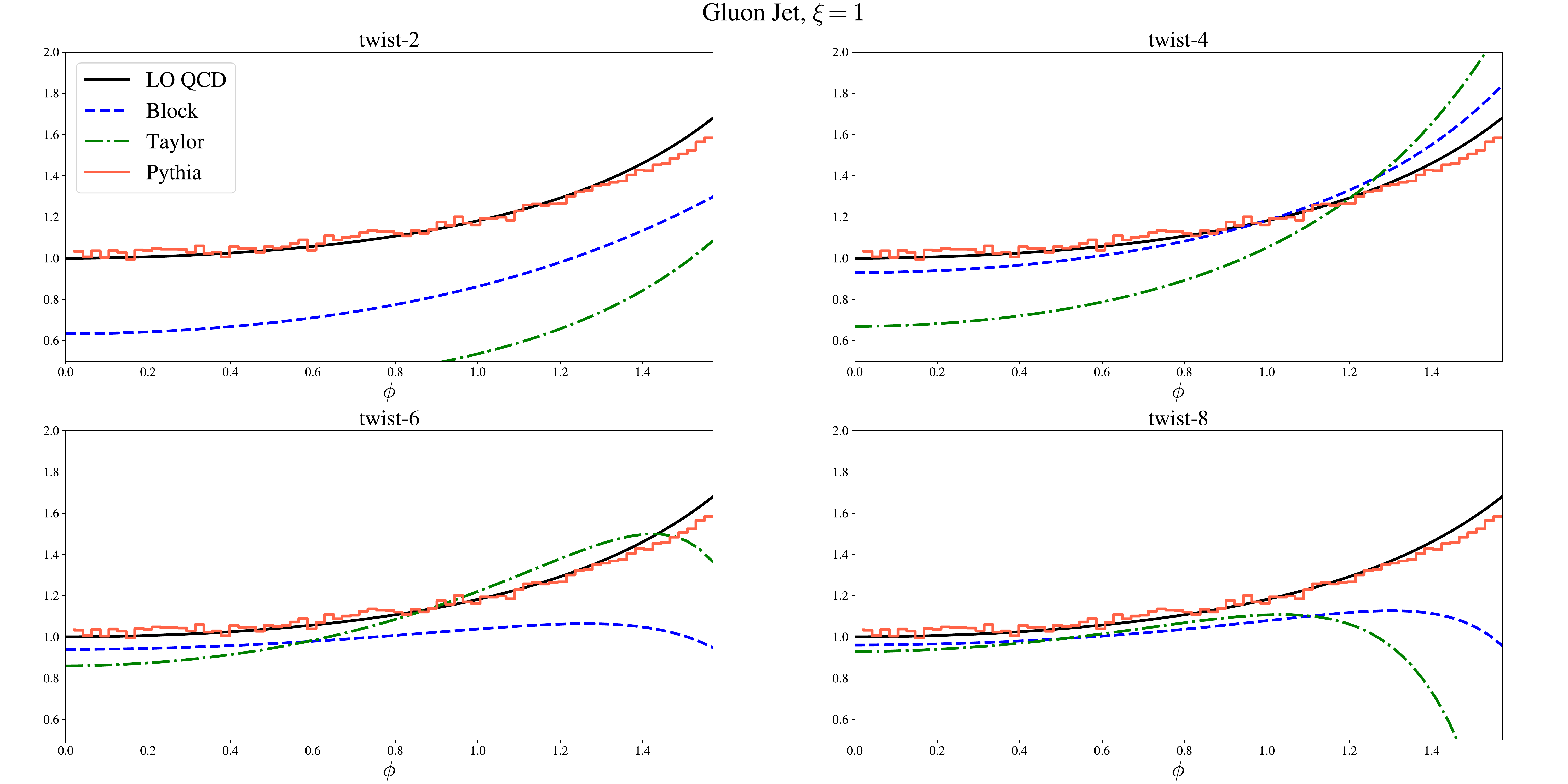}\label{fig:gluon_2d}
}\qquad
%\captionsetup{font={footnotesize}}
\end{center}
\caption{A slice of the leading order three-point corelator at $\xi=1.0$ for quark jets in (a) and gluon jets in (b). We compare the full QCD result, the expansion in celestial blocks, the naive Taylor expansion, and the result from Pythia. This slice contains the equilateral triangle configuration at $\phi=\pi/2$, which is far from the OPE limit. We observe poor convergence of the expansion in this limit.}
\label{fig:slice_1}
\end{figure}

We show the results for slices with $\xi=0.4, 0.6, 0.8, 1.0$, and for both quark and gluon jets in Figs.~\ref{fig:slice_04}-\ref{fig:slice_1}. In these plots we show the leading order QCD result, the twist expansion in celestial blocks, and the naive twist expansion. Furthermore, we also show the simulated results from the parton shower program Pythia \cite{Sjostrand:2014zea}, using the package \textsc{EnergyEnergyCorrelators}~\cite{Patrick} and public simulated data from \cite{Komiske:2019jim}. Pythia is based on iterated $1\to 2$ splittings, although it incorporates some kinematic power corrections. It also incorporates resummation, and therefore in particular as $\xi\to 0$, it cannot be directly compared with leading order QCD. Nevertheless, it is interesting as a benchmark for comparing against the leading order QCD result, to estimate the effect of higher order perturbative corrections. 

By looking at the different plots, we see a number of general features. First, we see that generically the expansion in celestial blocks converges much more rapidly than the naive Taylor expansion about the OPE limit, and describes better the $\phi$ dependence. This is expected, since the blocks incorporate higher order shape dependence that is dictated by symmetries. Furthermore, we typically find that the convergence is quite good already by twist-4, and converges rapidly even up to $\xi=0.8$. This suggests that the full shape dependence can be efficiently approximated using only a few blocks throughout most of the phase space, which we find to be quite promising.

We also observe some more interesting behavior at $\xi=1$, as seen in \Fig{fig:slice_1}. This slice contains the equilateral triangle configuration at $\xi=1, \phi=\pi/2$, which is as far away from the OPE limit as possible. It should therefore be viewed as on the boundary of the convergence of the light-ray OPE expansion. We have used a fixed conformal frame in doing our expansion, where for local operators, one expects convergence for $|z|<1$ \cite{Pappadopulo:2012jk}. Here we see that although the convergence is quite good at lower twists, it seems to begin to diverge at higher orders in the block expansion. This is particularly true in the regime corresponding to an equilateral triangle ($\phi=\pi/2$). This poor behavior of the expansion seems to arise at the same order in the twist expansion as the appearance of $\log(z)$ terms (equivalently, derivatives $\partial_\delta G_{\delta, j}(z,\bar z)$. It would be interesting to understand this further. It would also be interesting to try using radial coordinates \cite{Hogervorst:2013sma}, which at least in the case of local operators improves the regime of convergence. More generally, it would be interesting to develop an understanding of the convergence properties of the OPE for multi-point correlators similar to that for correlators of local operators in CFTs (see e.g. \cite{Pappadopulo:2012jk,Hogervorst:2013sma,Rychkov:2015lca}). We leave more detailed studies of these issues to future work. Nevertheless, we find in general that the convergence of the light-ray OPE is quite promising for approximating the full shape dependence.

Quite interestingly, except for at $\xi=0.4$ (where resummation is expected to play a role), we see quite remarkably good agreement between the leading order QCD result and Pythia. This motivates more detailed studies comparing analytic calculations with real LHC data.  There has recently been work on implementing triple collinear splitting functions into parton showers \cite{Li:2016yez,Gellersen:2021eci,Hoche:2017iem,Hoche:2017hno,Loschner:2021keu}. We believe that this is particularly interesting in light of the measurement of multi-particle correlators \cite{Komiske:2022enw}, and the recent improvements in the analytic studies of these observables. In particular, it would be interesting to study more systematically the non-gaussianties of the three-point correlator, analogous to studies for correlation function of local operators, see e.g. \cite{Rychkov:2016mrc} in the critical 3d Ising model.

We conclude this section by noting that we find it quite beautiful that the symmetry based techniques and elegant mathematical structures discussed in this paper apply directly to the physical observable measured experimentally, namely the energy pattern on the calorimeter. While symmetry based techniques have been widely applied in the study of scattering amplitudes, leading to remarkable progress (see e.g. \cite{Henn:2020omi} for a particularly nice perspective), there has always been a disconnect between the underlying theoretical world, and the world of physical observables at colliders, that has to be overcome for phenomenological applications. Here we have highlighted that by formulating questions in terms of the energy correlators, we can achieve an elegant underlying theoretical structure directly in a physical observable.

%%%%%%%%%%%%%%%%%%%%%%%%%%%%%%%%%%%%%%%%%%%%%%%%%%%%%%%%%%%%%%%%%%%%%%%%%%%%%%%%
\section{Soft Singularities and Zero Modes}\label{sec:zero}
%%%%%%%%%%%%%%%%%%%%%%%%%%%%%%%%%%%%%%%%%%%%%%%%%%%%%%%%%%%%%%%%%%%%%%%%%%%%%%%%

Throughout this paper, our analysis has relied only on the symmetries and mathematical structure of the three-point correlator. This was true both for the derivation of the celestial blocks, as well as for the Lorentzian inversion. While these techniques allow us to understand the structure of the light-ray OPE at high twists (high transverse spins), they do not provide any insight into the explicit light-ray operators appearing in the OPE,  their mixing structure, or how these operators correspond to aspects of the perturbative calculation. Indeed, one shortcoming of the Lorentzian inversion applied at leading order, is that it gives only averaged OPE data at each dimension. 

In \cite{Chen:2021gdk}, we performed an explicit analysis of the light-ray OPE at leading twist in QCD.  One clear direction in which our analysis could be extended is through the explicit construction of the higher twist light-ray operators appearing in the OPE, enabling one to perform the OPE of these operators, and match the data generated by the Lorentzian inversion formula. 

One of the reasons this is particularly interesting in the perturbative context is to gain an intuition for how different regions in momentum space for perturbative calculations are reproduced in the light-ray OPE language. For the case of local operators, this is extremely well understood. For example, logarithms in the calculation of correlation functions of local operators are generated from UV regions and are associated with anomalous dimensions. However, once things are light-transformed to the celestial sphere, this intuition is lost, and so one would like to understand precisely how the light-ray OPE encodes features of perturbative amplitudes. 

One particularly interesting feature of the OPE expansion of the three-point correlator is the appearance of $\log z \bar z$ terms at subleading twist, which results in derivatives of celestial blocks. While this behavior is familiar in the context of  perturbative calculations of correlators of local operators, where it arises from anomalous dimensions and/or mixing of operators, and hence UV behavior, it provides an interesting test case for understanding what is generating this behavior in the energy correlators, and how it is reproduced by the light-ray OPE. The explicit calculation of~\cite{Chen:2021gdk} showed that the origin of these $\log$s is the propagator $1/s_{123}^2$ in the bulk of the Minkowski space, when the energy fraction of the third unsqueezed particle is comparable with the OPE angle. In this sense, in this perturbative calculation, the $\log$ is closely related to a soft mode contribution of unsqueezed particle, i.e. from an IR behavior. This suggests some UV/IR mixing in the OPE intuition under the light-transform.

Here we try and further develop our understanding of this feature, and show how these terms are reproduced in terms of the collision of two blocks. The explicit understanding of these higher twist terms is quite challenging in either $\cN=4$ or QCD, and therefore we introduce a toy example in 4D $\phi^4$ theory. In particular, we consider the tree-level three-point number correlator $\langle \mathcal{N}(\vec{n}_1) \mathcal{N}(\vec{n}_2) \mathcal{N}(\vec{n}_3)\rangle$ in $\phi^4$ theory as an example,
\begin{equation}
\langle \mathcal{N}(\vec{n}_1) \mathcal{N}(\vec{n}_2) \mathcal{N}(\vec{n}_3) \rangle_{\phi} 
\sim \int_0^\infty d\xi_1 d\xi_2 d\xi_3\, \xi_1 \xi_2 \xi_3 \delta(1-\xi_1-\xi_2-\xi_3) \frac{1}{s_{123}^2}\,,
\end{equation}
where the $1 \to 3$ splitting function is simply $1$ in this case.
The number operator $\mathcal{N}(\vec{n})$ is a particle number counting ``detector" which, loosely speaking, is the light-ray version of the local operator $\phi \partial_\mu \phi$.

The reason to consider the number operator is that  the additional energy weighting suppress the effect of the soft contribution and postpones the $\log$ to higher twists. We calculate the collinear limit of the correlator $\langle \mathcal{N} \mathcal{N} \mathcal{N}\rangle_{\phi}$ and expand in the squeezed limit
\begin{equation}
\begin{split}
\langle \mathcal{N}(\vec{n}_1) \mathcal{N}(\vec{n}_2) \mathcal{N}(\vec{n}_3) \rangle_{\phi} 
&\sim
\left( -\frac{1}{3}\log |z| -\frac{1}{18}\right) + (z+\bar{z})\left( -\frac{1}{3}\log |z| -\frac{5}{36}\right) + \cdots \\
&=-\frac{1}{3}\partial_\delta G_{\delta=4,j=0}(z,\bar{z})-\frac{1}{18} G_{\delta=4,j=0}(z,\bar z)+\cdots,
\end{split}
\label{eq:numbercorrelator}
\end{equation}
from which we explicitly see the $\log |z|$ indeed appears at leading power. In the following, we will approach this $\log$ from two different perspectives: an expansion in terms of distributions, and regulating the spin quantum number, $J$. 

Before presenting the details, we want to emphasize that the structure is quite independent of the energy weightings on the squeezed pair so we can generalize to the special family with $J_3=1$ (i.e. setting $\mathbb{O}^{[J_3=1]}=\mathcal{N}$ in the following definition) among  general energy weighting correlators
\begin{equation}
\mathrm{EEEC}^{J_1,J_2,J_3}_\phi \! \equiv
\langle \mathbb{O}^{[J_1]}(\vec{n}_1) \mathbb{O}^{[J_2]}(\vec{n}_2) \mathbb{O}^{[J_3]}(\vec{n}_3) \rangle_{\phi}
\sim \!\! \int_0^\infty \!\!\!\!\! d\xi_1 d\xi_2 d\xi_3\, \xi_1^{J_1} \xi_2^{J_2} \xi_3^{J_3} \delta(1-\xi_1-\xi_2-\xi_3) \frac{1}{s_{123}^2}.
\end{equation}

First, let's consider the squeezed limit expansion of the propagator $1/s_{123}^2$ in terms of distributions. When considering the squeezed limit $u \to 0$ at the integrand level, the Taylor expansion in $u$ no longer makes sense in the region $\xi_3 \lesssim u$ for
\begin{equation}
\frac{E_J^4 x_L^2}{s_{123}^2} = \frac{1}{(\xi_1 \xi_2 u + \xi_2 \xi_3 v +\xi_1 \xi_3)^2} = \frac{1}{\left[ \xi_3 (1-\xi_3) (\xi_1^\prime + \xi_2^\prime v) +(1-\xi_3)^2 \xi_1^\prime \xi_2^\prime u\right]^2}.
\end{equation}
In the last step, we rewrite $\xi_{1,2}$ as the momentum fractions relative to their sum $\xi_{1,2}^\prime=\frac{\xi_{1,2}}{\xi_1+\xi_2}$, aiming for the factorization of the phase space 
\begin{equation}
\! \int_0^1 \!\!\!\! d\xi_1 d\xi_2 d\xi_3\, \xi_1 \xi_2 \xi_3 \delta(1-\xi_1-\xi_2-\xi_3) \! = \!\! \int_0^1 \!\!\!\! d\xi_1^\prime d\xi_2^\prime\, {\xi_1^\prime}^{J_1} {\xi_2^\prime}^{J_2} \delta(1-\xi_1^\prime-\xi_2^\prime) \!\! \int_0^1 \!\!\! d\xi_3\, \xi_3 (1-\xi_3)^{1+J_1+J_2}.
\end{equation}

For regular test functions satisfying $\sim O((1-\xi_3)^2)$, we can apply the following $b\rightarrow 0$ expansion with distributions
\begin{equation}
\frac{1}{[\xi_3(1-\xi_3) a +(1-\xi_3)^2 b]^2} = \frac{1}{b} \frac{\delta(\xi_3)}{a} 
+ \frac{\delta^\prime(\xi_3) }{a^2 (1-\xi_3)^2}\left(\log \frac{b}{a} +1\right) 
+ \frac{1}{a^2 (1-\xi_3)^2} \left[\frac{1}{\xi_3^2}\right]_{++} +O(b),
\end{equation}
where we define the double plus distribution as the functional on test functions supported on $(0,1)$
\begin{equation}
\left[\frac{1}{\xi_3^2}\right]_{++} (f) = \int_0^1 d\xi_3 \frac{1}{\xi_3^2}\left( f(\xi_3)-f(0)-\xi_3 f^\prime(0)\right).
\end{equation}
Intuitively, we can view the $\delta$-function and its derivative as the contribution from the zero mode while the double plus distribution is essentially a zero-mode-subtracted contribution. 
For the correlator $\langle \mathbb{O}^{[J_1]}\mathbb{O}^{[J_2]}\mathcal{N}\rangle_\phi$, the zero mode contribution at leading power is~(taking also $v \to 1$) 
\begin{equation}
\begin{split}
&\int_0^1 d\xi_1^\prime {\xi_1^\prime}^{J_1} (1-\xi_1^\prime)^{J_2} \int_0^1 d\xi_3 \; \xi_3 (1-\xi_3)^{J_1+J_2+1} \frac{\delta^\prime (\xi_3)}{(1-\xi_3)^2} \left(1+\log (\xi_1^\prime(1-\xi_1^\prime) u)\right) \\
=& -\frac{\Gamma(1+J_1)\Gamma(1+J_2)}{\Gamma(2+J_1+J_2)} \left( 1 + H_{J_1} + H_{J_2} -2 H_{J_1+J_2+1} + 2\log |z|\right)
\end{split}
\end{equation}
where $H_n$ is the standard harmonic number. The zero mode contribution for some low-lying values of $(J_1, J_2)$ are
\begin{equation}
(1)\, J_1=J_2=1: \frac{1}{9} -\frac{1}{3}\log |z|, \,
(2)\, J_1=1, J_2=2: \frac{1}{18}-\frac{1}{6}\log |z|,\,
(3)\, J_1=J_2=2: \frac{17}{900} -\frac{1}{15} \log |z|.
\label{eq:zeromode}
\end{equation}
The leading power zero-mode-subtracted contribution is 
\begin{equation}
\int_0^1 \!\! d\xi_1^\prime {\xi_1^\prime}^{J_1} (1-\xi_1^\prime)^{J_2} \!\! \int_0^1 \!\! d\xi_3 \; \xi_3 (1-\xi_3)^{J_1+J_2+1} \frac{1}{(1-\xi_3)^2}\left[\frac{1}{\xi_3^2}\right]_{++} \!\!\!=\! -\frac{\Gamma(1+J_1)\Gamma(1+J_2)}{\Gamma(2+J_1+J_2)} H_{J_1+J_2-1}, 
\end{equation}
with low-lying values being
\begin{equation}
(1)\; J_1=J_2=1:\; -\frac{1}{6}, \quad 
(2)\; J_1=1, \,J_2=2:\; -\frac{1}{8},\quad 
(3)\; J_1=J_2=2: \; -\frac{11}{180}.
\label{eq:nonzeromode}
\end{equation}
Clearly, adding \eqref{eq:zeromode} and \eqref{eq:nonzeromode} for the $J_1=J_2=1$ case reproduce exactly the leading power contribution of \eqref{eq:numbercorrelator}. 
This provides evidence that logarithms appearing at leading order in three-point lightray correlators are coming from the zero mode contribution.

One way to see how this gives rise to derivatives of the celestial blocks is to consider the general form $\langle \mathbb{O}^{[J_1]} \mathbb{O}^{[J_2]} \mathbb{O}^{[J_3]}\rangle$, keeping $J_3$ generic and regarding it as a regulator to the soft mode. Take $\mathrm{EEEC}^{1,1,J_3}_\phi = \langle \mathcal{N}(\vec{n}_1) \mathcal{N}(\vec{n}_2) \mathbb{O}^{[J_3]}(\vec{n}_3) \rangle_\phi$ for example, we find two types of blocks in the squeezed limit expansion classified by whether the block depends on $J_3$
\begin{align}
&\mathrm{EEEC}^{1,1,J_3}_\phi
\sim \left[ \frac{1}{6{\color{blue}(J_3-1)} J_3} G_{4,0} -\frac{J_3+2}{30(J_3-2){\color{blue}(J_3-1)}J_3} G_{6,0} +\cdots \right] \\
+ &  \left[ \frac{{\color{blue}\Gamma(1-J_3)} \Gamma(1+J_3)^3}{\Gamma(2J_3+2)} G_{2J_3+2,0} + 
\frac{{\color{blue}\Gamma(1-J_3)} \Gamma(1+J_3)\Gamma(2+J_3) \Gamma(3+J_3)}{2\Gamma(2J_3+4)} G_{2J_3+4,0}+\cdots\right].\nn
\end{align}
With the help of $J_3$ regularization, we notice that the block coefficients separately diverge in the $J_3\to 1$ limit when these two types of celestial blocks collide with each other. The poles then cancel between these two types of blocks and the derivatives of blocks (hence the $\log$s) are generated in the finite terms (in a similar manner as the discussion of $\log$s in perturbative CFT correlators in \cite{Fitzpatrick:2011dm}):
\begin{align}
&\mathrm{EEEC}^{1,1,J_3}_\phi  \sim
\left[\frac{1}{J_3-1}\left( \frac{1}{6} G_{4,0} +\frac{1}{10} G_{6,0}+\cdots\right) 
+\left({\color{red}-\frac{1}{6}} G_{4,0} + \frac{1}{30} G_{6,0}+\cdots\right)  +\cdots\right]\\
&+\left[ \frac{1}{J_3-1} \left(-\frac{1}{6} G_{4,0}-\frac{1}{10} G_{6,0}+\cdots\right)
+\left({\color{red}\frac{1}{9}}G_{4,0} {\color{red}-\frac{1}{3}}\partial_{\delta} G_{4,0} +\frac{11}{150}G_{6,0} -\frac{1}{5}\partial_{\delta}G_{6,0}+\cdots  \right) +\cdots
\right]. \nn
\end{align}
In addition, we present the case for $J_1=1,\,J_2=2$ 
\begin{align}
&\mathrm{EEEC}^{1,2,J_3}_\phi  \sim
\left[\frac{1}{J_3-1}\left( \frac{1}{12} G_{5,0} +\frac{2}{25} G_{7,0}+\cdots\right) 
+\left({\color{red}-\frac{1}{8}} G_{5,0} - \frac{1}{50} G_{7,0}+\cdots\right)  +\cdots\right]\\
&+\left[ \frac{1}{J_3-1} \left(-\frac{1}{12} G_{5,0}-\frac{2}{25} G_{7,0}+\cdots\right)
+\left({\color{red}\frac{1}{18}}G_{5,0} {\color{red}-\frac{1}{6}}\partial_{\delta} G_{5,0} +\frac{53}{750}G_{7,0} -\frac{4}{25}\partial_{\delta}G_{7,0}+\cdots  \right) +\cdots
\right]\,,\nn
\end{align}
and for and $J_1=J_2=2$ 
\begin{align}
&\hspace{-0.5cm}\mathrm{EEEC}^{2,2,J_3}_\phi  \sim
\left[\frac{1}{J_3-1}\left( \frac{1}{30} G_{6,0} +\frac{1}{21} G_{8,0}+\cdots\right) 
+\left({\color{red}-\frac{11}{180}} G_{6,0} - \frac{19}{630} G_{8,0}+\cdots\right)  +\cdots\right]\\
&+\left[ \frac{1}{J_3-1} \left(-\frac{1}{30} G_{6,0}-\frac{1}{21} G_{8,0}+\cdots\right)
+\left({\color{red}\frac{17}{900}}G_{6,0} {\color{red}-\frac{1}{15}}\partial_{\delta} G_{6,0} +\frac{379}{8820}G_{8,0} -\frac{2}{21}\partial_{\delta}G_{8,0}+\cdots  \right) +\cdots
\right]\,,\nn
\end{align}
from which find the same rational numbers (in red) as those in the discussion using distributions above. 

We therefore find that certain features of the light-ray OPE are driven by infrared (soft) features of the amplitudes. This may be specific to weak coupling perturbation theory, since it relies on the $1/s_{123}$ propagator structure, which would be smeared out at finite coupling. It would be particularly interesting to understand at an operator level in terms of explicit light-ray operators how this is reproduced. We believe this scalar example might be a useful playground for this.  In addition to the approaches considered here, one can also consider finite $\omega$ light-ray operators of \cite{Korchemsky:2021okt}, which also eliminate this behavior, but are non-analytic as $\omega\to 0$.

One of the reasons that we find this issue important to resolve, is that part of the intuition for the light-ray OPE in perturbation theory is that particle excitations with non-zero momentum behave as local operators on the celestial sphere. However, this is not true for zero-modes, which are not localized on the celestial sphere. In general, this can lead to problems, referred to as cross-talk \cite{Belitsky:2013xxa,Belitsky:2013bja,Korchemsky:2021okt}. In \cite{Kologlu:2019bco} certain conditions were proven to eliminate this cross-talk, however, these were proven for the OPE of two light-ray operators starting from a four-point function of light-ray operators. The behavior we observe here relies strongly on the presence of the third light-ray operators. We again emphasize that although we studied in the case of the correlator of number operators, this was only to enhance it to leading twist, but it occurs at subleading twist in the OPE of three energy operators. We believe that this is an important issue to resolve to completely understand the structure of the light-ray OPE in perturbation theory, but we leave this to future work. 

%%%%%%%%%%%%%%%%%%%%%%%%%%%%%%%%%%%%%%%%%%%%%%%%%%%%%%%%%%%%%%%%%%%%%%%%%%%%%%%%
\section{Conclusions}\label{sec:conc}
%%%%%%%%%%%%%%%%%%%%%%%%%%%%%%%%%%%%%%%%%%%%%%%%%%%%%%%%%%%%%%%%%%%%%%%%%%%%%%%%

In this paper we have used the results for the leading order perturbative three-point energy correlator in the collinear limit, computed in \cite{Chen:2019bpb}, as a playground to better understand aspects of the light-ray OPE in perturbation theory, with the hope of making it into a practical tool for perturbative QCD calculations. We have attempted to illustrate in this concrete setting a number of techniques that are commonly used in the CFT literature (conformal blocks, Lorentzian inversion, Casimir differential equations, ...), and showed that they can be used to study questions of phenomenological relevance in jet substructure, once jet substructure is formulated in terms of energy correlators.\footnote{The only natural conclusion from this reasoning is that jet substructure {\bf{should}} be formulated in terms of correlation functions!}

A particularly elegant feature of the energy correlation functions is that they are strongly constrained by symmetries, and can be expanded in eigenfunctions of the Lorentz group, referred to as celestial blocks. We derived the structure of the celestial blocks for the three-point correlator, both using a light-like source, and as an expansion about general kinematics, and illustrated how these celestial blocks interplay with standard QCD factorization.  In particular, using specific partonic channels of the QCD result, we showed that the light-ray OPE, and the celestial blocks have a transparent interpretation when the intermediate state coincides with a perturbative quark or gluon state. 

Due to the simple structure of the three-point energy correlator in the collinear limit, we then showed how one can use the Lorentzian inversion formula to read off the spectrum of light-ray operators appearing in the OPE. One of the interesting features of our analysis is that it illustrates (at least in perturbation theory) that the OPE data for correlation functions of light-ray operators is analytic in transverse spin. This strongly emphasize the central role of transverse spin \cite{1822249} in the study of correlation functions of light-ray operators. Since the study of correlators of local operators led to analyticity in spin, which in turn led to the study of correlators of light-ray operators, which we find to be analytic in transverse spin, it is natural to assume that there is some more general overarching structure into which analyticity in transverse spin fits. Furthermore, while we were able to mathematically perform the inversion, many questions remain about the physical interpretation of the behavior of light-ray operators as a function of transverse spin, such as the interpretation of the transverse spin Regge intercept, and the high transverse spin limit. We believe that these would be interesting to investigate further to better understand the space of light-ray operators and the light-ray OPE.

Using the leading order perturbative results for the three-point correlator on quark and gluon jets, we showed numerically how the expansion in terms of celestial blocks is an improvement over the naive Taylor expansion more commonly used in the studies of QCD observables at subleading power. While many powerful techniques exist in perturbative QCD for studying jet substructure observables and factorization, they have for the most part not emphasized the role of symmetries. We believe that it is in this respect that techniques from the CFT literature, which exploit underlying symmetries to significant gain, can have a particularly large impact, and deserve further study from the QCD community. We hope that our introduction to the use of these techniques in the simple setting of the three-point correlator is a first step in this direction.

Finally, we tried to further develop the intuition of the perturbative light-ray OPE by understanding the precise momentum regions that give rise to derivatives of the celestial blocks in the twist expansion. While derivatives of conformal blocks often appear in the perturbative expansion of local correlators, there they are associated with anomalous dimensions, i.e. UV behavior. Here we found examples where they are instead associated with soft behavior, in particular a zero mode of the third light-ray operator, and we showed how this behavior gave rise to the collision of two celestial blocks. One of the reasons that we find non-vanishing contributions from a zero-mode to be slightly concerning is that the perturbative intuition for the light-ray OPE arises from the fact that particle states of finite momentum are local on the celestial sphere, while zero-modes are not. For energy correlator observables, this contribution from the zero mode requires the presence of $\geq 3$ light-ray operators, and therefore first appears in the three-point correlator. We believe understanding exactly how this contribution is reproduced by explicit light-ray operators is important for better understanding the behavior of the light-ray OPE in perturbation theory.

We hope that the results of this paper are interesting both for the development of the light-ray OPE, as well as for its phenomenological applications to QCD, and jet substructure. We find it particularly exciting that by formulating jet substructure in terms of energy correlators, a wealth of beautiful theoretical techniques can be directly applied to studying the observables that are actually measured by experimentalists at collider experiments, namely the distribution of energy in their detectors. The fact that correlation functions of this energy distribution are so tightly constrained by symmetries is quite remarkable. We look forward to applications of the light-ray OPE to LHC phenomenology in the future!

%%%%%%%%%%%%%%%%%%%%%%%%%%%%%%%%%%%%%%%%%%
\begin{acknowledgments}
%%%%%%%%%%%%%%%%%%%%%%%%%%%%%%%%%%%%%%%%%%

We thank Cyuan-Han Chang, David Simmons-Duffin, Zhong-Jie Huang, Petr Kravchuk,  Xiaoyuan Zhang, and Alexander Zhiboedov for useful discussions. We also thank Cyuan-Han Chang and David Simmons-Duffin for communication prior to submission.
HC and HXZ were supported by National Science Foundation of China under contract No.~11975200.  
I.M is supported by start-up funds from Yale University. HXZ is grateful to the support from the Qiushi Science and Technologies Foundation.
JS acknowledges the support of Harvard University, where a portion of this work was completed, and funding from the Simons Collaboration on the Nonperturbative Bootstrap. 
  
%%%%%%%%%%%%%%%%%%%%%%%%%%%%%%%%%%%%%%%%%%
\end{acknowledgments}
%%%%%%%%%%%%%%%%%%%%%%%%%%%%%%%%%%%%%%%%%%

\appendix

%%%%%%%%%%%%%%%%%%%%%%%%%%%%%%%%%%%%%%%%%%
\section{Strong Coupling Data in $\cN=4$ \label{secStrong}}
%%%%%%%%%%%%%%%%%%%%%%%%%%%%%%%%%%%%%%%%%%

The n-point energy correlator was computed in $\cN=4$  by Hofman and Maldacena in \cite{Hofman:2008ar} in a strong coupling expansion, 
\begin{equation}
    \langle \mathcal{E}(\vec{n}_1) \cdots \mathcal{E}(\vec{n}_n)\rangle=\left(\frac{Q}{4\pi}\right)^n
    \left[
        1+\sum_{i<j}\frac{6\pi^2}{\lambda}\left[ (\vec{n}_i\cdot\vec{n}_j)^2-\frac{1}{3}\right]+\cO \left(\frac{1}{\lambda}\right)
    \right],
\end{equation}
which can be re-written for the case of $n=3$ in terms of the $(x_L,z,\bar{z})$ variables as
\begin{equation}
\begin{split}
    \langle \mathcal{E}(\vec{n}_1) \mathcal{E}(\vec{n}_2) \mathcal{E}(\vec{n}_3)\rangle
    =\left(\frac{Q}{4\pi}\right)^3 & \left[ 
        1 + \frac{6\pi^2}{\lambda} \left[ 2+x_L(-2-2 z \bar{z}-2(1-z)(1-\bar{z})) \right.\right.\\
       & \left.\left. + x_L^2 (1+(z\bar{z})^2+((1-z)(1-\bar{z})^2))\right]+\cO \left(\frac{1}{\lambda}\right)
    \right].
\end{split}
\end{equation}
The fact that the leading order result in the strong coupling expansion is constant indicates that the strong coupling tends to smear out the energy distribution to be completely uniform \cite{Strassler:2008bv,Hofman:2008ar,Hatta:2008tx}, with only small angular correlations appearing at subleading orders. 
In this section, we decompose this result into celestial blocks to illustrate the use of the celestial blocks at subleading order in the $x_L$ expansion.

Using the series form of the celestial blocks for the EEEC discussed in Sec. \ref{sec: eeec_subleading}, the relevant light-ray operator spectrum at LO at strong coupling is given by 
\begin{equation}
    1= f_{6,0;9,0}+\left(-f_{6,0;11,0}-\frac{3}{4}f_{8,0;11,0}\right)+\left(\frac{63}{110}f_{6,0;13,0} +\frac{9}{11} f_{8,0;13,0}+\frac{9}{28} f_{10,0;13,0} \right)+\cdots\,,
\end{equation}
where the first orders in the expansion of $f$ were given in \Eq{eq:sub_blocks}.
As expected, this does not contain any operators with transverse spin. The first transverse spin effects appear at NLO in the strong coupling expansion
\begin{equation}
    \begin{split}
       & \frac{6\pi^2}{\lambda}\sum_{1\leq i < j\leq 3}\left[\left(\vec{n}_i\cdot\vec{n}_j\right)^2-\frac{1}{3}\right]
        = \frac{6\pi^2}{\lambda}\left[2f_{6,0;9,0}+\left(-6f_{6,0;11,0}-\frac{9}{2}f_{8,0;11,0}\right)\right.\\
       & \hspace{4cm} \left. +\left(\frac{453}{55}f_{6,0;13,0}+{\color{red} \frac{3}{7}f_{8,2;13,0}}+\frac{120}{11}f_{8,0;13,0}+\frac{225}{49}f_{10,0;13,0}\right)+\cdots\right],
    \end{split}
\end{equation}
where we have highlighted the term with non-vanishing transverse spin in red. It would be interesting to understand the nature of the light-ray operators appearing in the energy correlator at strong coupling. 

\bibliography{spinning_gluon}{}

\providecommand{\href}[2]{#2}\begingroup\raggedright\begin{thebibliography}{100}

\bibitem{Cacciari:2005hq}
M.~Cacciari and G.~P. Salam, {\it {Dispelling the $N^{3}$ myth for the $k_t$
  jet-finder}},  {\em Phys. Lett. B} {\bf 641} (2006) 57--61,
  [\href{http://arxiv.org/abs/hep-ph/0512210}{{\tt hep-ph/0512210}}].

\bibitem{Cacciari:2008gp}
M.~Cacciari, G.~P. Salam, and G.~Soyez, {\it {The anti-$k_t$ jet clustering
  algorithm}},  {\em JHEP} {\bf 04} (2008) 063,
  [\href{http://arxiv.org/abs/0802.1189}{{\tt arXiv:0802.1189}}].

\bibitem{Cacciari:2011ma}
M.~Cacciari, G.~P. Salam, and G.~Soyez, {\it {FastJet User Manual}},  {\em Eur.
  Phys. J. C} {\bf 72} (2012) 1896, [\href{http://arxiv.org/abs/1111.6097}{{\tt
  arXiv:1111.6097}}].

\bibitem{Butterworth:2008iy}
J.~M. Butterworth, A.~R. Davison, M.~Rubin, and G.~P. Salam, {\it {Jet
  substructure as a new Higgs search channel at the LHC}},  {\em Phys. Rev.
  Lett.} {\bf 100} (2008) 242001, [\href{http://arxiv.org/abs/0802.2470}{{\tt
  arXiv:0802.2470}}].

\bibitem{Kaplan:2008ie}
D.~E. Kaplan, K.~Rehermann, M.~D. Schwartz, and B.~Tweedie, {\it {Top Tagging:
  A Method for Identifying Boosted Hadronically Decaying Top Quarks}},  {\em
  Phys. Rev. Lett.} {\bf 101} (2008) 142001,
  [\href{http://arxiv.org/abs/0806.0848}{{\tt arXiv:0806.0848}}].

\bibitem{Krohn:2009th}
D.~Krohn, J.~Thaler, and L.-T. Wang, {\it {Jet Trimming}},  {\em JHEP} {\bf 02}
  (2010) 084, [\href{http://arxiv.org/abs/0912.1342}{{\tt arXiv:0912.1342}}].

\bibitem{Larkoski:2017jix}
A.~J. Larkoski, I.~Moult, and B.~Nachman, {\it {Jet Substructure at the Large
  Hadron Collider: A Review of Recent Advances in Theory and Machine
  Learning}},  {\em Phys. Rept.} {\bf 841} (2020) 1--63,
  [\href{http://arxiv.org/abs/1709.04464}{{\tt arXiv:1709.04464}}].

\bibitem{Marzani:2019hun}
S.~Marzani, G.~Soyez, and M.~Spannowsky, {\em {Looking inside jets: an
  introduction to jet substructure and boosted-object phenomenology}},
  vol.~958.
\newblock Springer, 2019.

\bibitem{Cunqueiro:2021wls}
L.~Cunqueiro and A.~M. Sickles, {\it {Studying the QGP with Jets at the LHC and
  RHIC}},  \href{http://arxiv.org/abs/2110.14490}{{\tt arXiv:2110.14490}}.

\bibitem{Hofman:2008ar}
D.~M. Hofman and J.~Maldacena, {\it {Conformal collider physics: Energy and
  charge correlations}},  {\em JHEP} {\bf 05} (2008) 012,
  [\href{http://arxiv.org/abs/0803.1467}{{\tt arXiv:0803.1467}}].

\bibitem{Belitsky:2013xxa}
A.~Belitsky, S.~Hohenegger, G.~Korchemsky, E.~Sokatchev, and A.~Zhiboedov, {\it
  {From correlation functions to event shapes}},  {\em Nucl. Phys. B} {\bf 884}
  (2014) 305--343, [\href{http://arxiv.org/abs/1309.0769}{{\tt
  arXiv:1309.0769}}].

\bibitem{Li:2021zcf}
Y.~Li, I.~Moult, S.~S. van Velzen, W.~J. Waalewijn, and H.~X. Zhu, {\it
  {Extending Precision Perturbative QCD with Track Functions}},
  \href{http://arxiv.org/abs/2108.01674}{{\tt arXiv:2108.01674}}.

\bibitem{Chicherin:2020azt}
D.~Chicherin, J.~Henn, E.~Sokatchev, and K.~Yan, {\it {From correlation
  functions to event shapes in QCD}},
  \href{http://arxiv.org/abs/2001.10806}{{\tt arXiv:2001.10806}}.

\bibitem{Sveshnikov:1995vi}
N.~Sveshnikov and F.~Tkachov, {\it {Jets and quantum field theory}},  {\em
  Phys. Lett. B} {\bf 382} (1996) 403--408,
  [\href{http://arxiv.org/abs/hep-ph/9512370}{{\tt hep-ph/9512370}}].

\bibitem{Tkachov:1995kk}
F.~V. Tkachov, {\it {Measuring multi - jet structure of hadronic energy flow or
  What is a jet?}},  {\em Int. J. Mod. Phys. A} {\bf 12} (1997) 5411--5529,
  [\href{http://arxiv.org/abs/hep-ph/9601308}{{\tt hep-ph/9601308}}].

\bibitem{Korchemsky:1999kt}
G.~P. Korchemsky and G.~F. Sterman, {\it {Power corrections to event shapes and
  factorization}},  {\em Nucl. Phys. B} {\bf 555} (1999) 335--351,
  [\href{http://arxiv.org/abs/hep-ph/9902341}{{\tt hep-ph/9902341}}].

\bibitem{Bauer:2008dt}
C.~W. Bauer, S.~P. Fleming, C.~Lee, and G.~F. Sterman, {\it {Factorization of
  e+e- Event Shape Distributions with Hadronic Final States in Soft Collinear
  Effective Theory}},  {\em Phys. Rev. D} {\bf 78} (2008) 034027,
  [\href{http://arxiv.org/abs/0801.4569}{{\tt arXiv:0801.4569}}].

\bibitem{Belitsky:2013bja}
A.~Belitsky, S.~Hohenegger, G.~Korchemsky, E.~Sokatchev, and A.~Zhiboedov, {\it
  {Event shapes in $\mathcal{N} = 4$ super-Yang-Mills theory}},  {\em Nucl.
  Phys. B} {\bf 884} (2014) 206--256,
  [\href{http://arxiv.org/abs/1309.1424}{{\tt arXiv:1309.1424}}].

\bibitem{Kravchuk:2018htv}
P.~Kravchuk and D.~Simmons-Duffin, {\it {Light-ray operators in conformal field
  theory}},  {\em JHEP} {\bf 11} (2018) 102,
  [\href{http://arxiv.org/abs/1805.00098}{{\tt arXiv:1805.00098}}].

\bibitem{Chen:2020vvp}
H.~Chen, I.~Moult, X.~Zhang, and H.~X. Zhu, {\it {Rethinking jets with energy
  correlators: Tracks, resummation, and analytic continuation}},  {\em Phys.
  Rev. D} {\bf 102} (2020), no.~5 054012,
  [\href{http://arxiv.org/abs/2004.11381}{{\tt arXiv:2004.11381}}].

\bibitem{Larkoski:2013eya}
A.~J. Larkoski, G.~P. Salam, and J.~Thaler, {\it {Energy Correlation Functions
  for Jet Substructure}},  {\em JHEP} {\bf 06} (2013) 108,
  [\href{http://arxiv.org/abs/1305.0007}{{\tt arXiv:1305.0007}}].

\bibitem{Larkoski:2014gra}
A.~J. Larkoski, I.~Moult, and D.~Neill, {\it {Power Counting to Better Jet
  Observables}},  {\em JHEP} {\bf 12} (2014) 009,
  [\href{http://arxiv.org/abs/1409.6298}{{\tt arXiv:1409.6298}}].

\bibitem{Moult:2016cvt}
I.~Moult, L.~Necib, and J.~Thaler, {\it {New Angles on Energy Correlation
  Functions}},  {\em JHEP} {\bf 12} (2016) 153,
  [\href{http://arxiv.org/abs/1609.07483}{{\tt arXiv:1609.07483}}].

\bibitem{Dixon:2019uzg}
L.~J. Dixon, I.~Moult, and H.~X. Zhu, {\it {Collinear limit of the
  energy-energy correlator}},  {\em Phys. Rev. D} {\bf 100} (2019), no.~1
  014009, [\href{http://arxiv.org/abs/1905.01310}{{\tt arXiv:1905.01310}}].

\bibitem{Chen:2019bpb}
H.~Chen, M.-X. Luo, I.~Moult, T.-Z. Yang, X.~Zhang, and H.~X. Zhu, {\it {Three
  point energy correlators in the collinear limit: symmetries, dualities and
  analytic results}},  {\em JHEP} {\bf 08} (2020), no.~08 028,
  [\href{http://arxiv.org/abs/1912.11050}{{\tt arXiv:1912.11050}}].

\bibitem{Jaarsma:2022kdd}
M.~Jaarsma, Y.~Li, I.~Moult, W.~Waalewijn, and H.~X. Zhu, {\it {Renormalization
  Group Flows for Track Function Moments}},
  \href{http://arxiv.org/abs/2201.05166}{{\tt arXiv:2201.05166}}.

\bibitem{Chen:2020adz}
H.~Chen, I.~Moult, and H.~X. Zhu, {\it {Quantum Interference in Jet
  Substructure from Spinning Gluons}},
  \href{http://arxiv.org/abs/2011.02492}{{\tt arXiv:2011.02492}}.

\bibitem{Chen:2021gdk}
H.~Chen, I.~Moult, and H.~X. Zhu, {\it {Spinning Gluons from the QCD Light-Ray
  OPE}},  \href{http://arxiv.org/abs/2104.00009}{{\tt arXiv:2104.00009}}.

\bibitem{Holguin:2022epo}
J.~Holguin, I.~Moult, A.~Pathak, and M.~Procura, {\it {A New Paradigm for
  Precision Top Physics: Weighing the Top with Energy Correlators}},
  \href{http://arxiv.org/abs/2201.08393}{{\tt arXiv:2201.08393}}.

\bibitem{Komiske:2022enw}
P.~T. Komiske, I.~Moult, J.~Thaler, and H.~X. Zhu, {\it {Analyzing N-point
  Energy Correlators Inside Jets with CMS Open Data}},
  \href{http://arxiv.org/abs/2201.07800}{{\tt arXiv:2201.07800}}.

\bibitem{Moult:2018jzp}
I.~Moult and H.~X. Zhu, {\it {Simplicity from Recoil: The Three-Loop Soft
  Function and Factorization for the Energy-Energy Correlation}},  {\em JHEP}
  {\bf 08} (2018) 160, [\href{http://arxiv.org/abs/1801.02627}{{\tt
  arXiv:1801.02627}}].

\bibitem{Gao:2019ojf}
A.~Gao, H.~T. Li, I.~Moult, and H.~X. Zhu, {\it {Precision QCD Event Shapes at
  Hadron Colliders: The Transverse Energy-Energy Correlator in the Back-to-Back
  Limit}},  {\em Phys. Rev. Lett.} {\bf 123} (2019), no.~6 062001,
  [\href{http://arxiv.org/abs/1901.04497}{{\tt arXiv:1901.04497}}].

\bibitem{Moult:2019vou}
I.~Moult, G.~Vita, and K.~Yan, {\it {Subleading power resummation of rapidity
  logarithms: the energy-energy correlator in $ \mathcal{N} $ = 4 SYM}},  {\em
  JHEP} {\bf 07} (2020) 005, [\href{http://arxiv.org/abs/1912.02188}{{\tt
  arXiv:1912.02188}}].

\bibitem{Ebert:2020sfi}
M.~A. Ebert, B.~Mistlberger, and G.~Vita, {\it {The Energy-Energy Correlation
  in the back-to-back limit at N$^3$LO and N$^3$LL$^\prime$}},
  \href{http://arxiv.org/abs/2012.07859}{{\tt arXiv:2012.07859}}.

\bibitem{Li:2021txc}
H.~T. Li, Y.~Makris, and I.~Vitev, {\it {Energy-energy correlators in Deep
  Inelastic Scattering}},  \href{http://arxiv.org/abs/2102.05669}{{\tt
  arXiv:2102.05669}}.

\bibitem{Li:2020bub}
H.~T. Li, I.~Vitev, and Y.~J. Zhu, {\it {Transverse-Energy-Energy Correlations
  in Deep Inelastic Scattering}},  {\em JHEP} {\bf 11} (2020) 051,
  [\href{http://arxiv.org/abs/2006.02437}{{\tt arXiv:2006.02437}}].

\bibitem{Wilson:1969zs}
K.~G. Wilson, {\it {Nonlagrangian models of current algebra}},  {\em Phys.
  Rev.} {\bf 179} (1969) 1499--1512.

\bibitem{Kadanoff:1969zz}
L.~P. Kadanoff, {\it {Operator Algebra and the Determination of Critical
  Indices}},  {\em Phys. Rev. Lett.} {\bf 23} (1969) 1430--1433.

\bibitem{Balitsky:1987bk}
I.~Balitsky and V.~M. Braun, {\it {Evolution Equations for QCD String
  Operators}},  {\em Nucl. Phys. B} {\bf 311} (1989) 541--584.

\bibitem{Balitsky:1988fi}
I.~Balitsky and V.~M. Braun, {\it {Nonlocal Operator Expansion for Structure
  Functions of $e^+ e^-$ Annihilation}},  {\em Phys. Lett. B} {\bf 222} (1989)
  123--131.

\bibitem{Balitsky:1990ck}
I.~I. Balitsky and V.~M. Braun, {\it {The Nonlocal operator expansion for
  inclusive particle production in e+ e- annihilation}},  {\em Nucl. Phys. B}
  {\bf 361} (1991) 93--140.

\bibitem{Kologlu:2019mfz}
M.~Kologlu, P.~Kravchuk, D.~Simmons-Duffin, and A.~Zhiboedov, {\it {The
  light-ray OPE and conformal colliders}},
  \href{http://arxiv.org/abs/1905.01311}{{\tt arXiv:1905.01311}}.

\bibitem{1822249}
C.-H. Chang, M.~Kologlu, P.~Kravchuk, D.~Simmons-Duffin, and A.~Zhiboedov, {\it
  {Transverse spin in the light-ray OPE}},
  \href{http://arxiv.org/abs/2010.04726}{{\tt arXiv:2010.04726}}.

\bibitem{Belitsky:2013ofa}
A.~Belitsky, S.~Hohenegger, G.~Korchemsky, E.~Sokatchev, and A.~Zhiboedov, {\it
  {Energy-Energy Correlations in N=4 Supersymmetric Yang-Mills Theory}},  {\em
  Phys. Rev. Lett.} {\bf 112} (2014), no.~7 071601,
  [\href{http://arxiv.org/abs/1311.6800}{{\tt arXiv:1311.6800}}].

\bibitem{Henn:2019gkr}
J.~Henn, E.~Sokatchev, K.~Yan, and A.~Zhiboedov, {\it {Energy-energy
  correlation in $N$=4 super Yang-Mills theory at next-to-next-to-leading
  order}},  {\em Phys. Rev. D} {\bf 100} (2019), no.~3 036010,
  [\href{http://arxiv.org/abs/1903.05314}{{\tt arXiv:1903.05314}}].

\bibitem{Dixon:2018qgp}
L.~J. Dixon, M.-X. Luo, V.~Shtabovenko, T.-Z. Yang, and H.~X. Zhu, {\it
  {Analytical Computation of Energy-Energy Correlation at Next-to-Leading Order
  in QCD}},  {\em Phys. Rev. Lett.} {\bf 120} (2018), no.~10 102001,
  [\href{http://arxiv.org/abs/1801.03219}{{\tt arXiv:1801.03219}}].

\bibitem{Luo:2019nig}
M.-X. Luo, V.~Shtabovenko, T.-Z. Yang, and H.~X. Zhu, {\it {Analytic
  Next-To-Leading Order Calculation of Energy-Energy Correlation in
  Gluon-Initiated Higgs Decays}},  {\em JHEP} {\bf 06} (2019) 037,
  [\href{http://arxiv.org/abs/1903.07277}{{\tt arXiv:1903.07277}}].

\bibitem{Gao:2020vyx}
J.~Gao, V.~Shtabovenko, and T.-Z. Yang, {\it {Energy-energy correlation in
  hadronic Higgs decays: analytic results and phenomenology at NLO}},  {\em
  JHEP} {\bf 02} (2021) 210, [\href{http://arxiv.org/abs/2012.14188}{{\tt
  arXiv:2012.14188}}].

\bibitem{Karlberg:2021kwr}
A.~Karlberg, G.~P. Salam, L.~Scyboz, and R.~Verheyen, {\it {Spin correlations
  in final-state parton showers and jet observables}},  {\em Eur. Phys. J. C}
  {\bf 81} (2021), no.~8 681, [\href{http://arxiv.org/abs/2103.16526}{{\tt
  arXiv:2103.16526}}].

\bibitem{Dolan:2003hv}
F.~A. Dolan and H.~Osborn, {\it {Conformal partial waves and the operator
  product expansion}},  {\em Nucl. Phys. B} {\bf 678} (2004) 491--507,
  [\href{http://arxiv.org/abs/hep-th/0309180}{{\tt hep-th/0309180}}].

\bibitem{Caron-Huot:2017vep}
S.~Caron-Huot, {\it {Analyticity in Spin in Conformal Theories}},  {\em JHEP}
  {\bf 09} (2017) 078, [\href{http://arxiv.org/abs/1703.00278}{{\tt
  arXiv:1703.00278}}].

\bibitem{Simmons-Duffin:2017nub}
D.~Simmons-Duffin, D.~Stanford, and E.~Witten, {\it {A spacetime derivation of
  the Lorentzian OPE inversion formula}},  {\em JHEP} {\bf 07} (2018) 085,
  [\href{http://arxiv.org/abs/1711.03816}{{\tt arXiv:1711.03816}}].

\bibitem{Braun:2003rp}
V.~Braun, G.~Korchemsky, and D.~M\"uller, {\it {The Uses of conformal symmetry
  in QCD}},  {\em Prog. Part. Nucl. Phys.} {\bf 51} (2003) 311--398,
  [\href{http://arxiv.org/abs/hep-ph/0306057}{{\tt hep-ph/0306057}}].

\bibitem{Braun:2020zjm}
V.~M. Braun, Y.~Ji, and A.~N. Manashov, {\it {Two-photon processes in conformal
  QCD: Resummation of the descendants of leading-twist operators}},  {\em JHEP}
  {\bf 51} (2021) 51, [\href{http://arxiv.org/abs/2011.04533}{{\tt
  arXiv:2011.04533}}].

\bibitem{Polyakov:1970lyy}
A.~M. Polyakov, {\it {A Similarity hypothesis in the strong interactions. 1.
  Multiple hadron production in e+ e- annihilation}},  {\em Zh. Eksp. Teor.
  Fiz.} {\bf 59} (1970) 542--552.

\bibitem{Polyakov:1971gx}
A.~M. Polyakov, {\it {Similarity hypothesis in strong interactions. 2. cascade
  formation of hadrons and their energy distribution in e+ e- annihilation}},
  {\em Zh. Eksp. Teor. Fiz.} {\bf 60} (1971) 1572--1583.

\bibitem{Ferrara:1973yt}
S.~Ferrara, A.~F. Grillo, and R.~Gatto, {\it {Tensor representations of
  conformal algebra and conformally covariant operator product expansion}},
  {\em Annals Phys.} {\bf 76} (1973) 161--188.

\bibitem{Polyakov:1974gs}
A.~M. Polyakov, {\it {Nonhamiltonian approach to conformal quantum field
  theory}},  {\em Zh. Eksp. Teor. Fiz.} {\bf 66} (1974) 23--42.

\bibitem{Rychkov:2016iqz}
S.~Rychkov, {\em {EPFL Lectures on Conformal Field Theory in D\ensuremath{>}= 3
  Dimensions}}.
\newblock SpringerBriefs in Physics. 1, 2016.

\bibitem{Poland:2016chs}
D.~Poland and D.~Simmons-Duffin, {\it {The conformal bootstrap}},  {\em Nature
  Phys.} {\bf 12} (2016), no.~6 535--539.

\bibitem{Simmons-Duffin:2016gjk}
D.~Simmons-Duffin, {\it {The Conformal Bootstrap}},  in {\em {Theoretical
  Advanced Study Institute in Elementary Particle Physics}: {New Frontiers in
  Fields and Strings}}, pp.~1--74, 2017.
\newblock \href{http://arxiv.org/abs/1602.07982}{{\tt arXiv:1602.07982}}.

\bibitem{Poland:2018epd}
D.~Poland, S.~Rychkov, and A.~Vichi, {\it {The Conformal Bootstrap: Theory,
  Numerical Techniques, and Applications}},  {\em Rev. Mod. Phys.} {\bf 91}
  (2019) 015002, [\href{http://arxiv.org/abs/1805.04405}{{\tt
  arXiv:1805.04405}}].

\bibitem{Basham:1979gh}
C.~L. Basham, L.~S. Brown, S.~D. Ellis, and S.~T. Love, {\it {Energy
  Correlations in Perturbative Quantum Chromodynamics: A Conjecture for All
  Orders}},  {\em Phys. Lett. B} {\bf 85} (1979) 297--299.

\bibitem{Basham:1978zq}
C.~Basham, L.~Brown, S.~Ellis, and S.~Love, {\it {Energy Correlations in
  electron-Positron Annihilation in Quantum Chromodynamics: Asymptotically Free
  Perturbation Theory}},  {\em Phys. Rev. D} {\bf 19} (1979) 2018.

\bibitem{Basham:1978bw}
C.~Basham, L.~S. Brown, S.~D. Ellis, and S.~T. Love, {\it {Energy Correlations
  in electron - Positron Annihilation: Testing QCD}},  {\em Phys. Rev. Lett.}
  {\bf 41} (1978) 1585.

\bibitem{Basham:1977iq}
C.~L. Basham, L.~S. Brown, S.~D. Ellis, and S.~T. Love, {\it {Electron -
  Positron Annihilation Energy Pattern in Quantum Chromodynamics:
  Asymptotically Free Perturbation Theory}},  {\em Phys. Rev. D} {\bf 17}
  (1978) 2298.

\bibitem{Hofman:2016awc}
D.~M. Hofman, D.~Li, D.~Meltzer, D.~Poland, and F.~Rejon-Barrera, {\it {A Proof
  of the Conformal Collider Bounds}},  {\em JHEP} {\bf 06} (2016) 111,
  [\href{http://arxiv.org/abs/1603.03771}{{\tt arXiv:1603.03771}}].

\bibitem{Cordova:2017zej}
C.~Cordova, J.~Maldacena, and G.~J. Turiaci, {\it {Bounds on OPE Coefficients
  from Interference Effects in the Conformal Collider}},  {\em JHEP} {\bf 11}
  (2017) 032, [\href{http://arxiv.org/abs/1710.03199}{{\tt arXiv:1710.03199}}].

\bibitem{Belitsky:2014zha}
A.~Belitsky, S.~Hohenegger, G.~Korchemsky, and E.~Sokatchev, {\it {N=4
  superconformal Ward identities for correlation functions}},  {\em Nucl. Phys.
  B} {\bf 904} (2016) 176--215, [\href{http://arxiv.org/abs/1409.2502}{{\tt
  arXiv:1409.2502}}].

\bibitem{Korchemsky:2015ssa}
G.~Korchemsky and E.~Sokatchev, {\it {Four-point correlation function of
  stress-energy tensors in $ \mathcal{N}=4 $ superconformal theories}},  {\em
  JHEP} {\bf 12} (2015) 133, [\href{http://arxiv.org/abs/1504.07904}{{\tt
  arXiv:1504.07904}}].

\bibitem{Kologlu:2019bco}
M.~Kologlu, P.~Kravchuk, D.~Simmons-Duffin, and A.~Zhiboedov, {\it {Shocks,
  Superconvergence, and a Stringy Equivalence Principle}},
  \href{http://arxiv.org/abs/1904.05905}{{\tt arXiv:1904.05905}}.

\bibitem{Korchemsky:2019nzm}
G.~Korchemsky, {\it {Energy correlations in the end-point region}},  {\em JHEP}
  {\bf 01} (2020) 008, [\href{http://arxiv.org/abs/1905.01444}{{\tt
  arXiv:1905.01444}}].

\bibitem{Korchemsky:2021okt}
G.~Korchemsky, E.~Sokatchev, and A.~Zhiboedov, {\it {Generalizing event shapes:
  In search of lost collider time}},
  \href{http://arxiv.org/abs/2106.14899}{{\tt arXiv:2106.14899}}.

\bibitem{Korchemsky:2021htm}
G.~P. Korchemsky and A.~Zhiboedov, {\it {On the light-ray algebra in conformal
  field theories}},  \href{http://arxiv.org/abs/2109.13269}{{\tt
  arXiv:2109.13269}}.

\bibitem{Poland:2021xjs}
D.~Poland and V.~Prilepina, {\it {Recursion relations for 5-point conformal
  blocks}},  {\em JHEP} {\bf 10} (2021) 160,
  [\href{http://arxiv.org/abs/2103.12092}{{\tt arXiv:2103.12092}}].

\bibitem{Collins:1981ta}
J.~C. Collins and G.~F. Sterman, {\it {Soft Partons in {QCD}}},  {\em Nucl.
  Phys. B} {\bf 185} (1981) 172--188.

\bibitem{Bodwin:1984hc}
G.~T. Bodwin, {\it {Factorization of the Drell-Yan Cross-Section in
  Perturbation Theory}},  {\em Phys. Rev. D} {\bf 31} (1985) 2616. [Erratum:
  Phys.Rev.D 34, 3932 (1986)].

\bibitem{Collins:1985ue}
J.~C. Collins, D.~E. Soper, and G.~F. Sterman, {\it {Factorization for Short
  Distance Hadron - Hadron Scattering}},  {\em Nucl. Phys. B} {\bf 261} (1985)
  104--142.

\bibitem{Collins:1988ig}
J.~C. Collins, D.~E. Soper, and G.~F. Sterman, {\it {Soft Gluons and
  Factorization}},  {\em Nucl. Phys. B} {\bf 308} (1988) 833--856.

\bibitem{Collins:1989gx}
J.~C. Collins, D.~E. Soper, and G.~F. Sterman, {\it {Factorization of Hard
  Processes in QCD}},  {\em Adv. Ser. Direct. High Energy Phys.} {\bf 5} (1989)
  1--91, [\href{http://arxiv.org/abs/hep-ph/0409313}{{\tt hep-ph/0409313}}].

\bibitem{Collins:2011zzd}
J.~Collins, {\em {Foundations of perturbative QCD}}, vol.~32.
\newblock Cambridge University Press, 11, 2013.

\bibitem{Nayak:2005rt}
G.~C. Nayak, J.-W. Qiu, and G.~F. Sterman, {\it {Fragmentation, NRQCD and NNLO
  factorization analysis in heavy quarkonium production}},  {\em Phys. Rev. D}
  {\bf 72} (2005) 114012, [\href{http://arxiv.org/abs/hep-ph/0509021}{{\tt
  hep-ph/0509021}}].

\bibitem{Mitov:2012gt}
A.~Mitov and G.~Sterman, {\it {Final state interactions in single- and
  multi-particle inclusive cross sections for hadronic collisions}},  {\em
  Phys. Rev. D} {\bf 86} (2012) 114038,
  [\href{http://arxiv.org/abs/1209.5798}{{\tt arXiv:1209.5798}}].

\bibitem{Kang:2016ehg}
Z.-B. Kang, F.~Ringer, and I.~Vitev, {\it {Jet substructure using
  semi-inclusive jet functions in SCET}},  {\em JHEP} {\bf 11} (2016) 155,
  [\href{http://arxiv.org/abs/1606.07063}{{\tt arXiv:1606.07063}}].

\bibitem{Kang:2016mcy}
Z.-B. Kang, F.~Ringer, and I.~Vitev, {\it {The semi-inclusive jet function in
  SCET and small radius resummation for inclusive jet production}},  {\em JHEP}
  {\bf 10} (2016) 125, [\href{http://arxiv.org/abs/1606.06732}{{\tt
  arXiv:1606.06732}}].

\bibitem{Kang:2017frl}
Z.-B. Kang, F.~Ringer, and I.~Vitev, {\it {Inclusive production of small radius
  jets in heavy-ion collisions}},  {\em Phys. Lett. B} {\bf 769} (2017)
  242--248, [\href{http://arxiv.org/abs/1701.05839}{{\tt arXiv:1701.05839}}].

\bibitem{Aversa:1988fv}
F.~Aversa, P.~Chiappetta, M.~Greco, and J.~P. Guillet, {\it {Higher Order
  Corrections to QCD Jets}},  {\em Phys. Lett. B} {\bf 210} (1988) 225.

\bibitem{Aversa:1988mm}
F.~Aversa, M.~Greco, P.~Chiappetta, and J.~P. Guillet, {\it {HIGHER ORDER
  CORRECTIONS TO QCD JETS: GLUON-GLUON PROCESSES}},  {\em Phys. Lett. B} {\bf
  211} (1988) 465.

\bibitem{Aversa:1990uv}
F.~Aversa, M.~Greco, P.~Chiappetta, and J.~P. Guillet, {\it {Jet inclusive
  production to O $(alpha-s^{3)}$ : Comparison with data}},  {\em Phys. Rev.
  Lett.} {\bf 65} (1990) 401--403.

\bibitem{Aversa:1989xw}
F.~Aversa, M.~Greco, P.~Chiappetta, and J.~P. Guillet, {\it {Jet Production in
  Hadronic Collisions to O ($\alpha^- s^3$)}},  {\em Z. Phys. C} {\bf 46}
  (1990) 253.

\bibitem{Aversa:1988vb}
F.~Aversa, P.~Chiappetta, M.~Greco, and J.~P. Guillet, {\it {QCD Corrections to
  Parton-Parton Scattering Processes}},  {\em Nucl. Phys. B} {\bf 327} (1989)
  105.

\bibitem{Czakon:2021ohs}
M.~L. Czakon, T.~Generet, A.~Mitov, and R.~Poncelet, {\it {B-hadron
  hadro-production in NNLO QCD: application to LHC $t\bar{t}$ events with
  leptonic decays}},  \href{http://arxiv.org/abs/2102.08267}{{\tt
  arXiv:2102.08267}}.

\bibitem{Konishi:1979cb}
K.~Konishi, A.~Ukawa, and G.~Veneziano, {\it {Jet Calculus: A Simple Algorithm
  for Resolving QCD Jets}},  {\em Nucl. Phys.} {\bf B157} (1979) 45--107.

\bibitem{Bauer:2000ew}
C.~W. Bauer, S.~Fleming, and M.~E. Luke, {\it {Summing Sudakov logarithms in B
  ---> X(s gamma) in effective field theory}},  {\em Phys. Rev. D} {\bf 63}
  (2000) 014006, [\href{http://arxiv.org/abs/hep-ph/0005275}{{\tt
  hep-ph/0005275}}].

\bibitem{Bauer:2000yr}
C.~W. Bauer, S.~Fleming, D.~Pirjol, and I.~W. Stewart, {\it {An Effective field
  theory for collinear and soft gluons: Heavy to light decays}},  {\em Phys.
  Rev. D} {\bf 63} (2001) 114020,
  [\href{http://arxiv.org/abs/hep-ph/0011336}{{\tt hep-ph/0011336}}].

\bibitem{Bauer:2001ct}
C.~W. Bauer and I.~W. Stewart, {\it {Invariant operators in collinear effective
  theory}},  {\em Phys. Lett.} {\bf B516} (2001) 134--142,
  [\href{http://arxiv.org/abs/hep-ph/0107001}{{\tt hep-ph/0107001}}].

\bibitem{Bauer:2001yt}
C.~W. Bauer, D.~Pirjol, and I.~W. Stewart, {\it {Soft collinear factorization
  in effective field theory}},  {\em Phys. Rev. D} {\bf 65} (2002) 054022,
  [\href{http://arxiv.org/abs/hep-ph/0109045}{{\tt hep-ph/0109045}}].

\bibitem{Campbell:1997hg}
J.~M. Campbell and E.~W.~N. Glover, {\it {Double unresolved approximations to
  multiparton scattering amplitudes}},  {\em Nucl. Phys. B} {\bf 527} (1998)
  264--288, [\href{http://arxiv.org/abs/hep-ph/9710255}{{\tt hep-ph/9710255}}].

\bibitem{Catani:1998nv}
S.~Catani and M.~Grazzini, {\it {Collinear factorization and splitting
  functions for next-to-next-to-leading order QCD calculations}},  {\em Phys.
  Lett. B} {\bf 446} (1999) 143--152,
  [\href{http://arxiv.org/abs/hep-ph/9810389}{{\tt hep-ph/9810389}}].

\bibitem{Gehrmann-DeRidder:1997fom}
A.~Gehrmann-De~Ridder and E.~W.~N. Glover, {\it {A Complete O (alpha alpha-s)
  calculation of the photon + 1 jet rate in e+ e- annihilation}},  {\em Nucl.
  Phys. B} {\bf 517} (1998) 269--323,
  [\href{http://arxiv.org/abs/hep-ph/9707224}{{\tt hep-ph/9707224}}].

\bibitem{Ritzmann:2014mka}
M.~Ritzmann and W.~J. Waalewijn, {\it {Fragmentation in Jets at NNLO}},  {\em
  Phys. Rev. D} {\bf 90} (2014), no.~5 054029,
  [\href{http://arxiv.org/abs/1407.3272}{{\tt arXiv:1407.3272}}].

\bibitem{Ellis:1980wv}
R.~K. Ellis, D.~A. Ross, and A.~E. Terrano, {\it {The Perturbative Calculation
  of Jet Structure in e+ e- Annihilation}},  {\em Nucl. Phys. B} {\bf 178}
  (1981) 421--456.

\bibitem{Collins:1987cp}
J.~C. Collins, {\it {Spin Correlations in Monte Carlo Event Generators}},  {\em
  Nucl. Phys. B} {\bf 304} (1988) 794--804.

\bibitem{Knowles:1988hu}
I.~Knowles, {\it {A Linear Algorithm for Calculating Spin Correlations in
  Hadronic Collisions}},  {\em Comput. Phys. Commun.} {\bf 58} (1990) 271--284.

\bibitem{Knowles:1987cu}
I.~Knowles, {\it {Angular Correlations in \{QCD\}}},  {\em Nucl. Phys. B} {\bf
  304} (1988) 767--793.

\bibitem{Knowles:1988vs}
I.~Knowles, {\it {Spin Correlations in Parton - Parton Scattering}},  {\em
  Nucl. Phys. B} {\bf 310} (1988) 571--588.

\bibitem{Hamilton:2021dyz}
K.~Hamilton, A.~Karlberg, G.~P. Salam, L.~Scyboz, and R.~Verheyen, {\it {Soft
  spin correlations in final-state parton showers}},
  \href{http://arxiv.org/abs/2111.01161}{{\tt arXiv:2111.01161}}.

\bibitem{Yamazaki:2016vqi}
M.~Yamazaki, {\it {Comments on Determinant Formulas for General CFTs}},  {\em
  JHEP} {\bf 10} (2016) 035, [\href{http://arxiv.org/abs/1601.04072}{{\tt
  arXiv:1601.04072}}].

\bibitem{Karateev:2017jgd}
D.~Karateev, P.~Kravchuk, and D.~Simmons-Duffin, {\it {Weight Shifting
  Operators and Conformal Blocks}},  {\em JHEP} {\bf 02} (2018) 081,
  [\href{http://arxiv.org/abs/1706.07813}{{\tt arXiv:1706.07813}}].

\bibitem{langlands1989irreducible}
R.~P. Langlands, {\it Irreducible representations of real algebraic groups},
  {\em Representation theory and harmonic analysis on semisimple Lie groups}
  {\bf 31} (1989) 101--170.

\bibitem{Dirac:1936fq}
P.~A.~M. Dirac, {\it {Wave equations in conformal space}},  {\em Annals Math.}
  {\bf 37} (1936) 429--442.

\bibitem{Mack:1969rr}
G.~Mack and A.~Salam, {\it {Finite component field representations of the
  conformal group}},  {\em Annals Phys.} {\bf 53} (1969) 174--202.

\bibitem{Boulware:1970ty}
D.~G. Boulware, L.~S. Brown, and R.~D. Peccei, {\it {Deep-inelastic
  electroproduction and conformal symmetry}},  {\em Phys. Rev. D} {\bf 2}
  (1970) 293--298.

\bibitem{Cornalba:2009ax}
L.~Cornalba, M.~S. Costa, and J.~Penedones, {\it {Deep Inelastic Scattering in
  Conformal QCD}},  {\em JHEP} {\bf 03} (2010) 133,
  [\href{http://arxiv.org/abs/0911.0043}{{\tt arXiv:0911.0043}}].

\bibitem{Weinberg:2010fx}
S.~Weinberg, {\it {Six-dimensional Methods for Four-dimensional Conformal Field
  Theories}},  {\em Phys. Rev. D} {\bf 82} (2010) 045031,
  [\href{http://arxiv.org/abs/1006.3480}{{\tt arXiv:1006.3480}}].

\bibitem{Costa:2011mg}
M.~S. Costa, J.~Penedones, D.~Poland, and S.~Rychkov, {\it {Spinning Conformal
  Correlators}},  {\em JHEP} {\bf 11} (2011) 071,
  [\href{http://arxiv.org/abs/1107.3554}{{\tt arXiv:1107.3554}}].

\bibitem{Dolan:2011dv}
F.~A. Dolan and H.~Osborn, {\it {Conformal Partial Waves: Further Mathematical
  Results}},  \href{http://arxiv.org/abs/1108.6194}{{\tt arXiv:1108.6194}}.

\bibitem{Ferrara:1972kab}
S.~Ferrara, A.~F. Grillo, G.~Parisi, and R.~Gatto, {\it {Covariant expansion of
  the conformal four-point function}},  {\em Nucl. Phys. B} {\bf 49} (1972)
  77--98. [Erratum: Nucl.Phys.B 53, 643--643 (1973)].

\bibitem{Ferrara:1972uq}
S.~Ferrara, A.~F. Grillo, G.~Parisi, and R.~Gatto, {\it {The shadow operator
  formalism for conformal algebra. Vacuum expectation values and operator
  products}},  {\em Lett. Nuovo Cim.} {\bf 4S2} (1972) 115--120.

\bibitem{Ferrara:1974nf}
S.~Ferrara, A.~F. Grillo, R.~Gatto, and G.~Parisi, {\it {Analyticity properties
  and asymptotic expansions of conformal covariant green's functions}},  {\em
  Nuovo Cim. A} {\bf 19} (1974) 667--695.

\bibitem{Ferrara:1974ny}
S.~Ferrara, R.~Gatto, and A.~F. Grillo, {\it {Properties of Partial Wave
  Amplitudes in Conformal Invariant Field Theories}},  {\em Nuovo Cim. A} {\bf
  26} (1975) 226.

\bibitem{Dobrev:1977qv}
V.~K. Dobrev, G.~Mack, V.~B. Petkova, S.~G. Petrova, and I.~T. Todorov, {\em
  {Harmonic Analysis on the n-Dimensional Lorentz Group and Its Application to
  Conformal Quantum Field Theory}}, vol.~63.
\newblock 1977.

\bibitem{Karateev:2018oml}
D.~Karateev, P.~Kravchuk, and D.~Simmons-Duffin, {\it {Harmonic Analysis and
  Mean Field Theory}},  {\em JHEP} {\bf 10} (2019) 217,
  [\href{http://arxiv.org/abs/1809.05111}{{\tt arXiv:1809.05111}}].

\bibitem{Isachenkov:2016gim}
M.~Isachenkov and V.~Schomerus, {\it {Superintegrability of $d$-dimensional
  Conformal Blocks}},  {\em Phys. Rev. Lett.} {\bf 117} (2016), no.~7 071602,
  [\href{http://arxiv.org/abs/1602.01858}{{\tt arXiv:1602.01858}}].

\bibitem{Schomerus:2021ins}
V.~Schomerus, {\it {Conformal Hypergeometry and Integrability}},  11, 2021.
\newblock \href{http://arxiv.org/abs/2111.14864}{{\tt arXiv:2111.14864}}.

\bibitem{Correia:2020xtr}
M.~Correia, A.~Sever, and A.~Zhiboedov, {\it {An Analytical Toolkit for the
  S-matrix Bootstrap}},  \href{http://arxiv.org/abs/2006.08221}{{\tt
  arXiv:2006.08221}}.

\bibitem{Poschl:1933zz}
G.~Poschl and E.~Teller, {\it {Bemerkungen zur Quantenmechanik des
  anharmonischen Oszillators}},  {\em Z. Phys.} {\bf 83} (1933) 143--151.

\bibitem{Dolan:2000ut}
F.~A. Dolan and H.~Osborn, {\it {Conformal four point functions and the
  operator product expansion}},  {\em Nucl. Phys.} {\bf B599} (2001) 459--496,
  [\href{http://arxiv.org/abs/hep-th/0011040}{{\tt hep-th/0011040}}].

\bibitem{Altarelli:1977zs}
G.~Altarelli and G.~Parisi, {\it {Asymptotic Freedom in Parton Language}},
  {\em Nucl. Phys. B} {\bf 126} (1977) 298--318.

\bibitem{Moult:2016fqy}
I.~Moult, L.~Rothen, I.~W. Stewart, F.~J. Tackmann, and H.~X. Zhu, {\it
  {Subleading Power Corrections for N-Jettiness Subtractions}},  {\em Phys.
  Rev. D} {\bf 95} (2017), no.~7 074023,
  [\href{http://arxiv.org/abs/1612.00450}{{\tt arXiv:1612.00450}}].

\bibitem{Moult:2017jsg}
I.~Moult, L.~Rothen, I.~W. Stewart, F.~J. Tackmann, and H.~X. Zhu, {\it {N
  -jettiness subtractions for $gg\to H$ at subleading power}},  {\em Phys. Rev.
  D} {\bf 97} (2018), no.~1 014013,
  [\href{http://arxiv.org/abs/1710.03227}{{\tt arXiv:1710.03227}}].

\bibitem{Ebert:2018gsn}
M.~A. Ebert, I.~Moult, I.~W. Stewart, F.~J. Tackmann, G.~Vita, and H.~X. Zhu,
  {\it {Subleading power rapidity divergences and power corrections for
  q$_{T}$}},  {\em JHEP} {\bf 04} (2019) 123,
  [\href{http://arxiv.org/abs/1812.08189}{{\tt arXiv:1812.08189}}].

\bibitem{Ebert:2018lzn}
M.~A. Ebert, I.~Moult, I.~W. Stewart, F.~J. Tackmann, G.~Vita, and H.~X. Zhu,
  {\it {Power Corrections for N-Jettiness Subtractions at ${\cal
  O}(\alpha_s)$}},  {\em JHEP} {\bf 12} (2018) 084,
  [\href{http://arxiv.org/abs/1807.10764}{{\tt arXiv:1807.10764}}].

\bibitem{Moult:2019uhz}
I.~Moult, I.~W. Stewart, G.~Vita, and H.~X. Zhu, {\it {The Soft Quark
  Sudakov}},  {\em JHEP} {\bf 05} (2020) 089,
  [\href{http://arxiv.org/abs/1910.14038}{{\tt arXiv:1910.14038}}].

\bibitem{Albayrak:2019gnz}
S.~Albayrak, D.~Meltzer, and D.~Poland, {\it {More Analytic Bootstrap:
  Nonperturbative Effects and Fermions}},  {\em JHEP} {\bf 08} (2019) 040,
  [\href{http://arxiv.org/abs/1904.00032}{{\tt arXiv:1904.00032}}].

\bibitem{Caron-Huot:2020ouj}
S.~Caron-Huot, Y.~Gobeil, and Z.~Zahraee, {\it {The leading trajectory in the
  2+1D Ising CFT}},  \href{http://arxiv.org/abs/2007.11647}{{\tt
  arXiv:2007.11647}}.

\bibitem{Liu:2020tpf}
J.~Liu, D.~Meltzer, D.~Poland, and D.~Simmons-Duffin, {\it {The Lorentzian
  inversion formula and the spectrum of the 3d O(2) CFT}},  {\em JHEP} {\bf 09}
  (2020) 115, [\href{http://arxiv.org/abs/2007.07914}{{\tt arXiv:2007.07914}}].
  [Erratum: JHEP 01, 206 (2021)].

\bibitem{Atanasov:2022bpi}
A.~Atanasov, A.~Hillman, D.~Poland, J.~Rong, and N.~Su, {\it {Precision
  Bootstrap for the $\mathcal{N}=1$ Super-Ising Model}},
  \href{http://arxiv.org/abs/2201.02206}{{\tt arXiv:2201.02206}}.

\bibitem{Alday:2007mf}
L.~F. Alday and J.~M. Maldacena, {\it {Comments on operators with large spin}},
   {\em JHEP} {\bf 11} (2007) 019, [\href{http://arxiv.org/abs/0708.0672}{{\tt
  arXiv:0708.0672}}].

\bibitem{Komargodski:2012ek}
Z.~Komargodski and A.~Zhiboedov, {\it {Convexity and Liberation at Large
  Spin}},  {\em JHEP} {\bf 11} (2013) 140,
  [\href{http://arxiv.org/abs/1212.4103}{{\tt arXiv:1212.4103}}].

\bibitem{Fitzpatrick:2012yx}
A.~L. Fitzpatrick, J.~Kaplan, D.~Poland, and D.~Simmons-Duffin, {\it {The
  Analytic Bootstrap and AdS Superhorizon Locality}},  {\em JHEP} {\bf 12}
  (2013) 004, [\href{http://arxiv.org/abs/1212.3616}{{\tt arXiv:1212.3616}}].

\bibitem{Alday:2015ota}
L.~F. Alday and A.~Zhiboedov, {\it {Conformal Bootstrap With Slightly Broken
  Higher Spin Symmetry}},  {\em JHEP} {\bf 06} (2016) 091,
  [\href{http://arxiv.org/abs/1506.04659}{{\tt arXiv:1506.04659}}].

\bibitem{Alday:2015eya}
L.~F. Alday, A.~Bissi, and T.~Lukowski, {\it {Large spin systematics in CFT}},
  {\em JHEP} {\bf 11} (2015) 101, [\href{http://arxiv.org/abs/1502.07707}{{\tt
  arXiv:1502.07707}}].

\bibitem{Alday:2016njk}
L.~F. Alday, {\it {Large Spin Perturbation Theory for Conformal Field
  Theories}},  {\em Phys. Rev. Lett.} {\bf 119} (2017), no.~11 111601,
  [\href{http://arxiv.org/abs/1611.01500}{{\tt arXiv:1611.01500}}].

\bibitem{Kos:2013tga}
F.~Kos, D.~Poland, and D.~Simmons-Duffin, {\it {Bootstrapping the $O(N)$ vector
  models}},  {\em JHEP} {\bf 06} (2014) 091,
  [\href{http://arxiv.org/abs/1307.6856}{{\tt arXiv:1307.6856}}].

\bibitem{Alday:2017vkk}
L.~F. Alday and S.~Caron-Huot, {\it {Gravitational S-matrix from CFT dispersion
  relations}},  {\em JHEP} {\bf 12} (2018) 017,
  [\href{http://arxiv.org/abs/1711.02031}{{\tt arXiv:1711.02031}}].

\bibitem{Caron-Huot:2018kta}
S.~Caron-Huot and A.-K. Trinh, {\it {All tree-level correlators in AdS$_5\times
  S_5$ supergravity: hidden ten-dimensional conformal symmetry}},  {\em JHEP}
  {\bf 01} (2019) 196, [\href{http://arxiv.org/abs/1809.09173}{{\tt
  arXiv:1809.09173}}].

\bibitem{Alday:2019clp}
L.~F. Alday, J.~Henriksson, and M.~van Loon, {\it {An alternative to diagrams
  for the critical O(N) model: dimensions and structure constants to order
  1/N$^{2}$}},  {\em JHEP} {\bf 01} (2020) 063,
  [\href{http://arxiv.org/abs/1907.02445}{{\tt arXiv:1907.02445}}].

\bibitem{Henriksson:2020jwk}
J.~Henriksson, {\em {Analytic Bootstrap for Perturbative Conformal Field
  Theories}}.
\newblock PhD thesis, Oxford U., 2020.
\newblock \href{http://arxiv.org/abs/2008.12600}{{\tt arXiv:2008.12600}}.

\bibitem{Caron-Huot:2020nem}
S.~Caron-Huot and J.~Sandor, {\it {Conformal Regge Theory at Finite Boost}},
  {\em JHEP} {\bf 05} (2021) 059, [\href{http://arxiv.org/abs/2008.11759}{{\tt
  arXiv:2008.11759}}].

\bibitem{Ebert:2016gcn}
M.~A. Ebert and F.~J. Tackmann, {\it {Resummation of Transverse Momentum
  Distributions in Distribution Space}},  {\em JHEP} {\bf 02} (2017) 110,
  [\href{http://arxiv.org/abs/1611.08610}{{\tt arXiv:1611.08610}}].

\bibitem{Maldacena:2015waa}
J.~Maldacena, S.~H. Shenker, and D.~Stanford, {\it {A bound on chaos}},  {\em
  JHEP} {\bf 08} (2016) 106, [\href{http://arxiv.org/abs/1503.01409}{{\tt
  arXiv:1503.01409}}].

\bibitem{Korchemsky:1988si}
G.~P. Korchemsky, {\it {Asymptotics of the Altarelli-Parisi-Lipatov Evolution
  Kernels of Parton Distributions}},  {\em Mod. Phys. Lett. A} {\bf 4} (1989)
  1257--1276.

\bibitem{Korchemsky:1992xv}
G.~P. Korchemsky and G.~Marchesini, {\it {Structure function for large x and
  renormalization of Wilson loop}},  {\em Nucl. Phys. B} {\bf 406} (1993)
  225--258, [\href{http://arxiv.org/abs/hep-ph/9210281}{{\tt hep-ph/9210281}}].

\bibitem{Barrat:2021yvp}
J.~Barrat, A.~Gimenez-Grau, and P.~Liendo, {\it {Bootstrapping holographic
  defect correlators in $\mathcal{N}=4$ super Yang-Mills}},
  \href{http://arxiv.org/abs/2108.13432}{{\tt arXiv:2108.13432}}.

\bibitem{Barrat:2020vch}
J.~Barrat, P.~Liendo, and J.~Plefka, {\it {Two-point correlator of chiral
  primary operators with a Wilson line defect in $ \mathcal{N} $ = 4 SYM}},
  {\em JHEP} {\bf 05} (2021) 195, [\href{http://arxiv.org/abs/2011.04678}{{\tt
  arXiv:2011.04678}}].

\bibitem{Sjostrand:2014zea}
T.~Sj\"ostrand, S.~Ask, J.~R. Christiansen, R.~Corke, N.~Desai, P.~Ilten,
  S.~Mrenna, S.~Prestel, C.~O. Rasmussen, and P.~Z. Skands, {\it {An
  introduction to PYTHIA 8.2}},  {\em Comput. Phys. Commun.} {\bf 191} (2015)
  159--177, [\href{http://arxiv.org/abs/1410.3012}{{\tt arXiv:1410.3012}}].

\bibitem{Patrick}
P.~T. Komiske, {\it
  \textrm{https://github.com/pkomiske/EnergyEnergyCorrelators}},  2022.

\bibitem{Komiske:2019jim}
P.~T. Komiske, R.~Mastandrea, E.~M. Metodiev, P.~Naik, and J.~Thaler, {\it
  {Exploring the Space of Jets with CMS Open Data}},  {\em Phys. Rev. D} {\bf
  101} (2020), no.~3 034009, [\href{http://arxiv.org/abs/1908.08542}{{\tt
  arXiv:1908.08542}}].

\bibitem{Pappadopulo:2012jk}
D.~Pappadopulo, S.~Rychkov, J.~Espin, and R.~Rattazzi, {\it {OPE Convergence in
  Conformal Field Theory}},  {\em Phys. Rev. D} {\bf 86} (2012) 105043,
  [\href{http://arxiv.org/abs/1208.6449}{{\tt arXiv:1208.6449}}].

\bibitem{Hogervorst:2013sma}
M.~Hogervorst and S.~Rychkov, {\it {Radial Coordinates for Conformal Blocks}},
  {\em Phys. Rev. D} {\bf 87} (2013) 106004,
  [\href{http://arxiv.org/abs/1303.1111}{{\tt arXiv:1303.1111}}].

\bibitem{Rychkov:2015lca}
S.~Rychkov and P.~Yvernay, {\it {Remarks on the Convergence Properties of the
  Conformal Block Expansion}},  {\em Phys. Lett. B} {\bf 753} (2016) 682--686,
  [\href{http://arxiv.org/abs/1510.08486}{{\tt arXiv:1510.08486}}].

\bibitem{Li:2016yez}
H.~T. Li and P.~Skands, {\it {A framework for second-order parton showers}},
  {\em Phys. Lett. B} {\bf 771} (2017) 59--66,
  [\href{http://arxiv.org/abs/1611.00013}{{\tt arXiv:1611.00013}}].

\bibitem{Gellersen:2021eci}
L.~Gellersen, S.~H\"oche, and S.~Prestel, {\it {Disentangling soft and
  collinear effects in QCD parton showers}},
  \href{http://arxiv.org/abs/2110.05964}{{\tt arXiv:2110.05964}}.

\bibitem{Hoche:2017iem}
S.~H\"oche and S.~Prestel, {\it {Triple collinear emissions in parton
  showers}},  {\em Phys. Rev. D} {\bf 96} (2017), no.~7 074017,
  [\href{http://arxiv.org/abs/1705.00742}{{\tt arXiv:1705.00742}}].

\bibitem{Hoche:2017hno}
S.~H\"oche, F.~Krauss, and S.~Prestel, {\it {Implementing NLO DGLAP evolution
  in Parton Showers}},  {\em JHEP} {\bf 10} (2017) 093,
  [\href{http://arxiv.org/abs/1705.00982}{{\tt arXiv:1705.00982}}].

\bibitem{Loschner:2021keu}
M.~L\"oschner, S.~Pl\"atzer, and E.~S. Dore, {\it {Multi-Emission Kernels for
  Parton Branching Algorithms}},  \href{http://arxiv.org/abs/2112.14454}{{\tt
  arXiv:2112.14454}}.

\bibitem{Rychkov:2016mrc}
S.~Rychkov, D.~Simmons-Duffin, and B.~Zan, {\it {Non-gaussianity of the
  critical 3d Ising model}},  {\em SciPost Phys.} {\bf 2} (2017), no.~1 001,
  [\href{http://arxiv.org/abs/1612.02436}{{\tt arXiv:1612.02436}}].

\bibitem{Henn:2020omi}
J.~M. Henn, {\it {What can we learn about QCD and collider physics from N=4
  super Yang-Mills?}},  \href{http://arxiv.org/abs/2006.00361}{{\tt
  arXiv:2006.00361}}.

\bibitem{Fitzpatrick:2011dm}
A.~L. Fitzpatrick and J.~Kaplan, {\it {Unitarity and the Holographic
  S-Matrix}},  {\em JHEP} {\bf 10} (2012) 032,
  [\href{http://arxiv.org/abs/1112.4845}{{\tt arXiv:1112.4845}}].

\bibitem{Strassler:2008bv}
M.~J. Strassler, {\it {Why Unparticle Models with Mass Gaps are Examples of
  Hidden Valleys}},  \href{http://arxiv.org/abs/0801.0629}{{\tt
  arXiv:0801.0629}}.

\bibitem{Hatta:2008tx}
Y.~Hatta, E.~Iancu, and A.~H. Mueller, {\it {Jet evolution in the N=4 SYM
  plasma at strong coupling}},  {\em JHEP} {\bf 05} (2008) 037,
  [\href{http://arxiv.org/abs/0803.2481}{{\tt arXiv:0803.2481}}].

\end{thebibliography}\endgroup
\bibliographystyle{JHEP}

\end{document}